\documentclass[aps,pra,showpacs]{revtex4}
\usepackage[T1]{fontenc}
\usepackage[latin9]{inputenc}
\usepackage{textcomp}
\usepackage{mathrsfs}
\usepackage{amsmath}
\usepackage{amssymb}
\usepackage{graphicx}
\usepackage{esint}

\makeatletter

\DeclareRobustCommand{\greektext}{%
  \fontencoding{LGR}\selectfont\def\encodingdefault{LGR}}
\DeclareRobustCommand{\textgreek}[1]{\leavevmode{\greektext #1}}
\DeclareFontEncoding{LGR}{}{}
\DeclareTextSymbol{\~}{LGR}{126}
\newcommand{\lyxmathsym}[1]{\ifmmode\begingroup\def\b@ld{bold}
  \text{\ifx\math@version\b@ld\bfseries\fi#1}\endgroup\else#1\fi}

\@ifundefined{textcolor}{}
{%
 \definecolor{BLACK}{gray}{0}
 \definecolor{WHITE}{gray}{1}
 \definecolor{RED}{rgb}{1,0,0}
 \definecolor{GREEN}{rgb}{0,1,0}
 \definecolor{BLUE}{rgb}{0,0,1}
 \definecolor{CYAN}{cmyk}{1,0,0,0}
 \definecolor{MAGENTA}{cmyk}{0,1,0,0}
 \definecolor{YELLOW}{cmyk}{0,0,1,0}
}

\makeatother

\begin{document}

\title{Fermi Gases with Synthetic Spin-Orbit Coupling}

\author{Jing Zhang}

\affiliation{State Key Laboratory of Quantum Optics and Quantum Optics Devices,
Institute of Opto-Electronics, Shanxi University, Taiyuan 030006,
P. R. China}

\author{Hui Hu and Xia-Ji Liu}

\affiliation{Centre for Atom Optics and Ultrafast Spectroscopy, Swinburne University
of Technology, Melbourne 3122, Australia}

\author{Han Pu}

\affiliation{Department of Physics and Astronomy, and Rice Quantum Institute,
Rice University, Houston, TX 77251, USA}

\date{\today}
\begin{abstract}
We briefly review recent progress on ultracold atomic Fermi gases
with different types of synthetic spin-orbit coupling, including the
one-dimensional (1D) equal weight Rashba-Dresselhaus and two-dimensional
(2D) Rasbha spin-orbit couplings. Theoretically, we show how the single-body,
two-body and many-body properties of Fermi gases are dramatically
changed by spin-orbit coupling. In particular, the interplay between
spin-orbit coupling and interatomic interaction may lead to several
long-sought exotic superfluid phases at low temperatures, such as
anisotropic superfluid, topological superfluid and inhomogeneous superfluid.
Experimentally, only the first type - equal weight combination of
Rasbha and Dresselhaus spin-orbit couplings - has been realized very
recently using a two-photon Raman process. We show how to characterize
a normal spin-orbit coupled atomic Fermi gas in both non-interacting
and strongly-interacting limits, using particularly momentum-resolved
radio-frequency spectroscopy. The experimental demonstration of a
strongly-interacting spin-orbit coupled Fermi gas opens a promising
way to observe various exotic superfluid phases in the near future. 
\end{abstract}

\pacs{05.30.Fk, 03.75.Hh, 03.75.Ss, 67.85.-d}

\maketitle
\tableofcontents{}

\section{Introduction}

Modern physical theories describe reality in terms of fields, many
of which obey gauge symmetry. Gauge symmetry is the property of a field theory in which different configurations of the underlying fields --- which are not themselves directly observable --- result in identical observable quantities. Electromagnetism is an ideal example to illustrate this point. A system of stationary electric charges produces an electric field ${\bf E}$ (but no magnetic field). It is convenient to define a scalar potential $V$, a voltage, that is also determined by the charge distribution. The electric field at any position is given by the gradient of the scalar potential: ${\bf E}({\bf r})=\nabla V({\bf r})$. In this system, a global symmtry is readily perceived: if the scalar potential everywhere is changed by the same amount, i.e., $V({\bf r}) \rightarrow V({\bf r}) + V_0$, the resulting electric field is unchanged. A more non-trivial example is given by a system of moving charges which produces  both electric and magnetic field. In addition to the scalar potential, we now also introduce a vector potential ${\bf A}$, the curl of which gives the magnetic field: ${\bf B}({\bf r}) = \nabla \times {\bf A}({\bf r})$. This system obeys the local gauge sysmmetry: any local change in the scalar potential [$V({\bf r}) \rightarrow V({\bf r}) - \partial \Lambda /\partial t$ with $\Lambda({\bf r}, t)$ being an arbitrary function of position and time] can be combined with a compensating change in the vector potential [${\bf A}({\bf r}) \rightarrow {\bf A}({\bf r}) + \nabla \Lambda$] in such a way that the electric and magnetic fields are invariant.    

Maxwell's classical theory of electromagnetism is the first gauge theory with local symmetry. A related symmetry can be demonstrated in the quantum theory of electromagnetic interactions, which describes the interaction between charged
particles. From first sight, Maxwell's theory should not directly
describe the center-of-mass motion of neutral atoms. However, a beautiful
series of experiments carried out at NIST \cite{lin1,lin2,lin3} demonstrated
that artificial gauge fields can be generated in cold atomic vapours
using laser fields, such that neutral atoms can be used to simulate
charged particles moving in electromagnetic fields \cite{note}. How to engineer
artificial gauge fields is reviewed by Spielman in an article published
in the previous volume of this book series \cite{SpielmanReview}.

It would not be very interesting if all light-induced gauge fields
could do is to make neutral atoms mimic the behavior of charged particles.
Indeed, artificial gauge field can be made non-Abelian, i.e., the
Cartesian components of the field do not commute with each other.
By contrast, the familiar electromagnetic fields are Abelian since
their Cartesian components are represented by $c$-numbers, thus
commuting with each other. A special feature of non-Abelian gauge
field is that it can induce spin-orbit coupling. The concept of spin-orbit
coupling (SOC) is encountered, for example, in the study of atomic
structure, where the coupling between the electron's orbital
motion and its intrinsic spin gives rise to the fine structure of
atomic spectrum. In the current context, SOC refers to the coupling
between the internal pseudo-spin degrees of freedom and the external
motional degrees of freedom of the atom. That such SOC can be induced
by laser fields can be easily understood as follows: The laser light
induces transitions between atomic internal states, and in the meantime
imparts photon's linear momentum to the atom. Thus the internal and
the external degrees of freedom are coupled via their interaction
with the photon.

SOC in cold atoms was first realized in a system of $^{87}$Rb condensate
by the NIST group in 2011 \cite{lin4}. Since then, several groups
have achieved SOC in both bosonic \cite{Jing_bosonic,chen1,engels,chen2}
and fermionic quantum gases \cite{Jing_fermionic,MIT_fermionic,Jing_RF,Jing_FeshbachMolecule,NIST-fermionic}.
SOC not only dramatically changes the single-particle dispersion relation,
but is also the key ingredient underlying many interesting many-body
phenomena and new materials such as topological insulators \cite{TI}
and quantum spin Hall effects \cite{QSH}. Due to the exquisite controllability
of atomic systems, one can naturally expect that SOC in cold atoms
will give rise to novel quantum states of matter and may lead to a
deeper understanding of related phenomena in other systems. For this
reason, spin-orbit coupled quantum gases have received tremendous
attention over the past few years, and they no doubt represent one
of the most active frontiers of cold atom research.

In this chapter, we will review the physics of spin-orbit coupled
Fermi gas, both theoretically and experimentally. Although we will
mainly focus on the research from our own groups, results from others
will also be mentioned.

\section{Theory of spin-orbit coupled Fermi gas}

We consider a spin-1/2 Fermi gas with SOC subject to attractively
interaction between unlike spins. One great advantage of the atomic
system is its unprecedented controllability. The interatomic interaction
can be precisely tuned using the Feshbach resonance technique \cite{ChinRMP},
which has already led to the discovery of the BEC-BCS crossover
from a Bose-Einstein condensate (BEC) to a Bardeen-Cooper-Schrieffer
(BCS) superfluid \cite{GiorginiRMP}. Different forms of SOC, many
of which do not exist in natural materials, can also be engineered.
The interplay between interatomic interactions and different forms
of SOC may give rise to a number of intriguing physical phenomena.
Here let us make some general remarks concerning the distinct features
that can be brought out by SOC in a Fermi gas: 
\begin{itemize}
\item SOC alters the single-particle dispersion which may lead to degenerate
single-particle ground state, and may render the topology of the Fermi
surface non-trivial \cite{vj1}. 
\item In the presence of attractive $s$-wave interaction, two fermions
may form pairs. In general such pairs contain both singlet and triplet
components \cite{vj2,hu1,zhai1,jiang,vj3,dong1} and have anisotropic
(i.e., direction-dependent) effective mass \cite{hu1,zhai1,jiang}.
In the many-body setting, a spin-orbit coupled superfluid Fermi gas
contains both singlet and triplet pairing correlation \cite{vj1,hu1,jiang,sademelo1}
and therefore may be regarded as an anisotropic superfluid \cite{hu1}. 
\item SOC may greatly enhance the pairing instability and hence dramatically
increases the superfluid transition temperature \cite{hu1,zhai1,he1}. 
\item SOC, together with effective Zeeman fields, may generate exotic pairing
\cite{zheng1,ShenoyFF,zheng2,yi1,liu1,dong2,hu2,zhou,iskin1} and/or
topologically non-trivial superfluid state \cite{Zhang2008,zhang1,liang,Zhu2011,liu2,wan,sademelo2,liu3,he2,Wei2012,iskin2,zhang2,impurity,impurity1d,chen,zhang3,yi2,liu4}.
At the boundaries of topologically trivial and non-trivial regimes,
exotic quasi-particle states (e.g., Majorana mode) may be created. 
\end{itemize}
In the remaining part of this section, we will discuss two particular
types of SOC. The first is the equal-weight Rashba-Dresselhaus
SOC \cite{soc} which is the only one that has been experimentally realized so
far. The second is the Rashba SOC which is of particular interest
as it occurs naturally in certain semiconductor materials. However,
before we do that, in the next subsection we first summarize the theoretical
framework and explain the basics of momentum- or spatially-resolved
radio-frequency (rf) spectroscopy, which turns out to be a very useful
experimental tool for characterizing spin-orbit coupled interacting
Fermi gases. For those readers who are interested in the physical
consequences of a detailed type of SOC, this technical part may be
skipped in their first reading.

\subsection{Theoretical framework}
\label{general}

In current experimental setups of ultracold atomic Fermi gases, the
interactions between atoms are often tuned to be as strong as possible,
in order to have an experimentally accessible superfluid transition
temperature. With such strong interactions, there is a significant
portion of Cooper pairs formed by two fermionic atoms with unlike
spin. Theoretically, therefore, it is very crucial to treat atoms
and Cooper pairs on an equal footing. Without SOC, a minimum theoretical
framework for this purpose is the many-body \textit{T}-matrix theory
or pair-fluctuation theory \cite{sademeloNSR,randeria1,HLDEPL,randeria2,stoof,HLDNJP}.
In this subsection, we introduce briefly the essential idea of the
pair-fluctuation theory using the functional path-integral approach and
generalize the theory to include SOC \cite{jiang}. Under this theoretical
framework, both two- and many-body physics can be discussed in a unified
fashion \cite{jiang}. We also discuss the mean-field Bogoliubov-de
Gennes equation, which represents a powerful tool for the study of trapped,
inhomogeneous Fermi superfluids at low temperatures \cite{liu2,liu3,Wei2012,iskin2,impurity,impurity1d}.

\subsubsection{Functional path-integral approach}

Consider, for example, a three-dimensional (3D) spin-1/2 Fermi gas
with mass $m$. The second-quantized Hamiltonian reads, 
\begin{eqnarray}
\mathcal{H} & = & \int d\mathbf{r}\left[\psi^{\dag}\left(\hat{\xi}_{\mathbf{k}}+V_{{\rm SO}}\right)\psi+U_{0}\,\psi_{\uparrow}^{\dag}\left(\mathbf{r}\right)\psi_{\downarrow}^{\dag}\left(\mathbf{r}\right)\psi_{\downarrow}\left(\mathbf{r}\right)\psi_{\uparrow}\left(\mathbf{r}\right)\right],\label{eq: Hami}
\end{eqnarray}
where $\hat{\xi}_{\mathbf{k}}\equiv\hat{\mathbf{k}}^{2}/(2m)-\mu=-\nabla^{2}/(2m)-\mu$
with the chemical potential $\mu$, $\psi\left(\mathbf{r}\right)=[\psi_{\uparrow}\left(\mathbf{r}\right),\psi_{\downarrow}\left(\mathbf{r}\right)]^{T}$
describes collectively the fermionic annihilation operator $\psi_{\sigma}\left(\mathbf{r}\right)$
for spin-$\sigma$ atom, and $V_{{\rm SO}}(\mathbf{\hat{k}})$ represents
the spin-orbit coupling whose explicit form we do not specify here.
The momentum $\hat{k}_{\alpha}\equiv-i\partial_{\alpha}$ ($\alpha=x,y,z$)
should be regarded as the operators in real space. For notational
simplicity, we take $\hbar=1$ throughout this paper. The last term
in Eq.~(\ref{eq: Hami}) represents the two-body contact $s$-wave
interaction between unlike spins. The use of the contact interatomic
interaction leads to an ultraviolet divergence at large momentum or
high energy. To overcome such a divergence, we express the interaction
strength $U_{0}$ in terms of the \textit{s}-wave scattering length
$a_{s}$, 
\begin{equation}
\frac{1}{U_{0}}=\frac{m}{4\pi a_{s}}-\frac{1}{V}\sum_{{\bf k}}\frac{m}{\mathbf{k}^{2}},
\end{equation}
where $V$ is the volume of the system.

The partition function of the system can be written as \cite{stoof}
\begin{equation}
\mathcal{Z}=\int\mathcal{D}[\psi\left(\mathbf{r},\tau\right),\bar{\psi}\left(\mathbf{r},\tau\right)]\exp\left\{ -\mathcal{S}\left[\psi\left(\mathbf{r},\tau\right),\bar{\psi}\left(\mathbf{r},\tau\right)\right]\right\} ,
\end{equation}
where the action 
\begin{equation}
\mathcal{S}\left[\psi,\bar{\psi}\right]=\int_{0}^{\beta}d\tau\left[\int d\mathbf{r}\sum_{\sigma}\bar{\psi}_{\sigma}\left(\mathbf{r},\tau\right)\partial_{\tau}\psi_{\sigma}\left(\mathbf{r},\tau\right)+\mathcal{H}\left(\psi,\bar{\psi}\right)\right]\,.
\end{equation}
is written as an integral over imaginary time $\tau$. Here $\beta=1/(k_{B}T)$
is the inverse temperature and $\mathcal{H}\left(\psi,\bar{\psi}\right)$
is obtained by replacing the field operators $\psi^{\dag}$ and $\psi$
with the Grassmann variables $\bar{\psi}$ and $\psi$, respectively.
We can use the Hubbard-Stratonovich transformation to transform the
quartic interaction term into a quadratic form as: 
\begin{equation}
e^{-U_{0}\int d{\bf r} d\tau\bar{\psi}_{\uparrow}\bar{\psi}_{\downarrow}\psi_{\downarrow}\psi_{\uparrow}}=\int\mathcal{D}\left[\Delta,\bar{\Delta}\right]\,\exp\left\{ \int_{0}^{\beta}d\tau\int d\mathbf{r}\left[\frac{\left\vert \Delta\left(\mathbf{r},\tau\right)\right\vert ^{2}}{U_{0}}+\left(\bar{\Delta}\psi_{\downarrow}\psi_{\uparrow}\mathbf{+}\Delta\bar{\psi}_{\uparrow}\bar{\psi}_{\downarrow}\right)\right]\right\} \,,
\end{equation}
from which the pairing field $\Delta\left(\mathbf{r},\tau\right)$
is defined.

Let us now introduce the 4-dimensional Nambu spinor $\Phi\left(\mathbf{r,}\tau\right)\equiv\lbrack\psi_{\uparrow},\psi_{\downarrow}\mathbf{,}\bar{\psi}_{\uparrow},\bar{\psi}_{\downarrow}]^{T}$
and rewrite the action as, 
\begin{equation}
\mathcal{Z}=\int\mathcal{D}[\Phi,\bar{\Phi}\mathbf{;}\,\Delta,\bar{\Delta}]\,\exp\left\{ \int\! d{\bf r}' \!\!\int\! d\mathbf{r}\!\!\int_{0}^{\beta}\! d\tau'\!\!\int_{0}^{\beta}\! d\tau \left[\frac{1}{2}\bar{\Phi}(\mathbf{r},\tau)G^{-1}\Phi(\mathbf{r'},\tau')+\frac{\left\vert \Delta\right\vert ^{2}}{U_{0}}\delta(\mathbf{r}-\mathbf{r'})\delta(\tau-\tau')\right]-\beta\sum_{\mathbf{k}}\hat{\xi}_{\mathbf{k}}\right\} ,
\end{equation}
where the $4\times4$ single-particle Green function is given by,
\begin{equation}
G^{-1}=\left[\begin{array}{cc}
-\partial_{\tau}-\hat{\xi}_{\mathbf{k}}-V_{{\rm SO}}(\mathbf{\hat{k}}) & i\Delta\hat{\sigma}_{y}\\
-i\bar{\Delta}\hat{\sigma}_{y} & -\partial_{\tau}+\hat{\xi}_{\mathbf{k}}+V_{{\rm SO}}^{T}(-\mathbf{\hat{k}})
\end{array}\right]\delta(\mathbf{r}-\mathbf{r'})\delta(\tau-\tau')\,,\label{eq: GreenFunction}
\end{equation}
with the Pauli matrices $\hat{\sigma}_{i}$ ($i=0,x,y,z$) describing
the spin degrees of freedom. The Nambu spinor representation treats
equally the particle and the hole excitations. As a result, a zero-point
energy appears in the last term of the action. Integrating out the
original fermionic fields, we may rewrite the partition function as
\begin{equation}
\mathcal{Z}=\int\mathcal{D}[\Delta,\bar{\Delta}]\,\exp\left\{ -\mathcal{S}_{{\rm eff}}\left[\Delta,\bar{\Delta}\right]\right\} \,,
\end{equation}
where the effective action is given by 
\begin{eqnarray}
\mathcal{S}_{{\rm eff}}\left[\Delta,\bar{\Delta}\right] & = & \int_{0}^{\beta}d\tau\int d\mathbf{r}\left[-\frac{\left\vert \delta\Delta\left(\mathbf{r},\tau\right)\right\vert ^{2}}{U_{0}}\right]-\frac{1}{2}\text{Tr}\ln\left[-G^{-1}\right]+\beta\sum_{\mathbf{k}}\hat{\xi}_{\mathbf{k}}.
\end{eqnarray}
where the trace is taken over all the spin, spatial, and temporal degrees
of freedom.

To proceed, we restrict ourselves to the Gaussian fluctuation and expand $\Delta\left(\mathbf{r},\tau\right)=\Delta_{0}(\mathbf{r})+\delta\Delta\left(\mathbf{r},\tau\right)$.
The effective action is then decomposed accordingly as $\mathcal{S}_{{\rm eff}}=\mathcal{S}_{0}+\delta\mathcal{S}$,
where the saddle-point action is 
\begin{equation}
\mathcal{S}_{0}=-\int_{0}^{\beta}d\tau\int d\mathbf{r}\frac{\left|\Delta_{0}\left(\mathbf{r}\right)\right|^{2}}{U_{0}}-\frac{1}{2}\text{Tr}\ln\left[-G{}_{0}^{-1}\right]+\beta\sum_{\mathbf{k}}\hat{\xi}_{\mathbf{k}}\,\label{eq: S0}
\end{equation}
and the pair-fluctuating action takes the form 
\begin{equation}
\delta\mathcal{S}=\int_{0}^{\beta}d\tau\int d\mathbf{r}\left[-\frac{\left\vert \delta\Delta\left(\mathbf{r},\tau\right)\right\vert ^{2}}{U_{0}}+\frac{1}{2}\left(\frac{1}{2}\right)\text{Tr}\left(G{}_{0}\Sigma\right)^{2}\right]\,
\end{equation}
with 
\begin{equation}
\Sigma=\left(\begin{array}{cc}
0 & i\delta\Delta\hat{\sigma}_{y}\\
-i\delta\bar{\Delta}\hat{\sigma}_{y} & 0
\end{array}\right).
\end{equation}
Here $G_{0}^{-1}$ is the inverse mean-field Green function and has
the same form as $G^{-1}$ in Eq.~(\ref{eq: GreenFunction}) with
$\Delta(\mathbf{r},\tau)$ replaced by $\Delta_{0}(\mathbf{r})$.
We note that the static pairing field $\Delta_{0}(\mathbf{r})$ can
be either homogeneous or inhomogeneous. In the latter case, a typical
form is $\Delta_{0}(\mathbf{r})=\Delta_{0}e^{i\mathbf{q}\cdot\mathbf{r}}$,
referred to as the Fulde-Ferrell superfluid \cite{Fulde1964}, in which the Cooper pairs
condense into a state with nonzero center-of-mass momentum \textbf{q}.

Let us now focus on a homogeneous system, where the momentum is a
good quantum number so that we take $\xi_{\mathbf{k}}=\hat{\xi_{\mathbf{k}}}$
and $V_{{\rm SO}}(\mathbf{k})=V_{{\rm SO}}(\mathbf{\hat{k}})$. The
fluctuating part of the effective action may be formally written in
terms of the many-body particle-particle vertex function $\Gamma\left({\bf q},i\nu_{n}\right)$
\cite{stoof}, 
\begin{equation}
\delta\mathcal{S}=k_{B}T\sum_{Q=(\mathbf{q},i\nu_{n})}\left[-\Gamma^{-1}\left(Q\right)\right]\delta\Delta(Q)\delta\bar{\Delta}(Q)\,,
\end{equation}
where $Q\equiv(\mathbf{q},i\nu_{n})$ and $\nu_{n}$ is the bosonic
Matsubara frequency. By integrating out the quadratic term in $\delta S$,
we obtain the contribution from the Gaussian pair fluctuations to the
thermodynamic potential as \cite{stoof} 
\begin{equation}
\delta\Omega=k_{B}T\sum_{{\bf q},i\nu_{n}}\ln\left[-\Gamma^{-1}\left({\bf q},i\nu_{n}\right)\right].\label{eq:deltaOmega}
\end{equation}
Within the Gaussian pair fluctuation approximation, naïvely, the vertex
function may be interpreted as the Green function of ``Cooper pairs''.
This idea is supported by Eq. (\ref{eq:deltaOmega}), as the thermodynamic
potential $\Omega_{B}$ of a free bosonic Green function $\mathcal{G}_{B}$
is formally given by $\Omega_{B}=k_{B}T\sum_{{\bf q},i\nu_{n}}\ln[-\mathcal{G}_{B}^{-1}(\mathbf{q},i\nu_{n})]$.
At this point, the advantage of using pair-fluctuation theory becomes
evident. For the fermionic degree of freedom, we simply work out the
single-particle Green function $G_{0}$ and the related mean-field
thermodynamic potential $\Omega_{0}=k_{B}T\mathcal{S}_{0}$. An example
will be provided later on in the study of the Fulde-Ferrell superfluidity.
While for Cooper pairs, we calculate the vertex function and the fluctuating
thermodynamic potential $\delta\Omega$. In this way, we may obtain
a satisfactory description of strongly-interacting Fermi systems \cite{HLDEPL,randeria2,HLDNJP}.

In the normal state where the pairing field vanishes, i.e., $\Delta_{0}=0$,
we may obtain the explicit expression of the vertex function. In this
case, the inverse Green function $G_{0}^{-1}$ has a diagonal form
and can be easily inverted to give \cite{jiang}: 
\begin{equation}
G{}_{0}\left(K\right)=\left(\begin{array}{cc}
\left[i\omega_{m}-\xi_{\mathbf{k}}-V_{{\rm SO}}\left(\mathbf{k}\right)\right]^{-1} & 0\\
0 & \left[i\omega_{m}+\xi_{\mathbf{k}}+V_{{\rm SO}}^{T}\left(-\mathbf{k}\right)\right]{}^{-1}
\end{array}\right)\equiv\left[\begin{array}{cc}
\mathcal{G}_{0}\left(K\right) & 0\\
0 & \mathcal{\tilde{G}}_{0}\left(K\right)
\end{array}\right],
\end{equation}
where $K\equiv({\bf k},i\omega_{m})$ and $\omega_{m}$ is the fermionic
Matsubara frequency. Here we have introduced the $2\times2$ particle
Green function $\mathcal{G}_{0}\left(K\right)$ and hole Green function
$\mathcal{\tilde{G}}_{0}\left(K\right)$, which are related to each
other by $\tilde{{\cal G}_{0}}(K)=-[{\cal G}_{0}(-K)]^{T}$. It is
straightforward to show that,

\begin{equation}
\Gamma^{-1}\left(Q\right)=\frac{1}{U_{0}}+\frac{k_{B}T}{2}\sum_{K=(\mathbf{k},i\omega_{m})}\left[\mathcal{G}_{0}\left(K\right)\left(i\hat{\sigma}_{y}\right)\mathcal{\tilde{G}}_{0}\left(K-Q\right)\left(i\hat{\sigma}_{y}\right)\right].\label{eq:VertexFunction}
\end{equation}
The detailed expression of the vertex function depends on the type
of SOC. In the study of Rashba SOC, we will give an example that shows how
to calculate the vertex function.

\subsubsection{Two-particle physics from the particle-particle vertex function}

The vertex function can describe the pairing instability of Cooper
pairs both on the Fermi surface and in the vacuum. In the latter case,
it describes exactly the two-particle state. The corresponding two-body
inverse vertex function $\Gamma_{{\rm 2b}}^{-1}\left(Q\right)$ can
be obtained from the many-body inverse vertex function by discarding
the Fermi distribution function and by setting chemical potential
$\mu=0$ \cite{footnote1}. One important question concerning the
two-body state is whether there exist bound states. For a given momentum
$\mathbf{q}$, the bound state energy $E(\mathbf{q})$ can be determined
from the two-particle vertex function using the following relation
($i\nu_{n}\rightarrow\omega+i0^{+}$) \cite{hu1,jiang}: 
\begin{equation}
{\rm Re}\left\{ \Gamma_{{\rm 2b}}^{-1}\left[\mathbf{q};\omega=E\left(\mathbf{q}\right)\right]\right\} =0\,.\label{eq:EnergyEquationFrom2BVertexFunction}
\end{equation}
A true bound state must satisfy $E(\mathbf{q})$ $<2E_{{\rm min}}$
where $E_{{\rm min}}$ is the single-particle ground state energy.

It is straightforward but lengthy to calculate the two-particle vertex
function for any type of SOC. Here, we quote only the energy equation
obtained using Eq. (\ref{eq:EnergyEquationFrom2BVertexFunction})
for the most general form of SOC \cite{dong2}, 
\begin{equation}
V_{\textrm{SO}}\left(\mathbf{\hat{k}}\right)=\sum_{i=x,y,z}\left(\lambda_{i}\hat{k}_{i}+h_{i}\right)\hat{\sigma}_{i}\,,\label{eq: GeneralSOC}
\end{equation}
where $\lambda_{i}$ is the strength of SOC in the direction $i=(x,y,z)$
and $h_{i}$ denotes the effective Zeeman field. The eigenenergy $E(\mathbf{q})_{{\bf }}$
of a two-body eigenstate with momentum ${\bf q}$ satisfies the equation:
\begin{equation}
\frac{m}{4\pi a_{s}}=\frac{1}{V}\sum_{{\bf k}}\left[\left({\mathcal{E}_{{\bf k,q}}-\frac{4\mathcal{E}_{{\bf k,q}}^{2}({\bf \boldsymbol{\lambda}}\cdot{\bf k})^{2}-4\left[\sum_{i=x,y,z}\lambda_{i}k_{i}(\lambda_{i}q_{i}+2h_{i})\right]^{2}}{\mathcal{E}_{{\bf k,q}}\left[\mathcal{E}_{{\bf k,q}}^{2}-\sum_{i=x,y,z}(\lambda_{i}q_{i}+2h_{i})^{2}\right]}}\right)^{-1}+\frac{1}{2\epsilon_{{\bf k}}}\right]\,,\label{eq: Energy2B}
\end{equation}
where $\mathcal{E}_{{\bf k,q}}\equiv E_{{\bf }}(\mathbf{q})-\epsilon_{\mathbf{\frac{q}{2}}+{\bf k}}-\epsilon_{\mathbf{\frac{q}{2}}-{\bf k}}$
and $\epsilon_{{\bf k}}=k^{2}/(2m)$. We note that, in general, the
lowest-energy two-particle state may occur at a finite momentum
$\mathbf{q}$. That is, the two-particle bound state could have a
nonzero center-of-mass momentum. Later, we shall see that this unusual
property has nontrivial consequences in the many-body setting. Another
peculiar feature of the two-particle bound state is that the pairs
may have an effective mass larger than $2m$. For example, for the bound
state with zero center-of-mass momentum $\mathbf{q}=0$, it would have
a quadratic dispersion for small $\mathbf{p}$, 
\begin{equation}
E(\mathbf{p})=E(\mathbf{0})+\frac{p_{x}^{2}}{2M_{x}}+\frac{p_{y}^{2}}{2M_{y}}+\frac{p_{z}^{2}}{2M_{z}}.
\end{equation}
The effective mass of the bound state $M_{i}$ ($i=x,y,z$) can then be determined
directly from this dispersion relation.

Another approach to study the two-particle state with SOC , more familiar
to most readers, is to use the following ansatz for the two-particle
wave function \cite{vj2,zhai1,hu3,peng}, 
\begin{equation}
\left|\Phi_{2B}\right\rangle =\frac{1}{\sqrt{\mathcal{C}}}\sum_{{\bf k}}\left[\psi_{\uparrow\downarrow}\left({\bf k}\right)c_{\mathbf{\frac{q}{2}}+{\bf k}\uparrow}^{\dagger}c_{\mathbf{\frac{q}{2}}-{\bf k}\downarrow}^{\dagger}+\psi_{\downarrow\uparrow}\left({\bf k}\right)c_{\mathbf{\frac{q}{2}}+{\bf k}\downarrow}^{\dagger}c_{\mathbf{\frac{q}{2}}-{\bf k}\uparrow}^{\dagger}+\psi_{\uparrow\uparrow}\left({\bf k}\right)c_{\frac{\mathbf{q}}{2}+{\bf k}\uparrow}^{\dagger}c_{\frac{\mathbf{q}}{2}-{\bf k}\uparrow}^{\dagger}+\psi_{\downarrow\downarrow}\left({\bf k}\right)c_{\frac{\mathbf{q}}{2}+{\bf k}\downarrow}^{\dagger}c_{\frac{\mathbf{q}}{2}-{\bf k}\downarrow}^{\dagger}\right]\left|\text{vac}\right\rangle ,\label{eq: PhiB}
\end{equation}
where $c_{{\bf k}\uparrow}^{\dagger}$ and $c_{{\bf k}\downarrow}^{\dagger}$
are creation field operators of spin-up and spin-down atoms with momentum
${\bf k}$ and ${\cal C}$ is the normalization factor. We note that,
in the presence of SOC, the wave function of the two-particle state
has both spin singlet and triplet components. Then, using the Schrödinger
equation ${\cal H}\,\left|\Phi_{2B}(\mathbf{q})\right\rangle =E(\mathbf{q})\,\left|\Phi_{2B}(\mathbf{q})\right\rangle $,
we can straightforwardly derive the equations for coefficients $\psi_{\sigma\sigma'}$
appearing in the above two-body wave function and then the energy
equation for $E(\mathbf{q})$. For the general form of SOC, Eq. (\ref{eq: GeneralSOC}),
it leads to exactly the same energy equation (\ref{eq: Energy2B})
\cite{dong2}.

Each of the two approaches mentioned above has its own advantages. The
vertex function approach is useful to understand the relationship between the
two-body physics and the many-body physics. For example, it can be used
to obtain the two-particle bound state in the presence of a Fermi
surface. The latter approach of using the two-particle Schrödinger equation
naturally yields the two-particle wave function. Both approaches have
been used extensively in the literature.

\subsubsection{Many-body T-matrix theory}
\label{tm}

The functional path-integral approach gives the simplest version of
the many-body \textit{T}-matrix theory, where the \emph{bare} Green
function has been used in the vertex function. Here, for completeness,
we mention briefly another partially self-consistent \textit{T}-matrix
scheme for a normal spin-orbit coupled Fermi gas, by taking one bare
and one fully dressed Green function in the vertex function \cite{Jing_RF,he1}.
In this scheme, we have the Dyson equation, 
\begin{equation}
{\cal G}(K)=\left[{\cal G}_{0}^{-1}(K)-\Sigma(K)\right]{}^{-1},\label{eq:DysonEq}
\end{equation}
where the self-energy is given by 
\begin{equation}
\Sigma(K)=k_{B}T\sum_{Q=({\bf q},i\nu_{n})}t(Q)(i\hat{\sigma}_{y})\tilde{{\cal G}_{0}}(K-Q)(i\hat{\sigma}_{y})\label{eq:SelfEnergy}
\end{equation}
and $\tilde{{\cal G}_{0}}(K)\equiv-[{\cal G}_{0}(-K)]^{T}$. Here
$t(Q)\equiv U_{0}/[1+U_{0}\chi\left(Q\right)]$ is the (scalar) \textit{T}-matrix
with a two-particle propagator 
\begin{equation}
\chi\left(Q\right)=\frac{k_{B}T}{2}\sum_{K=({\bf k},i\omega_{m})}\text{Tr}\left[{\cal G}(K)\left(i\hat{\sigma}_{y}\right)\tilde{{\cal G}}_{0}(K-Q)\left(i\hat{\sigma}_{y}\right)\right],\label{eq:PairPropagator}
\end{equation}
where the trace is taken over the spin degree of freedom only. Note that
a fully self-consistent \textit{T}-matrix theory may also be obtained
by replacing in Eqs.~(\ref{eq:SelfEnergy}) and (\ref{eq:PairPropagator})
the bare Green function $\tilde{{\cal G}}_{0}(K-Q)$ with the fully
dressed Green function $\tilde{{\cal G}}(K-Q)$. We note also that
Eqs. (\ref{eq:DysonEq})-(\ref{eq:PairPropagator}) provide a natural
generalization of the well-known many-body \textit{T}-matrix theory
\cite{HLDNJP}, by including the effect of SOC, where the particle
or hole Green function, ${\cal G}(K)$ or $\tilde{{\cal G}}(K)$,
now becomes a $2\times2$ matrix.

In general, the partially self-consistent \textit{T}-matrix equations
are difficult to solve \cite{HLDNJP}. At a \emph{qualitative} level,
we may adopt a pseudogap decomposition advanced by the Chicago group
\cite{pseudogap} and approximate the \textit{T}-matrix $t(Q)=t_{sc}(Q)+t_{pg}(Q)$ to be the sum of two parts. Here $t_{sc}(Q)=-(\Delta_{sc}^{2}/T)\delta\left(Q\right)$ is the contribution from the superfluid with $\Delta_{sc}$ being the superfluid order parameter, and $t_{pg}(Q)$ represents the contribution from 
un-condensed pairs which give rise to a pseudogap
\begin{equation}
\Delta_{pg}^{2}\equiv-k_{B}T\sum_{Q\neq0}t_{pg}(Q) \,.
\end{equation}
The full pairing order parameter is given by $\Delta_{0}^{2}=\Delta_{sc}^{2}+\Delta_{pg}^{2}$.
Accordingly, we have the self-energy $\Sigma(K)=\Sigma_{sc}(K)+\Sigma_{pg}(K)$,
where 
\begin{equation}
\Sigma_{sc}=-\Delta_{sc}^{2}(i\sigma_{y})\tilde{{\cal G}_{0}}(K)(i\sigma_{y})
\end{equation}
and 
\begin{equation}
\Sigma_{pg}=-\Delta_{pg}^{2}(i\sigma_{y})\tilde{{\cal G}_{0}}(K)(i\sigma_{y}).
\end{equation}
We note that, at zero temperature the pseudogap approximation is simply
the standard mean-field BCS theory, in which $\Sigma(K)=-\Delta_{0}^{2}(i\sigma_{y})\tilde{{\cal G}_{0}}(K)(i\sigma_{y})$.
Above the superfluid transition, however, it captures the essential
physics of fermionic pairing and therefore should be regarded as an
improved theory beyond mean-field. To calculate the pseudogap $\Delta_{pg}$,
we approximate 
\[
t_{pg}^{-1}(Q\simeq0)={\cal Z}\left[i\nu_{n}-\Omega_{{\bf q}}+\mu_{pair}\right],
\]
where the residue ${\cal Z}$ and the effective dispersion of pairs
$\Omega_{{\bf q}}=q^{2}/2M^{*}$ are to be determined by expanding
$\chi\left(Q\right)$ about $Q=0$ in the case that the Cooper pairs
condense into a zero-momentum state. The form of $t_{pg}(Q)$ leads
to 
\[
\Delta_{pg}^{2}(T)={\cal Z}^{-1}\sum_{{\bf q}}f_{B}(\Omega_{{\bf q}}-\mu_{pair}),
\]
where $f_{B}(x)\equiv1/(e^{x/k_{B}T}-1)$ is the bosonic distribution
function. We finally obtain two coupled equations, the gap equation
$1/U_{0}+\chi\left(Q=0\right)={\cal Z}\mu_{pair}$ and the number
equation $n=k_{B}T\sum_{K}$Tr${\cal G}(K)$, from which the superfluid
order parameter $\Delta_{sc}$ and the chemical potential $\mu$ can be determined. This pseudogap
method has been used to study the thermodynamics and momentum-resolved
rf spectroscopy of interacting Fermi gases with different types of
SOC \cite{Jing_RF,he1}.

\subsubsection{Bogoliubov-de Gennes equation for trapped Fermi systems}

All cold atom experiments are performed with some trapping potentials, $V_{T}(\mathbf{r})$. For such inhomogeneous systems, it
is difficult to directly consider pair fluctuations. In most cases,
we focus on the mean-field theory by using the saddle-point thermodynamic
potential Eq. (\ref{eq: S0}) and minimizing it to determine the order
parameter $\Delta_{0}(\mathbf{r})$. This amounts to diagonalizing
the $4\times4$ single-particle Green function $G_{0}^{-1}(\mathbf{r},\tau;\mathbf{r}',\tau')$
with the standard Bogoliubov transformation, 
\begin{equation}
\alpha_{\eta}=\int d{\bf r} \sum_{\sigma}\left[u_{\sigma\eta}\left(\mathbf{r}\right)\psi_{\sigma}\left(\mathbf{r}\right)+\nu_{\sigma\eta}\left(\mathbf{r}\right)\psi_{\sigma}^{\dagger}\left(\mathbf{r}\right)\right],
\end{equation}
where $\alpha_{\eta}$ is the field operator for Bogoliubov quasiparticle
with energy $E_{\eta}$ and Nambu spinor wave function $\Phi_{\eta}(\mathbf{r})\equiv[u_{\uparrow\eta}\left(\mathbf{r}\right),u_{\downarrow\eta}\left(\mathbf{r}\right),v_{\uparrow\eta}\left(\mathbf{r}\right),v_{\downarrow\eta}\left(\mathbf{r}\right)]^{T}$,
which satisfies the following Bogoliubov-de Gennes (BdG) equation,
\begin{equation}
\left[\begin{array}{cc}
-\nabla^{2}/(2m)-\mu+V_{T}(\mathbf{r})+V_{{\rm SO}}(\mathbf{\hat{k}}) & -i\Delta_{0}\left(\mathbf{r}\right)\hat{\sigma}_{y}\\
i\Delta_{0}^{*}\left(\mathbf{r}\right)\hat{\sigma}_{y} & \nabla^{2}/(2m)+\mu-V_{T}(\mathbf{r})-V_{{\rm SO}}^{T}(-\mathbf{\hat{k}})
\end{array}\right]\Phi_{\eta}\left(\mathbf{r}\right)=E_{\eta}\Phi_{\eta}\left(\mathbf{r}\right).\label{bdgeq}
\end{equation}
The BdG Hamiltonian in the above equation includes the pairing gap
function $\Delta_{0}\left(\mathbf{r}\right)$ that should be determined
self-consistently. For this purpose, we may take the inverse Bogoliubov
transformation and obtain 
\begin{equation}
\psi_{\sigma}\left(\mathbf{r}\right)=\sum_{\eta}\left[u_{\sigma\eta}\left(\mathbf{r}\right)\alpha_{\eta}+\nu_{\sigma\eta}^{*}\left(\mathbf{r}\right)\alpha_{\eta}^{\dagger}\right].
\end{equation}
The gap function $\Delta_{0}\left(\mathbf{r}\right)=-U_{0}\left\langle \psi_{\downarrow}(\mathbf{r})\psi_{\uparrow}(\mathbf{r})\right\rangle $
is then given by, 
\begin{equation}
\Delta_{0}(\mathbf{r})=-\frac{U_{0}}{2}\sum_{\eta}\left[u_{\uparrow\eta}\left(\mathbf{r}\right)v_{\downarrow\eta}^{*}\left(\mathbf{r}\right)f\left(E_{\eta}\right)+u_{\downarrow\eta}\left(\mathbf{r}\right)v_{\uparrow\eta}^{*}\left(\mathbf{r}\right)f\left(-E_{\eta}\right)\right],\label{gapeq}
\end{equation}
where $f\left(E\right)\equiv 1/[e^{ E/(k_BT)}+1]$ is the Fermi distribution
function at temperature $T$. Accordingly, the total density takes
the form, 
\begin{equation}
n\left(\mathbf{r}\right)=\frac{1}{2}\sum_{\sigma\eta}\left[\left|u_{\sigma\eta}\left(\mathbf{r}\right)\right|^{2}f\left(E_{\eta}\right)+\left|v_{\sigma\eta}\left(\mathbf{r}\right)\right|^{2}f\left(-E_{\eta}\right)\right].\label{numeq}
\end{equation}
The chemical potential $\mu$ can be determined using the number equation,
$N=\int d\mathbf{r}n\left(\mathbf{r}\right)$. 
This BdG approach has been used to investigate topological superfluids
in harmonically trapped spin-orbit coupled Fermi gases in 1D and 2D
\cite{liu2,liu3,Wei2012,iskin2,impurity,impurity1d}. It will be discussed
in greater detail in later sections.

It is important to note that, the use of Nambu spinor representation
enlarges the Hilbert space of the system. As a result, there is an
intrinsic particle-hole symmetry in the Bogoliubov solutions: For
any ``particle'' solution with wave function $\Phi_{\eta}^{(p)}(\mathbf{r})=[u_{\uparrow\eta}\left(\mathbf{r}\right),u_{\downarrow\eta}\left(\mathbf{r}\right),v_{\uparrow\eta}\left(\mathbf{r}\right),v_{\downarrow\eta}\left(\mathbf{r}\right)]^{T}$
and energy $E_{\eta}^{(p)}\geq0$, we can always find a partner
``hole'' solution with wave function $\Phi_{\eta}^{(h)}(\mathbf{r})=[v_{\uparrow\eta}^{*}\left(\mathbf{r}\right),v_{\downarrow\eta}^{*}\left(\mathbf{r}\right),u_{\uparrow\eta}^{*}\left(\mathbf{r}\right),u_{\downarrow\eta}^{*}\left(\mathbf{r}\right)]^{T}$
and energy $E_{\eta}^{(h)}=-E_{\eta}^{(p)}\leq0$. These two solutions
correspond exactly to the same physical state. To remove this redundancy,
we have added an extra factor of 1/2 in the expressions for pairing
gap function Eq. (\ref{gapeq}) and total density Eq. (\ref{numeq}).
As we shall see, this particle-hole symmetry is essential to the understanding of
the appearance of exotic Majorana fermions - particles that are their
own antiparticles - in topological superfluids.

\subsubsection{Momentum- or spatially-resolved radio-frequency spectrum}
\label{rf}

Radio-frequency (rf) spectroscopy, including both momentum-resolved
and spatially-resolved rf-spectroscopy, is a powerful tool to characterize
interacting many-body systems. It has been widely used to study fermionic
pairing in a two-component atomic Fermi gas near Feshbach resonances
in the BEC-BCS crossover \cite{Chin2004,Achirotzek2008,Schunck2008,Stewart2008,Zhang2012}.
Most recently, it has also been used to detect new quasiparticles
known as repulsive polarons \cite{Kohstall2012,Koschorreck2012},
which occur when ``impurity'' fermionic particles interact repulsively
with a fermionic environment.

The underlying mechanism of rf-spectroscopy is rather simple. The rf field drives transitions
between one of the hyperfine states (say, $\left|\downarrow\right\rangle $)
and an empty hyperfine state $\left|3\right\rangle $ which lies above
it by an energy $\omega_{3\downarrow}$. The Hamiltonian describing this rf-coupling
may be written as, 
\begin{equation}
{\cal V}_{\textrm{rf}}=V_{0}\int d{\bf r}\left[\psi_{3}^{\dagger}\left({\bf r}\right)\psi_{\downarrow}\left({\bf r}\right)+\psi_{\downarrow}^{\dagger}\left({\bf r}\right)\psi_{3}\left({\bf r}\right)\right],
\end{equation}
where 
$V_{0}$ is the strength of the rf drive. For a weak rf field, the
number of transferred atoms may be calculated using linear response
theory. At this point, it is important to note that a final state
effect might be present, which is caused by the interaction
between atoms in the final third state and those in the initial spin-up
or spin-down state. This final state effect is significant for $^{6}$Li
atoms; while for $^{40}$K atoms, it is not important \cite{GiorginiRMP}.

For momentum-resolved rf spectroscopy \cite{Stewart2008}, the momentum
distribution of the transferred atoms can be obtained by absorption imaging after
a time-of-flight. This gives rise to the information about the single-particle
spectral function of spin-down atoms of the original Fermi system,
${\cal A}_{\downarrow\downarrow}(\mathbf{k},\omega)$. In the absence
of the final-state effect, the rf transfer strength $\Gamma({\bf k},\omega)$
at a given momentum is given by, 
\begin{equation}
\Gamma({\bf k},\omega)={\cal A}_{\downarrow\downarrow}({\bf k},\epsilon_{{\bf k}}-\mu-\omega+\omega_{3\downarrow})f(\epsilon_{{\bf k}}-\mu-\omega+\omega_{3\downarrow})\,.
\end{equation}
Here, we have assumed that the atoms in the third state
have the dispersion relation $\epsilon_{{\bf k}}=k^{2}/(2m)$ in free
space and have taken the coupling strength $V_{0}=1$. Experimentally,
we can either measure the momentum-resolved rf spectroscopy along
a particular direction, say, the \textit{x}-direction, by integrating along the two perpendicular directions
\begin{equation}
\Gamma(k_{x},\omega)\equiv\sum_{k_{y},k_{z}}\Gamma({\bf k},\omega),
\end{equation}
or after integrating along the remaining direction, obtain the fully integrated rf spectrum $\Gamma(\omega)\equiv\sum_{{\bf k}}\Gamma({\bf k},\omega)$.
We note that, in the extremely weakly interacting BCS and BEC regimes,
where the physics is dominated by single-particle or two-particle
physics, respectively, we may use the Fermi golden rule to calculate
the momentum-resolved rf spectroscopy. This will be discussed in greater
detail in the relevant subsections. We note also that momentum-resolved
rf spectroscopy is precisely an ultracold atomic analogue of the well-known
angle-resolved photoemission spectroscopy (ARPES) widely used in solid-state
experiments.

Alternatively, we may use rf spectroscopy to probe the local information
about the original Fermi system. This was first demonstrated in measuring
the pairing gap by using phase-contrast imaging within the local density
approximation for a trapped Fermi gas \cite{Achirotzek2008}. A
more general idea is to use a specifically designed third state, which
has a very flat dispersion relation \cite{Jiang2011}. This leads
to a spatially-resolved rf spectroscopy, which measures precisely
the local density of states of the Fermi system, 
\begin{equation}
\rho_{\sigma}\left(\mathbf{r},\omega\right)=\frac{1}{2}\sum_{\eta}\left[\left|u_{\sigma\eta}\left(\mathbf{r}\right)\right|^{2}\delta\left(\omega-E_{\eta}\right)+\left|v_{\sigma\eta}\left(\mathbf{r}\right)\right|^{2}\delta\left(\omega+E_{\eta}\right)\right].\label{eq: ldos}
\end{equation}
It could be regarded as a cold-atom scanning tunneling microscopy
(STM). As we shall see, the spatially-resolved rf spectroscopy will
provide a useful although indirect measurement of the long-sought
Majorana fermion in atomic topological superfluids.

\subsection{1D equal-weight Rashba-Dresselhaus spin-orbit coupling}

\begin{figure}[h]
\begin{centering}
\includegraphics[clip,width=0.7\textwidth]{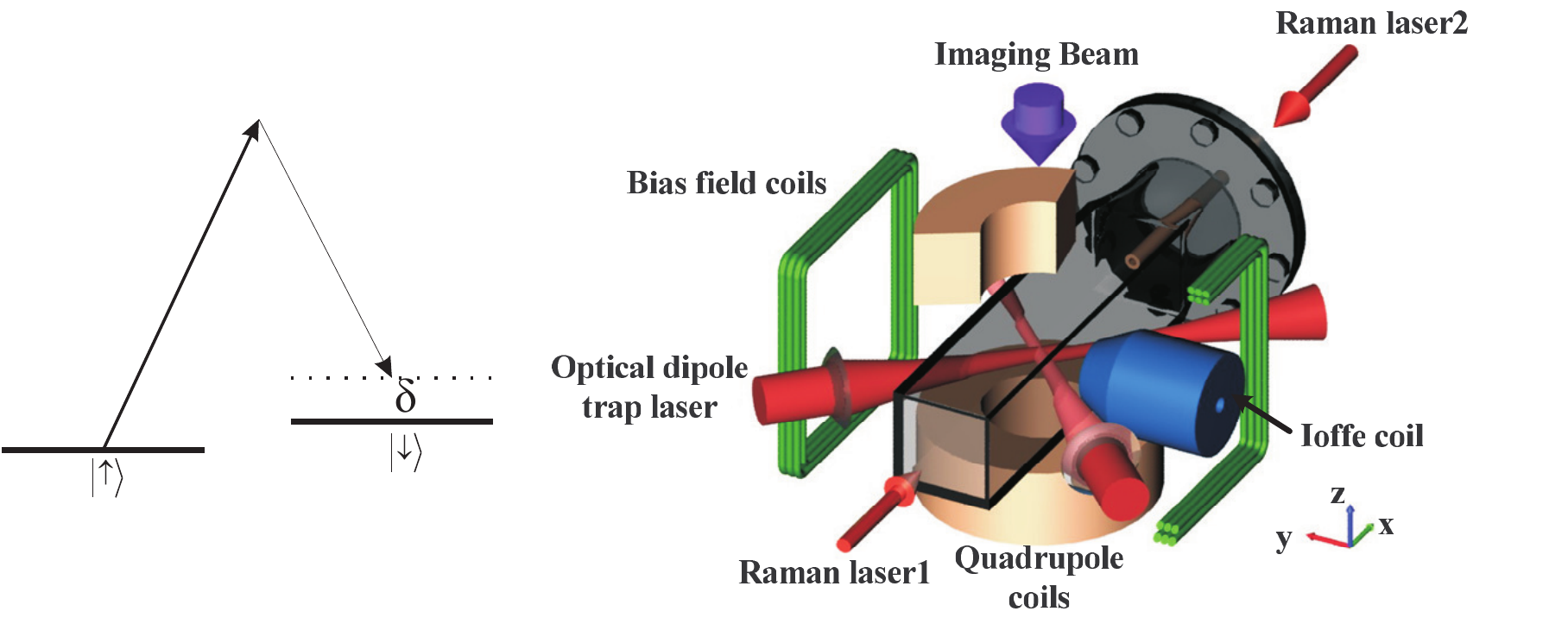} 
\par\end{centering}

\caption{ Left panel: schematic of the Raman transition that produced the equal-weight
Rashba-Dresselhaus SOC. The two atomic states are labeled as $|\uparrow\rangle$
and $|\downarrow\rangle$. $\delta$ is the two-photon Raman detuning.
Right panel: schematic of the experimental setup where a pair of Raman
beams counter-propagate along the ${x}$-axis. Right figure taken
from Ref. \cite{Jing_fermionic}.}

\label{fig1} 
\end{figure}

Let us now discuss the two specific types of SOC. One simple scheme to
create SOC in cold atoms is through a Raman transition that couples
two hyperfine ground states of the atom, as schematically shown in
Fig. \ref{fig1}. The Raman process is described
by the following single-particle Hamiltonian in the first-quantization
representation 
\begin{equation}
\mathcal{H}_{0}=\frac{{\bf \hat{p}^{2}}}{2m}+\frac{1}{2}\left(\begin{array}{cc}
\delta & \Omega\, e^{i2k_{r}x}\\
\Omega\, e^{-i2k_{r}x} & -\delta
\end{array}\right),
\end{equation}
where ${\bf \hat{p}}$ is the momentum operator of the atom, $2k_{r}\hat{x}$
is the photon recoil momentum taken to be along the \textit{x}-axis, $\delta$
and $\Omega$ are the two-photon detuning and the coupling strength
of the Raman beams, respectively. The Hamiltonian acts on the Hilbert
space expanded by the spin-up and spin-down basis, $\left|\uparrow\right\rangle $
and $\left|\downarrow\right\rangle $. By applying a unitary transformation with
\begin{equation}
U=\left(\begin{array}{cc}
e^{ik_{r}x} & 0\\
0 & e^{-ik_{r}x}
\end{array}\right)\,,\label{gauge}
\end{equation}
the Hamiltonian $\mathcal{H}_{0}$ can be recast into the following
form: 
\begin{equation}
\mathcal{H}_{\textrm{SO}}=U^{\dagger}\mathcal{H}_{0}U=\frac{\left(\hat{k}_{x}+k_{r}\hat{\sigma}_{z}\right)^{2}}{2m}+\frac{\left(\hat{k}_{y}^{2}+\hat{k}_{z}^{2}\right)}{2m}+\frac{\Omega}{2}\hat{\sigma}_{x}+\frac{\delta}{2}\hat{\sigma}_{z}\,.\label{HSO}
\end{equation}
Here, $\hat{\bf k}=(\hat{k}_{x}, \hat{k}_y, \hat{k}_z)$ denotes the quasi-momentum operator of the atom: When $\hat{\bf k}$ is applied to the transformed wave function, it gives the atomic quasi-momentum ${\bf k}$ that is related to the real momentum ${\bf p}$ 
as $\hat{\bf p}=(\hat{\bf k}\pm k_{r} \hat{x})$ with $\pm$ for spin-up and
down, respectively. From this expression, it is sometimes convenient
to regard both $\Omega$ and $\delta$ as the strengths of effective
Zeeman fields.

We note that after a pseudo-spin rotation ($\sigma_{z}\rightarrow\sigma_{x}$,
$\sigma_{x}\rightarrow-\sigma_{z}$), Hamiltonian (\ref{HSO}) can
be cast into the general form of SOC in Eq. (\ref{eq: GeneralSOC})
with ${\bf \boldsymbol{\lambda}}=(k_{r}^{2}/m,0,0)$ and ${\boldsymbol{h}}=(\delta/2,0,-\Omega/2)$.
It is clear that the SOC is along a specific direction. Actually,
it is an equal-weight combination of the well-known Rashba and Dresselhaus
SOCs in solid-state physics \cite{soc}. For this reason, hereafter we would refer
to it as 1D equal-weight Rashba-Dresselhaus SOC. We may also refer
to the detuning $\delta$ as the in-plane Zeeman field since it is
aligned along the same direction as the SOC. Accordingly, we call the
coupling strength $\Omega$ as the out-of-plane Zeeman field. As we
shall see, depending on $\delta$ and $\Omega$, the spin-orbit coupled
Fermi system can display distinct quantum superfluid phases at low
temperatures.

\subsubsection{Single-particle spectrum}
\label{sp}

\begin{figure}[h]
\begin{centering}
\includegraphics[clip,width=0.7\textwidth]{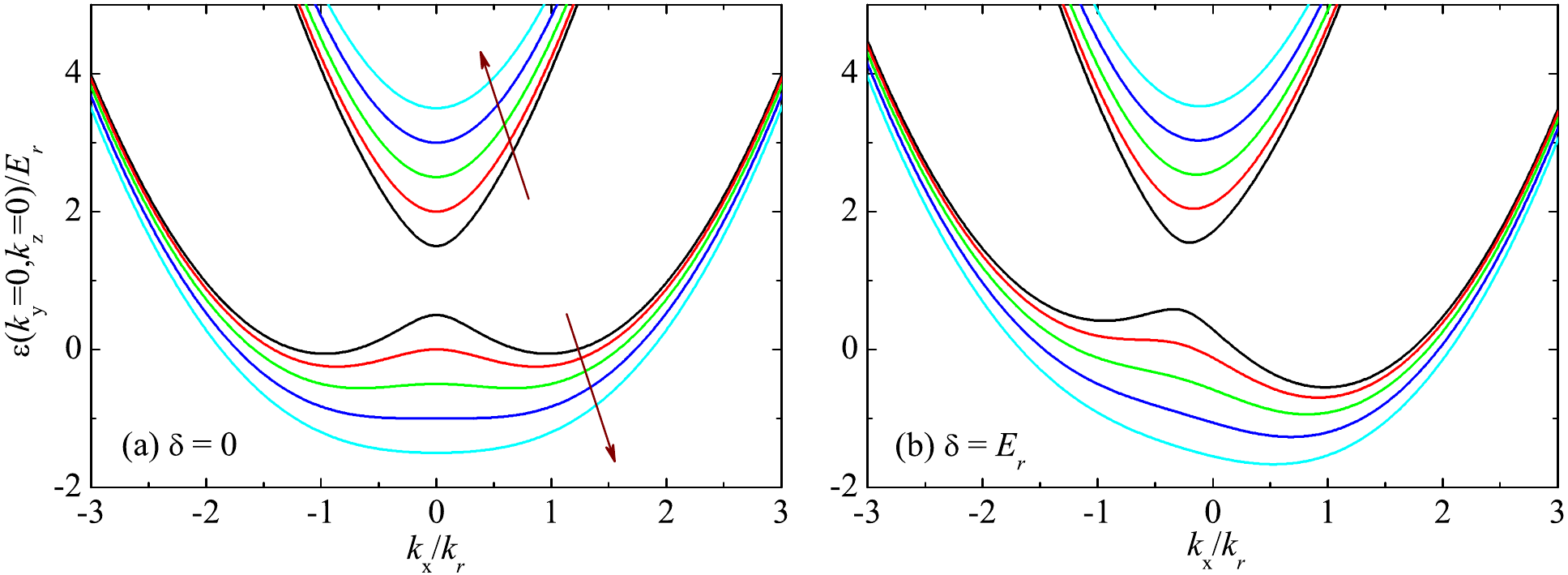} 
\par\end{centering}

\caption{ Single particle spectrum of a Fermi gas with 1D equal-weight Rashba-Dresselhaus
SOC, with (a) or without detuning (b). In each panel, we increase
the coupling strength of the Raman beams from $E_{r}$ to $5E_{r}$,
with a step of $E_{r}$, as indicated by the arrows. }

\label{fig2} 
\end{figure}

The single-particle spectrum can be easily obtained by diagonalizing
the Hamiltonian (\ref{HSO}), which is given by 
\begin{equation}
E{}_{\mathbf{k}\pm}=E_{r}+\frac{\mathbf{k}^{2}}{2m}\pm\sqrt{\left(\frac{\Omega}{2}\right)^{2}+\left(\lambda k_{x}+\frac{\delta}{2}\right)^{2}},
\end{equation}
where we have defined a recoil energy $E_{r}\equiv k_{r}^{2}/(2m)$
and an SOC strength $\lambda\equiv k_{r}/m$. The spectrum contains
two branches as shown in Fig. \ref{fig2}. For small $\Omega$, the
lower branch exhibits a double-well structure. The double wells are
symmetric (asymmetric) for $\delta=0$ ($\delta\neq0$). For large
$\Omega$, the two wells in the lower branch merge into a single one.
It is important to emphasize that in each branch atoms stay at a mixed
spin state with both spin-up and down components.

\begin{figure}[h]
\begin{centering}
\includegraphics[clip,width=0.35\textwidth]{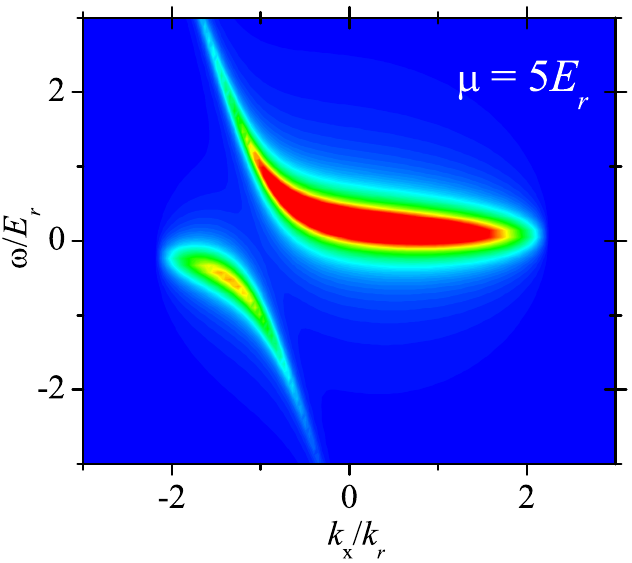}
\includegraphics[clip,width=0.35\textwidth]{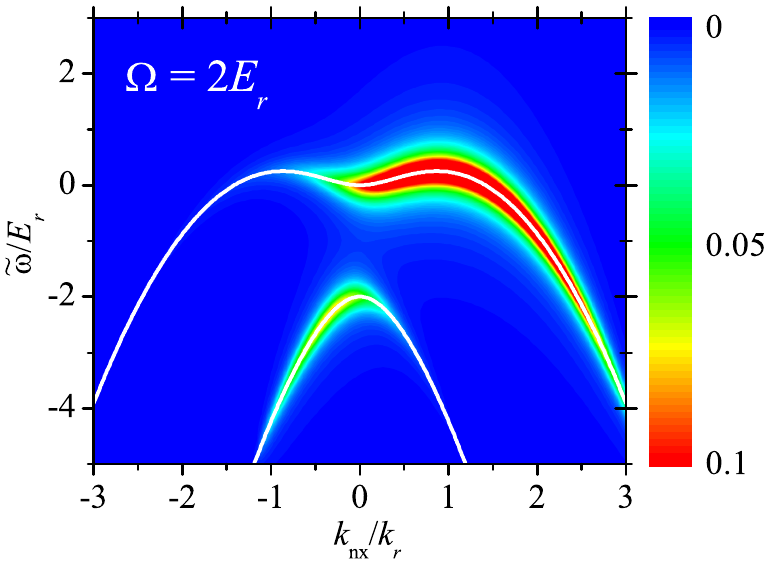} 
\par\end{centering}

\caption{ Theoretical simulation on momentum-resolved rf spectroscopy of a
Fermi gas with 1D equal-weight Rashba-Dresselhaus SOC. Left panel:
simulated experimental spectroscopy $\Gamma(k_{x},\omega)$. Right
panel: the spectroscopy $\Gamma(k_{nx}\equiv k_{x}+k_{r},\tilde{\omega}=\omega+k_{x}^{2}/2m)$.
Here, the intensity of the contour plot shows the number of transferred
atoms, increasingly linearly from 0 (blue) to its maximum value (red).
We have set $\omega_{3\downarrow}=0$ and used a Lorentzian distribution
to replace the Delta function. Figure taken from Ref. \cite{liu5}
with modification. }

\label{fig3} 
\end{figure}

The single-particle spectrum can be easily measured by using momentum-resolved
rf spectroscopy, as already shown at Shanxi University and MIT \cite{Jing_fermionic,MIT_fermionic}.
In this case, the number of transferred atoms can be calculated by
using the Fermi's golden rule \cite{liu5}: 
\begin{equation}
\Gamma\left(k_{x},\omega\right)=\sum_{i,f}\left|\left\langle \Phi_{f}\right|{\cal V}_{rf}\left|\Phi_{i}\right\rangle \right|^{2}f\left(E_{i}-\mu\right)\delta\left[\omega-\omega_{3\downarrow}-\left(E_{f}-E_{i}\right)\right],\label{eq: spRF}
\end{equation}
where the summation is over all possible initial single-particle
states $\Phi_{i}$ (with energy $E_{i}$ and a given wavevector $k_{x}$)
and final states $\Phi_{f}$ (with energy $E_{f}$), and the Dirac
$\delta$-function ensures energy conservation during the rf transition. In practice, the $\delta$-function is replaced by a function with finite width (e.g., $\delta(x) \rightarrow (\gamma/\pi) (x^2 + \gamma^2)^{-1}$ where $\gamma$ accounts for the energy resolution of the measurement). 
The single-particle wave function $\Phi_{i}$ is known from the diagonalization
of the Hamiltonian (\ref{HSO}) and the transfer element $\left\langle \Phi_{f}\right|{\cal V}_{rf}\left|\Phi_{i}\right\rangle $
is then easy to determine. The left panel of Fig. \ref{fig3} shows
the predicted momentum-resolved spectroscopy $\Gamma\left(k_{x},\omega\right)$
at $\delta=0$ and $\Omega=2E_{r}$. The chemical potential is tuned
($\mu=5E_{r}$) in such a way that there are significant populations
in both energy branches. The simulated spectrum is not straightforward
to understand, because of the final free-particle dispersion relation
in the energy conservation in Eq. (\ref{eq: spRF}) and also the recoil
momentum shift ($k_{r}$) arising from the unitary transformation Eq. (\ref{gauge}).
Therefore, it is useful to define 
\begin{equation}
\tilde{\Gamma}\left(k_{nx},\tilde{\omega}\right)\equiv\Gamma\left(k_{x}+k_{r},\omega+\frac{k_{x}^{2}}{2M}\right),
\end{equation}
for which, the energy conservation takes the form $\delta[\tilde{\omega}+E_{i}(k_{x})]$.
As shown on the right panel of Fig. \ref{fig3}, the single-particle
spectrum is now clearly visible.

\begin{figure}[h]
\begin{centering}
\includegraphics[clip,width=0.35\textwidth]{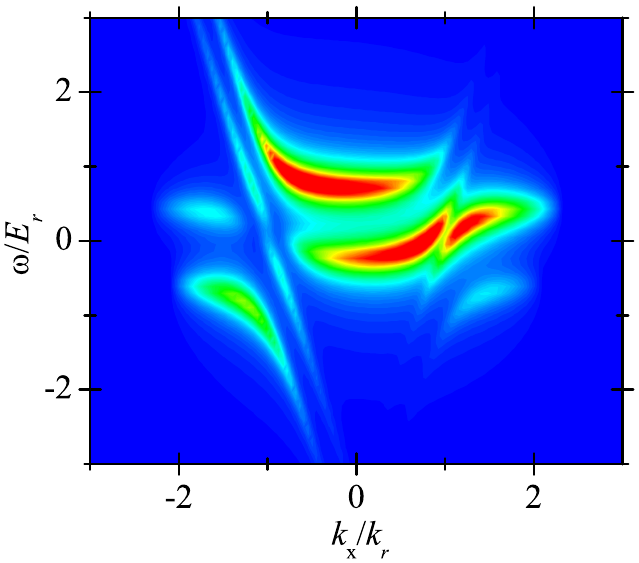}
\includegraphics[clip,width=0.35\textwidth]{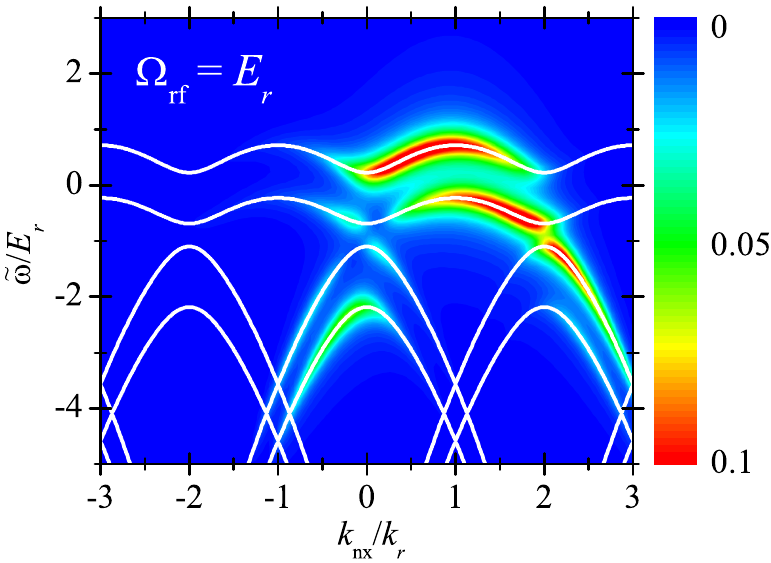} 
\par\end{centering}

\caption{Theoretical simulation on momentum-resolved rf spectroscopy of a Fermi
gas with 1D equal-weight Rashba-Dresselhaus SOC and an additional
spin-orbit lattice. The left and right panels show $\Gamma(k_{x},\omega)$
and $\Gamma(k_{nx}\equiv k_{x}+k_{r},\tilde{\omega}=\omega+k_{x}^{2}/2m)$,
respectively. The white lines on the right panel are the calculated
energy band structure. The spin-orbit lattice depth is $\Omega_{\textrm{rf}}=E_{r}$
and the other parameters are the same as in Fig. \ref{fig3}. Figure
taken from Ref. \cite{liu5} with modification. }

\label{fig4} 
\end{figure}

Experimentally, the single-particle properties of the Fermi gas
can also be easily tuned, for example, by using an additional rf field
to couple spin-up and down states \cite{MIT_fermionic}. After the
gauge transformation, it introduces a term $(\Omega_{\mathscr{rf}}/2)[\cos(2k_{r}x)\hat{\sigma}_{x}+\sin(2k_{r}x)\hat{\sigma}_{y}]$
in the spin-orbit Hamiltonian Eq.~(\ref{HSO}), which behaves like
a spin-orbit lattice and leads to the formation of energy bands.
In Fig.~\ref{fig4}, we show the simulation of momentum-resolved
rf spectroscopy under such an rf spin-orbit lattice. The energy band
structure is apparent. We refer to Ref. \cite{liu5} for more details
on the theoretical simulations, in particular the simulations in a
harmonic trap. The relevant measurements will be discussed in greater
detail later in the section on experiments.

\subsubsection{Two-body physics }
\label{2b}

We now turn to consider the interatomic interaction. The interplay
between interatomic interaction and SOC can lead to a number of intriguing
phenomena, even at the two-particle level. Let us first solve numerically
the energy $E(\mathbf{q})$ of the two-particle states by using the
general eigenenergy equation Eq. (\ref{eq: Energy2B}). A true bound
state must satisfy $E(\mathbf{q})<2E_{{\rm min}}$, where $E_{{\rm min}}$
is the single-particle ground state energy.

\begin{figure}[h]
\begin{centering}
\includegraphics[clip,width=0.35\textwidth]{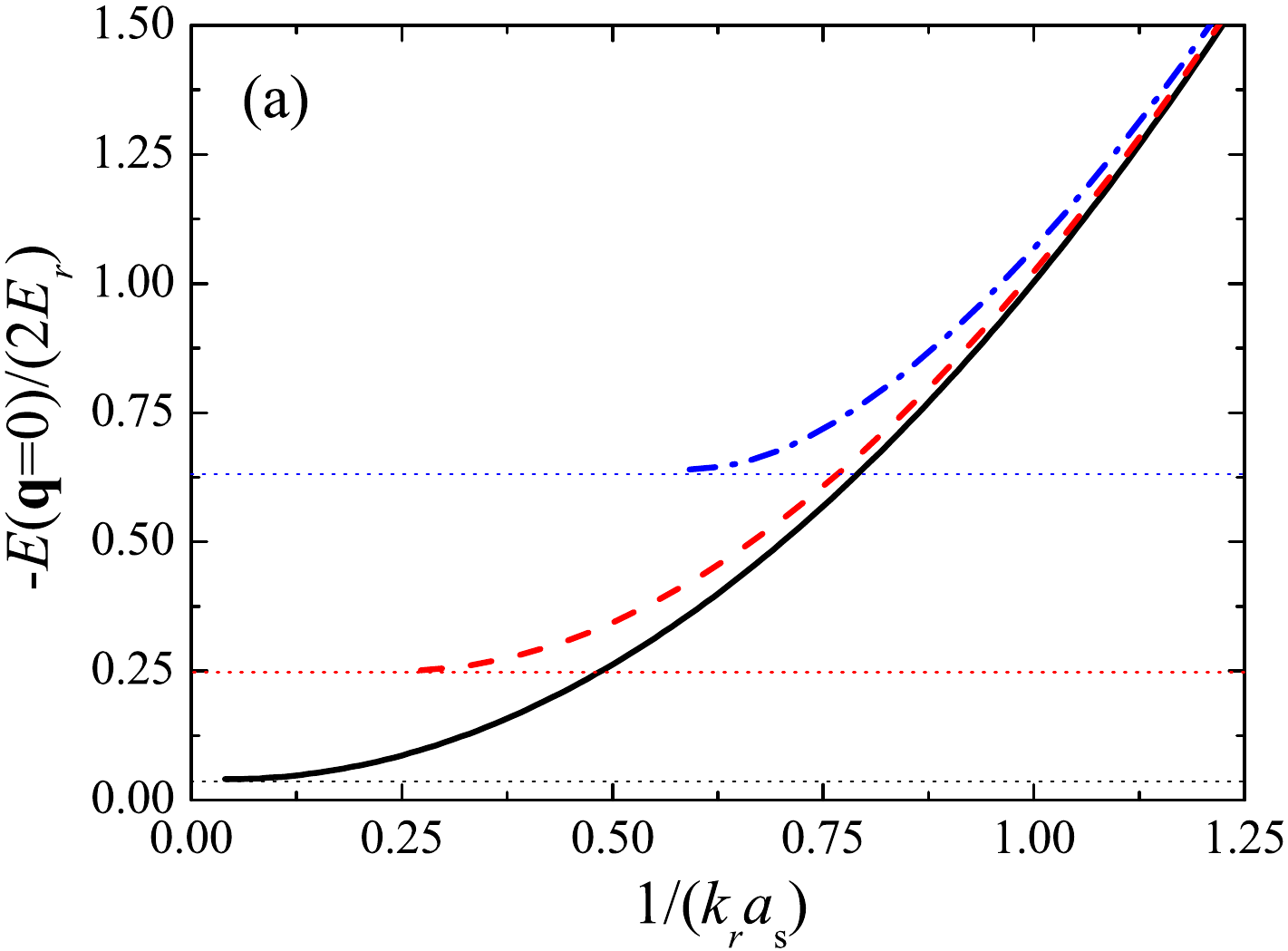} \includegraphics[clip,width=0.35\textwidth]{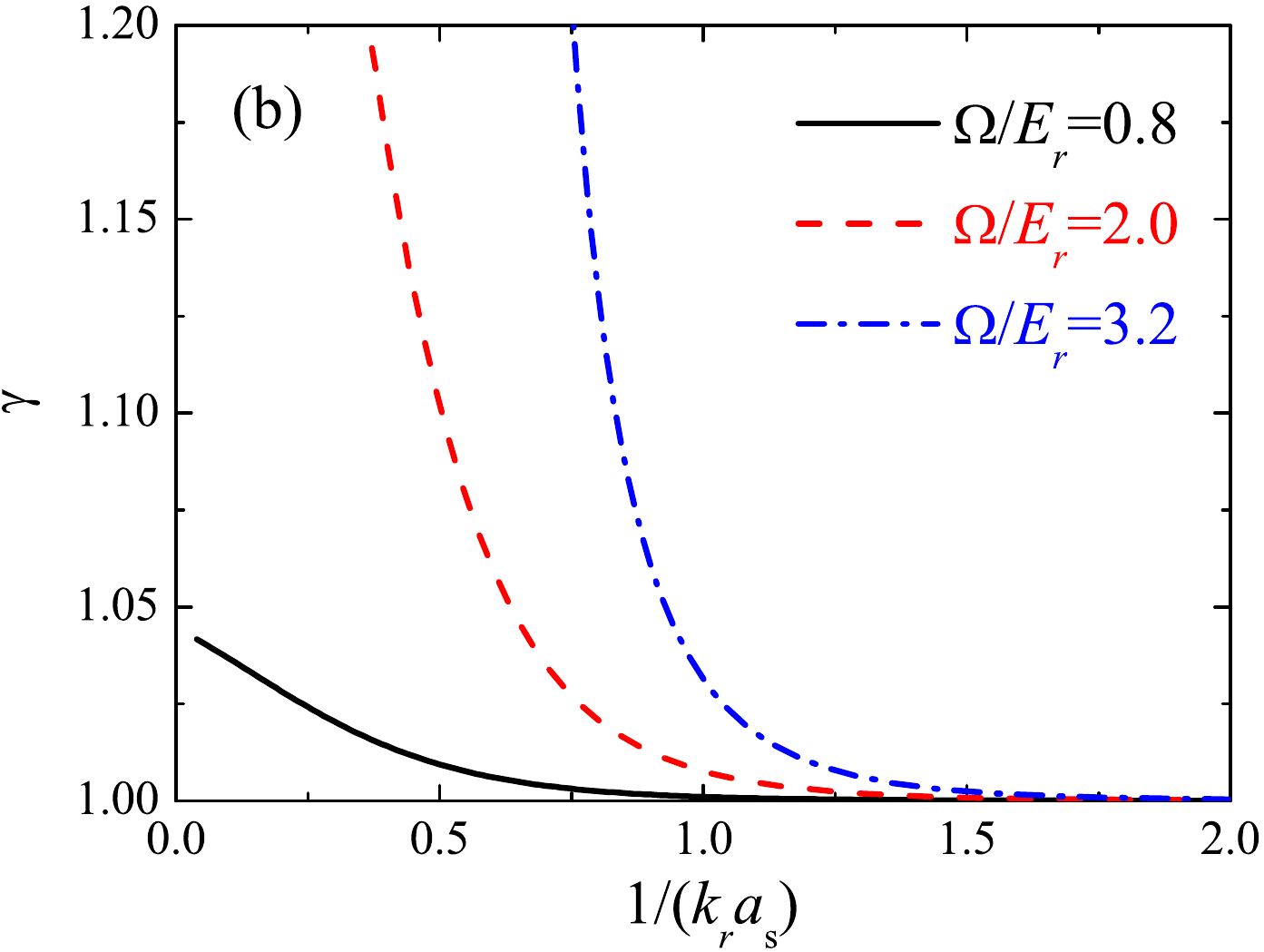} 
\par\end{centering}

\caption{ Energy $-E(q=0)$ (a) and effective mass ratio $\gamma=M_{x}/(2m)$
(b) of the two-particle ground bound state in the presence of 1D equal-weight
Rashba-Dresselhaus SOC, at zero detuning $\delta=0$ and at three
coupling strengths of Raman beams: $\Omega=0.8E_{r}$ (solid line),
$2E_{r}$ (dashed line), and $3.2E_{r}$ (dot-dashed line). The horizontal
dotted lines in (a) correspond to the threshold energies $-2E_{\textrm{min}}$
where the bound states disappear. Figure taken from Ref. \cite{peng}
with modification.}

\label{fig5} 
\end{figure}

At zero detuning $\delta=0$, the two-particle ground state has 
zero center-of-mass momentum $\mathbf{q}=0$ \cite{peng}. In Fig.
\ref{fig5}(a), we show its energy as a function of the dimensionless
interaction parameter $1/(k_{r}a_{s})$. In the presence of 1D equal-weight
Rashba-Dresselhaus SOC, a two-particle bound state occurs on the BEC
side with a positive \textit{s}-wave scattering length $a_{s}>0$.
The effective out-of-plane Zeeman field $\Omega$ acts as a pair-breaker
and pushes the threshold scattering length to the BEC limit. In other
words, the position of the Feshbach resonance, originally located at
$a_{s}=\pm\infty$, now shifts to the BEC side with at lower 
magnetic field strengths \cite{NIST-fermionic}. By calculating the dispersion
relation $E(\mathbf{q})$ around $\mathbf{q}=0$, we are able to determine
the effective mass, as shown in Fig. \ref{fig5}(b). It is interesting
that the effective mass along the direction of SOC is greatly altered.
It becomes much larger than $2m$ towards the threshold scattering
length. In the deep BEC limit, $1/(k_{r}a_{s})\rightarrow\infty$,
where two atoms form a tightly bound molecule, the mass is less affected
by the SOC or the effective Zeeman field, as we may anticipate.

\begin{figure}[h]
\begin{centering}
\includegraphics[clip,width=0.8\textwidth]{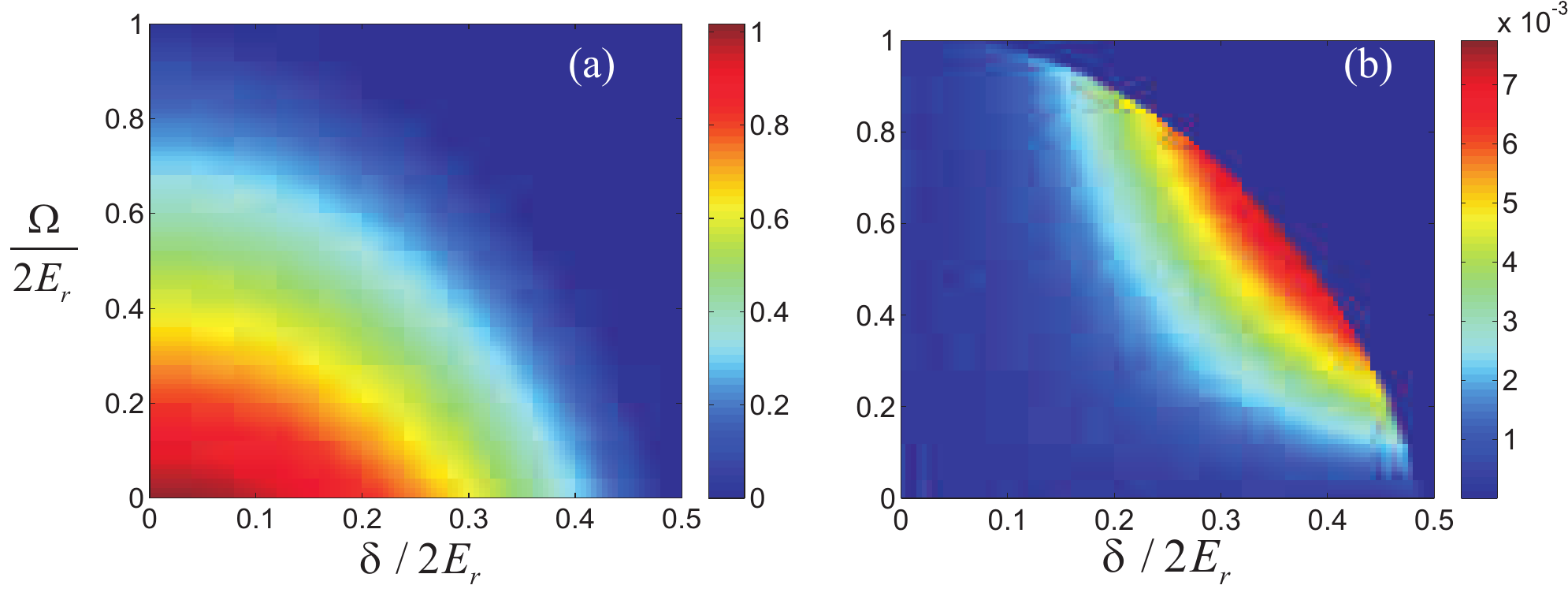} 
\par\end{centering}

\caption{ Binding energy $E_{b}=2E_{{\rm min}}-E_{{\bf q}_{0}}$ and the magnitude
of the lowest-energy bound state momentum $q_{0}$ as functions of
$\delta$ and $\Omega$. The coloring in (a) represents $E_{b}/E_{r}$,
and that in (b) represents $q_{0}/k_{r}$. In the upper right corner of both (a) and (b), there exist
no bound states. The scattering length is given by $1/(k_{r}a_{s})=1$.
Figure taken from Ref. \cite{dong1} with modifications.}

\label{fig6} 
\end{figure}

At nonzero detuning $\delta \neq 0$, the result shows that the two-particle
bound state will have its lowest energy at a finite center-of-mass
momentum ${\bf q}_{0}=(q_{0},0,0)$ \cite{dong1,ShenoyFF}. Fig. \ref{fig6}
shows the binding energy and the magnitude of ${\bf q}_0$
of the lowest-energy bound state. That the two-particle ground states possessing a finite momentum implies that the Cooper pairs, which
is a many-body counterpart of two-particle bound state, may acquire
finite center-of-mass momentum and therefore
condense into an inhomogeneous superfluid state. This possibility
will be addressed in greater detail later. We note that with the 
typical parameters, i.e., $\Omega\sim E_{r}$ and $\delta\sim E_{r}$,
$q_{0}$ is small and less than 1\% of the recoil momentum $k_{r}$,
as shown in Fig. \ref{fig6}(b). However, its magnitude can be significantly
enhanced by many-body effect. For Cooper pairs in the ground state,
$q_{0}$ can be tuned to be comparable with $k_{r}$ or the Fermi
wavevector $k_{F}$ \cite{liu1}.

\begin{figure}[h]
\begin{centering}
\includegraphics[clip,width=0.3\textwidth]{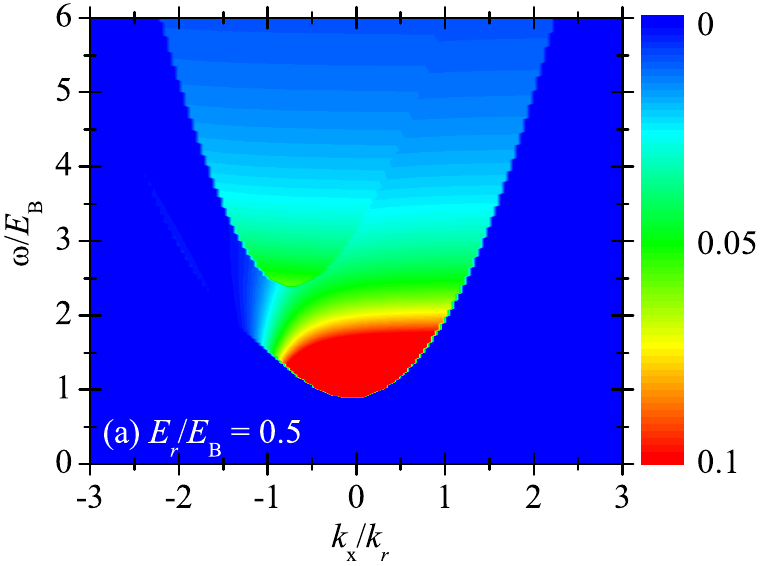} \includegraphics[clip,width=0.36\textwidth]{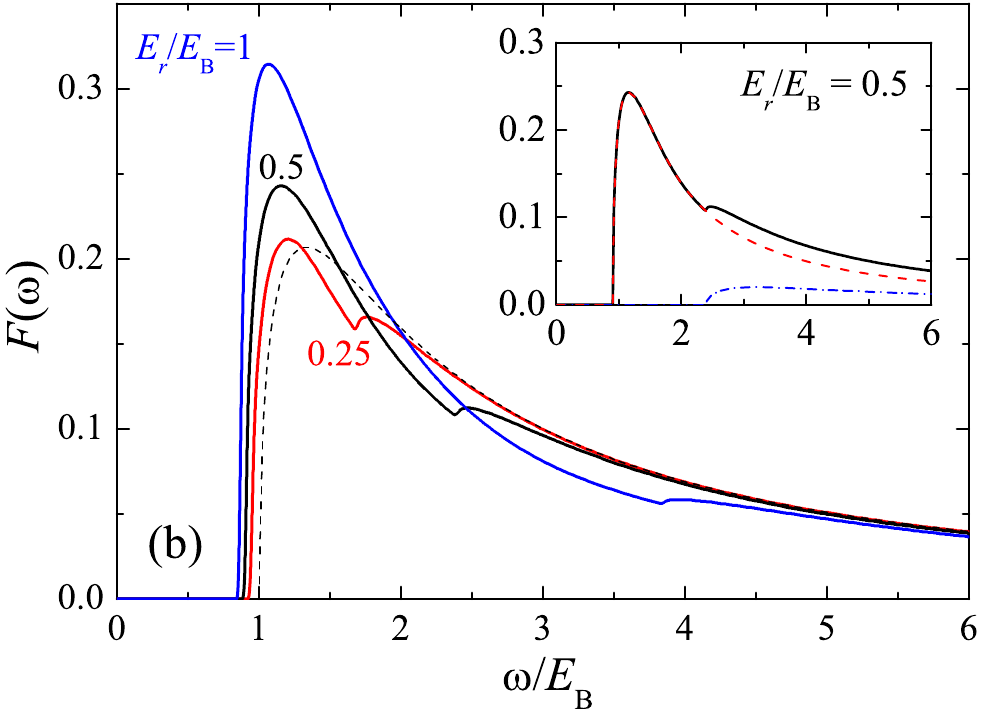} 
\par\end{centering}

\caption{ (a) Momentum-resolved rf spectroscopy (a) and integrated rf spectroscopy
(b) of the two-particle bound state at $\delta=0$ and $\Omega=2E_{r}$.
The energy of rf photon $\omega$ is measured in units of a binding
energy $E_{B}\equiv1/(ma_{s}^{2})$ and we have set $\omega_{3\downarrow}=0$.
In the right panel, the dashed line in the main figure plots the rf
line-shape in the absence of SOC: $F(\omega)=(2/\pi)\sqrt{\omega-E_{B}}/\omega^{2}$.
The inset highlights the different contribution from the two final
states, as described in the text. Figure taken from Ref. \cite{hu3}
with modification.}

\label{fig7} 
\end{figure}

Ideally, momentum-resolved rf-spectroscopy can be used to probe the
two-particle bound state discussed above. We can perform a
numerical simulation of the spectroscopy by using again the Fermi's
golden rule. Let us assume that a bound molecule is initially at rest
in the state $\left|\Phi_{2B}\right\rangle $ with energy $E_{i}$.
An rf photon with energy $\omega$ will break the molecule and transfer
the spin-down atom to the third state $\left|3\right\rangle $. In
the case that there is no final-state effect, the final state $\left|\Phi_{f}\right\rangle $
consists of a free atom in $|3\rangle$ and a remaining atom in the
spin-orbit system. According to the Fermi's golden rule, the rf strength
$\Gamma(\omega)$ of breaking molecules and transferring atoms is
proportional to the Franck-Condon factor \cite{Chin2005}, 
\begin{equation}
F\left(\omega\right)=\left|\left\langle \Phi_{f}\right|{\cal V}_{rf}\left|\Phi_{2B}\right\rangle \right|^{2}\delta\left[\omega-\omega_{3\downarrow}-(E_{f}-E_{i})\right].\label{FC}
\end{equation}
The integrated Franck-Condon factor satisfies the sum rule, $\int_{-\infty}^{+\infty}F\left(\omega\right)d\omega=1$.
A closed expression of $F\left(\omega\right)$ is derived in Refs.
\cite{hu3} and \cite{peng}, by carefully analyzing the initial two-particle
bound state $\left|\Phi_{2B}\right\rangle $ and the final state $\left|\Phi_{f}\right\rangle $.
Furthermore, by resolving the momentum of transferred atoms, we are
able to obtain the momentum-resolved Franck-Condon factor $F(k_{x},\omega)$.

Figs. \ref{fig7}(a) and \ref{fig7}(b) illustrate respectively the momentum-resolved and the
integrated rf spectrum of the two-particle ground state at zero
detuning $\delta=0$. One can easily resolve two different
responses in the spectrum due to two different final states, as
the remaining spin-up atom in the original spin-orbit system can occupy
either the upper or the lower energy branch. Indeed, in the integrated
rf spectrum, we can separate clearly the different contributions
from the two final states, as highlighted in the inset. This gives
rise to two peaks in the integrated spectrum. We note that the
lower peak exhibits a red shift as the SOC strength increases,
due to the decrease of the binding energy. It is also straightforward
to calculate the rf spectrum of the two-particle bound state at
nonzero detuning $\delta\neq0$ (not shown in the figure). However,
the spectrum remains essentially unchanged, due to the fact that
the center-of-mass momentum $q_{0}$ is quite small with typical experimental
parameters.

\subsubsection{Momentum-resolved radio-frequency spectrum of the superfluid phase}

Consider now the many-body state. As we mentioned earlier, since the
two-particle wave function contains both spin singlet and triplet
components, we anticipate that the superfluid phase at low temperatures
would involve both \textit{s}-wave pairing and high-partial-wave pairing.
Therefore, in general it is an anisotropic superfluid. This is to
be discussed later in detail for 2D Rashba SOC. Here, we are interested
in the phase diagram and the experimental probe of a 3D Fermi gas
with 1D equal-weight Rashba-Dresselhaus SOC. First, let us concentrate
on the case with zero detuning $\delta=0$, by using the many-body
\textit{T}-matrix theory within the pseudogap approximation \cite{Jing_RF}.

\begin{figure}[h]
\begin{centering}
\includegraphics[clip,width=0.5\textwidth]{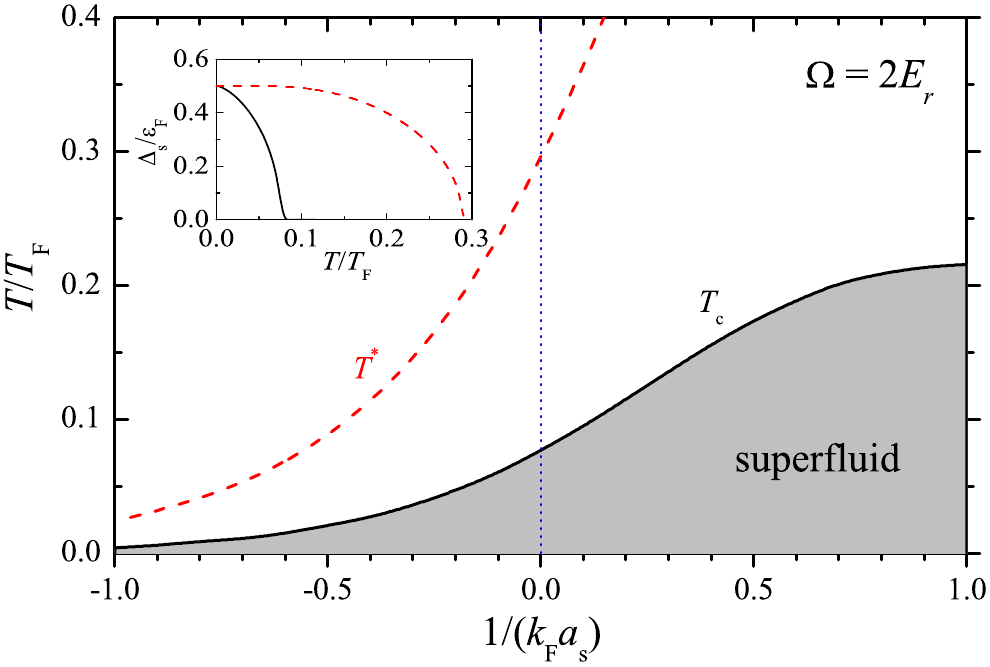} 
\par\end{centering}

\caption{ (a) Phase diagram of a spin-orbit coupled Fermi gas at the BEC-BCS
crossover at $\Omega=2E_{r}$ and $k_{F}=k_{r}$. The main figure
and inset show the superfluid transition temperature and the superfluid
order parameter at resonance, respectively, predicted by using our
\textit{T}-matrix theory (solid line) and the BCS mean-field theory
(dashed line). Figure taken from Ref. \cite{Jing_RF} with modification.}

\label{fig8} 
\end{figure}

Focusing on the vicinity of the Feshbach resonance where $a_{s}\rightarrow\pm\infty$,
in Fig. \ref{fig8} we show the superfluid transition temperature
$T_{c}$ and the pair breaking (pseudogap) temperature $T^{*}$ of
the spin-orbit coupled Fermi gas at $\Omega=2E_{r}$ and $k_{F}=k_{r}$.
The pseudogap temperature is calculated using the standard BCS mean-field
theory without taking into account the preformed pairs (i.e., $\Delta_{pg}=0$)
\cite{sademeloNSR,pseudogap}. We find that the region of superfluid
phase is strongly suppressed by SOC. In particular, at resonance the
superfluid transition temperature is about $T_{c}\simeq0.08T_{F}$,
which is significantly smaller than the experimentally determined
$T_{c}\simeq0.167(13)T_{F}$ for a unitary Fermi gas \cite{EoSMIT}.
Thus, it seems to be a challenge to observe a novel spin-orbit coupled
fermionic superfluid in the present experimental scheme.

\begin{figure}[h]
\begin{centering}
\includegraphics[clip,width=0.7\textwidth]{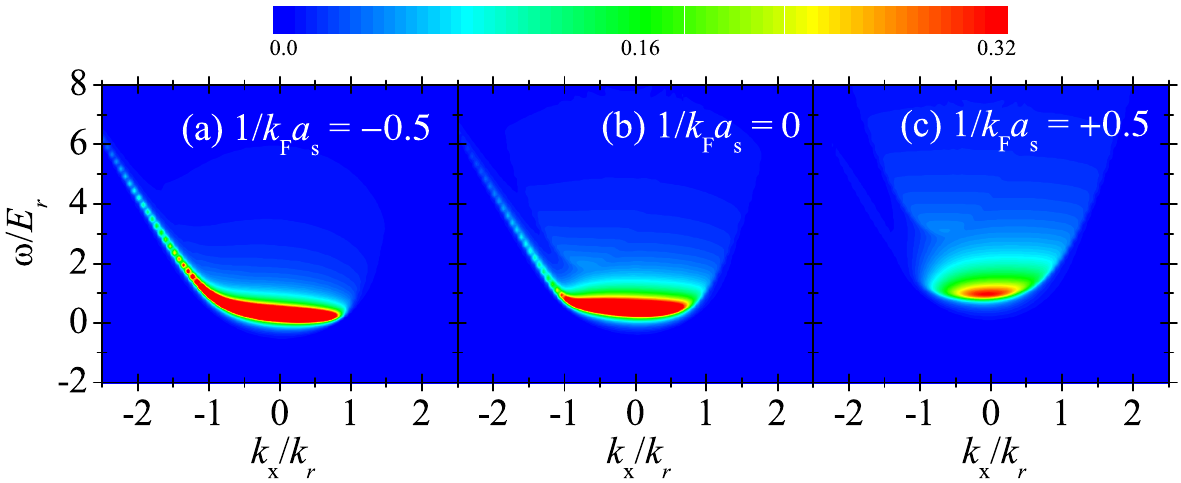} 
\par\end{centering}

\caption{ Zero-temperature momentum-resolved rf-spectroscopy of a spin-orbit
coupled Fermi gas across the Feshbach resonance, at the parameters
$\Omega=2E_{r}$ and $k_{F}=k_{r}$. Figure taken from Ref. \cite{Jing_RF}
with modification.}

\label{fig9} 
\end{figure}

In Figs. \ref{fig9}(a)-\ref{fig9}(c), we show the zero-temperature
momentum-resolved rf spectrum across the resonance. On the BCS
side ($1/k_{F}a_{s}=-0.5$), the spectrum is dominated by the
response from atoms and shows a characteristic high-frequency tail
at $k_{x}<0$ \cite{Jing_fermionic,MIT_fermionic,liu5}, see for example,
the left panel of Fig.~\ref{fig3}. We note that the density of the
Fermi cloud, chosen here following the real experimental parameters
\cite{Jing_fermionic}, is low and therefore only the lower energy
branch is occupied at low temperatures. Towards the BEC limit ($1/k_{F}a_{s}=+0.5$),
the spectrum may be understood from the picture of well-defined
bound pairs and shows a clear two-fold anisotropic distribution, as
we already mentioned in Fig. \ref{fig7}(a) \cite{hu3}. The spectrum
at the resonance is complicated and might be attributed to many-body
fermionic pairs. It is interesting that the response from many-body
pairs has a similar tail at high frequency as that from atoms. The
change of the rf spectrum across the resonance is continuous, in accordance
with a smooth BEC-BCS crossover.

\subsubsection{Fulde-Ferrell superfluidity}

The nature of superfluidity can be greatly changed by a nonzero detuning
$\delta\neq0$. As we discussed earlier in the two-body part, in this
case, the Cooper pairs may carry a nonzero center-of-mass momentum
and therefore condense into an inhomogeneous superfluid state, characterized
by the order parameter $\Delta_{0}(\mathbf{r})=\Delta_{0}e^{i\mathbf{q}\cdot\mathbf{r}}$.
This exotic superfluid has been proposed by Fulde and Ferrell \cite{Fulde1964},
soon after the discovery of the seminal BCS theory. Its existence
has attracted tremendous theoretical and experimental efforts over
the past five decades \cite{Radzihovsky2010}. Remarkably, to date
there is still no conclusive experimental evidence for FF superfluidity.
Here, we show that the superfluid phase of a 3D Fermi gas with 1D
equal-weight Rashba-Dresselhaus SOC and finite in-plane effective Zeeman
field $\delta$ is precisely the long-sought FF superfluid \cite{liu1}.
The same issue has also been addressed very recently by Vijay Shenoy
\cite{ShenoyFF}. We note that the FF superfluid can appear in other settings
with different types of SOC and dimensionality \cite{zheng1,zheng2,yi1,dong2,hu2,zhou,Barzykin2002}.

Theoretically, to determine the FF superfluid state, we solve the
BdG equation (\ref{bdgeq}) with $V_{T}(\mathbf{r})=0$ by using the
following ansatz for quasiparticle wave functions 
\begin{equation}
\Phi_{\mathbf{k\eta}}(\mathbf{x})=\frac{e^{i\mathbf{k\cdot}\mathbf{x}}}{\sqrt{V}}\left[u_{\mathbf{k\eta\uparrow}}e^{+iqx/2},u_{\mathbf{k\eta\downarrow}}e^{+iqx/2},v_{\mathbf{k\eta\uparrow}}e^{-iqx/2},v_{\mathbf{k}\eta\downarrow}e^{-iqx/2}\right]^{T}.
\end{equation}
The center-of-mass momentum $\mathbf{q}$ is assumed to be along the
\textit{x}-direction, inspired from the two-body solution \cite{dong1}.
The mean-field thermodynamic potential $\Omega_{0}$ at temperature
$T$ in Eq. (\ref{eq: S0}) is then given by 
\begin{equation}
\frac{\Omega_{0}}{V}=\frac{1}{2V}\left[\sum_{\mathbf{k}}\left(\xi_{\mathbf{k}+\mathbf{q}/2}+\xi_{\mathbf{k}-\mathbf{q}/2}\right)-\sum_{\mathbf{k\eta}}E_{\mathbf{k}\eta}\right]-\frac{k_{B}T}{V}\sum_{\mathbf{k\eta}}\ln\left(1+e^{-E_{\mathbf{k}\eta}/k_{B}T}\right)-\frac{\Delta_{0}^{2}}{U_{0}},\label{eq:Omega}
\end{equation}
where $E_{\mathbf{k}\eta}$ ($\eta=1,2,3,4$) is the quasiparticle
energy. Here, the summation over the quasiparticle energy must be
restricted to $E_{\mathbf{k}\eta}\geq0$ because of an inherent particle-hole
symmetry in the Nambu spinor representation. For a given set of parameters
(i.e, the temperature $T$, interaction strength $1/k_{F}a_{s}$,
etc.), different mean-field phases can be determined using the self-consistent
stationary conditions: $\partial\Omega/\partial\Delta=0$, $\partial\Omega/\partial q=0$,
as well as the conservation of total atom number, $N=-\partial\Omega/\partial\mu$.
At finite temperatures, the ground state has the lowest free energy
$F=\Omega+\mu N$. In the following, we consider the resonance case
with a divergent scattering length $1/k_{F}a_{s}=0$ and set $T=0.05T_{F}$,
where $T_{F}$ is the Fermi temperature. According to the typical
number of atoms in experiments \cite{Jing_fermionic,MIT_fermionic},
we take the Fermi wavevector $k_{F}=k_{r}$.

\begin{figure}[h]
\begin{centering}
\includegraphics[clip,width=0.8\textwidth]{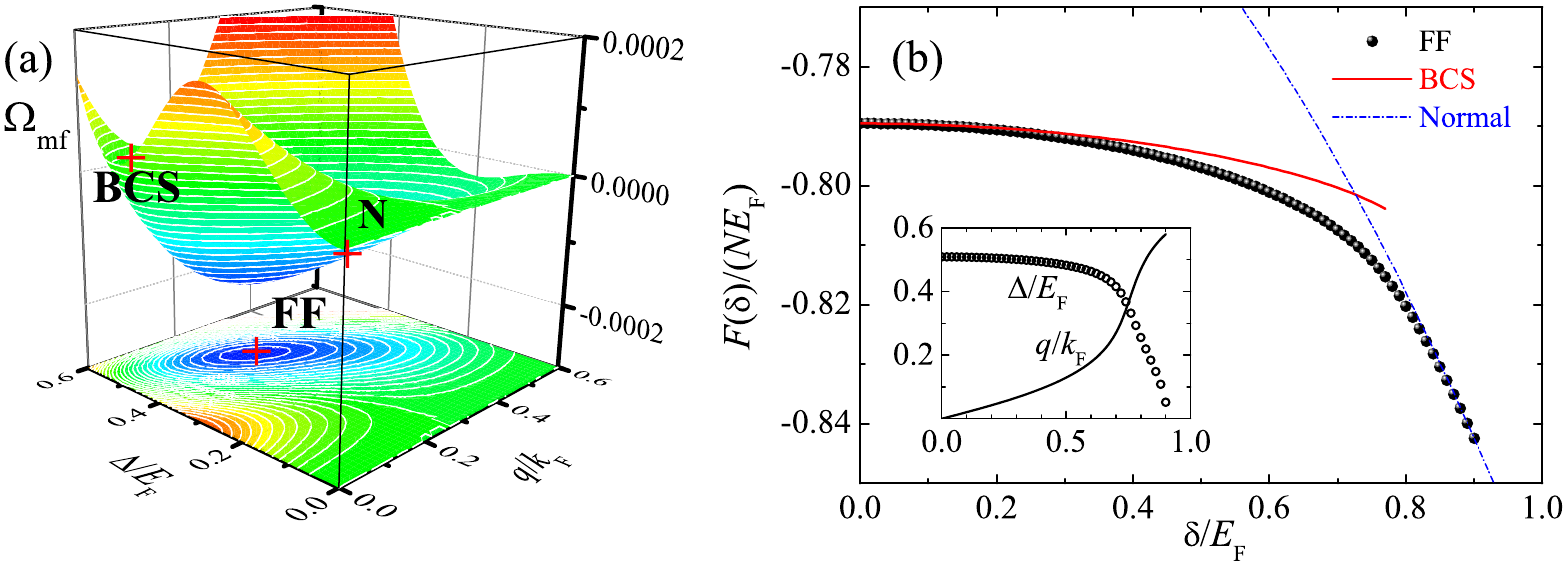} 
\par\end{centering}

\caption{ (a) Landscape of the thermodynamic potential, $\Omega_{\textrm{mf}}=[\Omega_{0}(\Delta,q)-\Omega_{0}(0,0)]/(NE_{F})$,
at $\Omega=2E_{F}$ and $\delta=0.68E_{F}$. The chemical potential
is fixed to $\mu=-0.471E_{F}$. The competing ground states include
(i) a normal Fermi gas with $\Delta_{0}=0$; (ii) a fully paired BCS
superfluid with $\Delta_{0}\neq0$ and $q=0$; and (iii) a finite
momentum paired FF superfluid with $\Delta_{0}\neq0$ and $q\neq0$.
(b) The free energy of different competing states as a function of
the detuning at $\Omega=2E_{F}$. The inset shows the detuning dependence
of the order parameter and momentum of the FF superfluid state. Figure
taken from Ref. \cite{liu1} with modification.}

\label{fig10} 
\end{figure}

\begin{figure}[h]
\begin{centering}
\includegraphics[clip,width=0.5\textwidth]{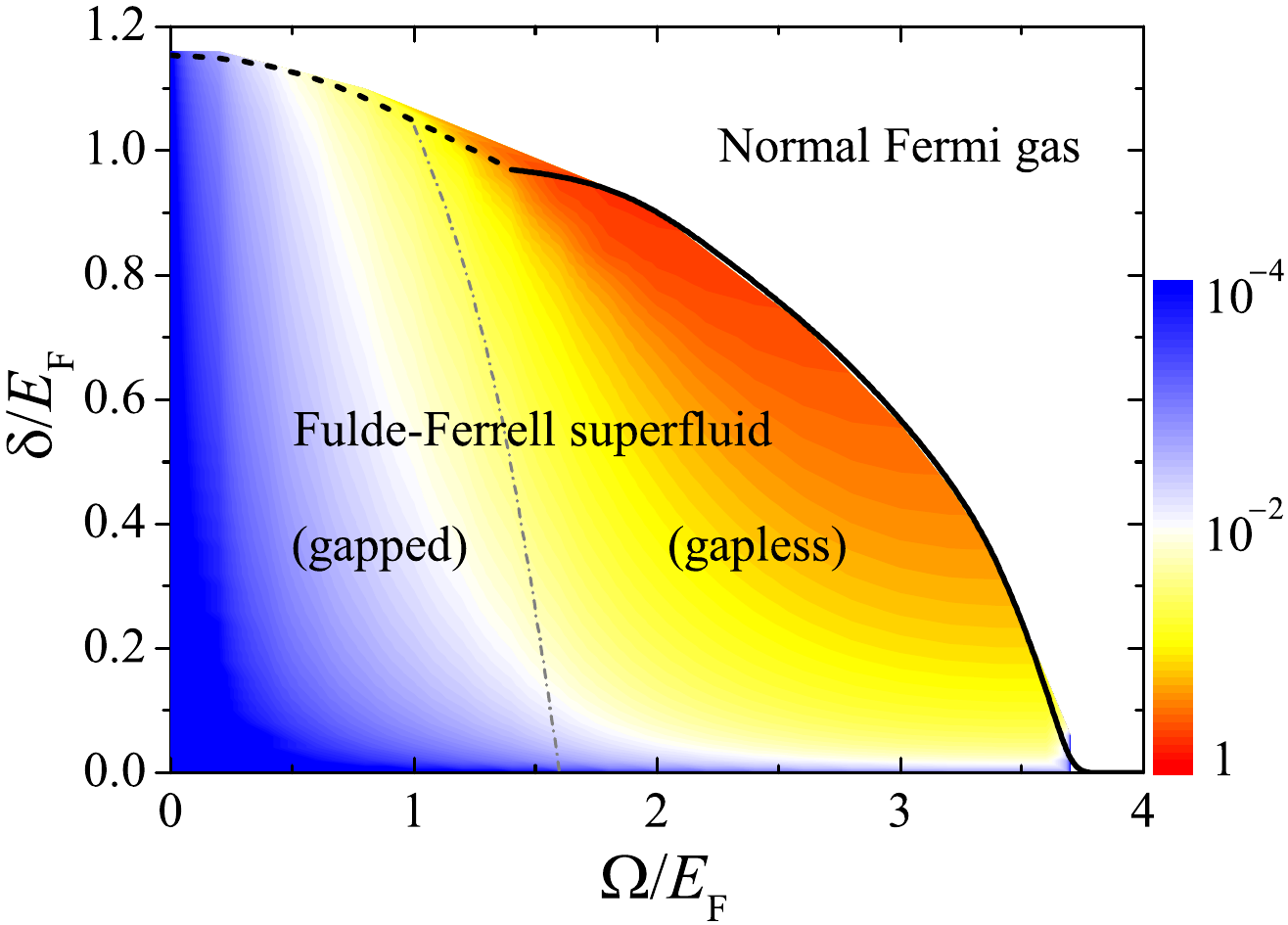} 
\par\end{centering}

\caption{ Phase diagram as a function of $\delta$ and $\Omega$, at $T=0.05T_{F}$.
By increasing $\delta$, the Fermi cloud changes from a FF superfluid
to a normal gas, via first-order (dashed line) and second-order (solid
line) transitions at low and high $\Omega$, respectively. The FF
superfluid can be either gapped or gapless, as separated by the dot-dashed
line. The coloring represents the magnitude of the centre-of-mass
momentum of Cooper pairs, $q/k_{F}$. The BCS superfluid occurs at
$\Omega=0$ or $\delta=0$ only. Figure taken from Ref. \cite{liu1}
with modification.}

\label{fig11} 
\end{figure}

In general, for any set of parameters there are three competing ground
states that are stable against phase separation (i.e., $\partial^{2}\Omega_{0}/\partial\Delta_{0}^{2}\geq0$),
as shown in Fig. \ref{fig10}(a): normal gas ($\Delta_{0}=0$), BCS
superfluid ($\Delta_{0}\neq0$ and $q=0$), and FF superfluid ($\Delta_{0}\neq0$
and $q\neq0$). Remarkably, in the presence of spin-orbit coupling
the FF superfluid is always more favorable in energy than the standard
BCS pairing state at finite detuning (Fig. \ref{fig10}(b)). It is
easy to check that the superfluid density of the BCS pairing state
in the SOC direction becomes negative (i.e., $\partial\Omega_{0}/\partial q<0)$,
signaling the instability towards an FF superfluid. Therefore, experimentally
the Fermi gas would always condense into an FF superfluid at finite
two-photon detuning. In Fig. \ref{fig11}, we report a low-temperature
phase diagram that could be directly observed in current experiments.
The FF superfluid occupies the major part of the phase diagram.

\begin{figure}[h]
\begin{centering}
\includegraphics[clip,width=0.6\textwidth]{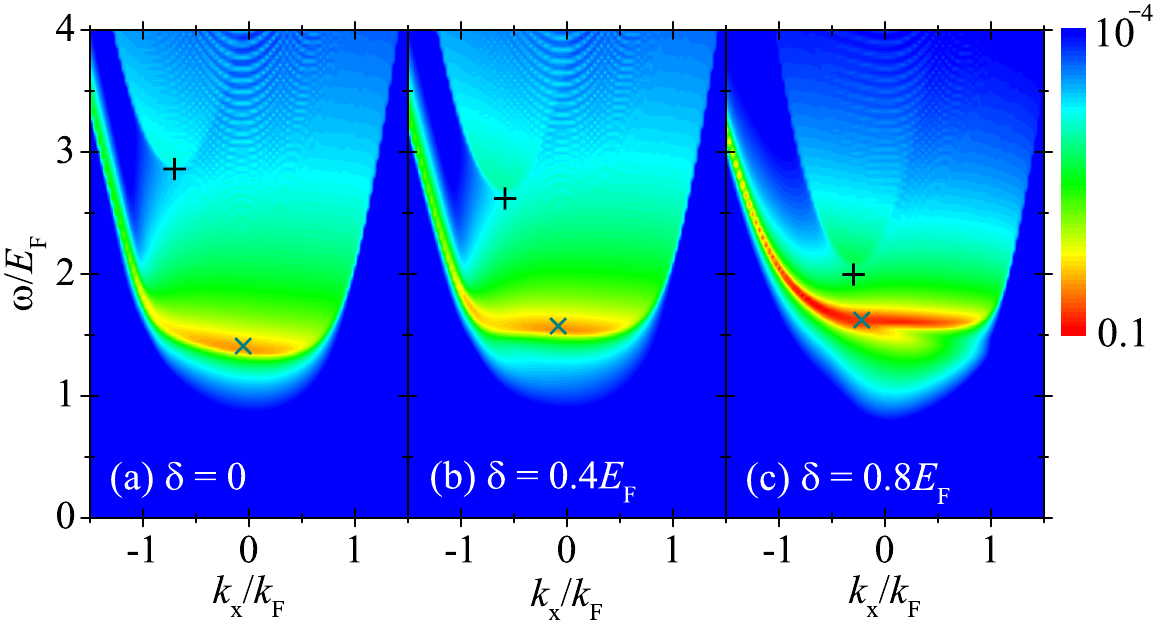} 
\par\end{centering}

\caption{Logarithmic contour plot of momentum-resolved rf spectroscopy: number
of transferred atoms $\Gamma(k_{x},\omega)$ at $\Omega=2E_{F}$ and
at three detunings: (a) $\delta=0$ and $q=0$, (b) $\delta=0.4E_{F}$
and $q\simeq0.1k_{F}$, and (c) $\delta=0.8E$ and $q\simeq0.6k_{F}$.
Figure taken from Ref. \cite{liu1} with modification.}

\label{fig12} 
\end{figure}

The experimental probe of an FF superfluid is a long-standing challenge.
Here, unique to cold atoms, momentum-resolved rf spectroscopy may
provide a smoking-gun signal of the FF superfluidity. The basic idea
is that, since Cooper pairs carry a finite center-of-mass momentum
$\mathbf{q}$, the transferred atoms in the rf transition acquire
an overall momentum $\mathbf{q}/2$. As a result, there would be a
$\mathbf{q}/2$ shift in the measured spectrum. In Fig. \ref{fig12},
we show the momentum-resolved rf spectrum $\Gamma(k_{x},\omega)$
on a logarithmic scale. As we discussed earlier in the two-body part,
there are two contributions to the spectrum, corresponding to
two different final states \cite{hu3}. These two contributions are
well separated in the frequency domain, with peak positions indicated
by the symbols ``$+$'' and ``$\times$'', respectively. Interestingly,
at finite detuning with a sizable FF momentum $\mathbf{q}$, the peak
positions of the two contributions are shifted roughly in opposite
directions by an amount $\mathbf{q}/2$. This provides clear evidence
for observing the FF superfluid.

\subsubsection{1D topological superfluidity}

Arguably, the most remarkable aspect of SOC is that it provides a
feasible routine to realize topological superfluids \cite{Zhang2008},
which have attracted tremendous interest over the past few years \cite{Qi2011}.
In addition to providing a new quantum phase of matter, topological
superfluids can host exotic quasiparticles at their boundaries, known as Majorana fermions - particles that are their own antiparticles
\cite{Majorana1937,Wilczek2009}. Due to their non-Abelian exchange
statistics, Majorana fermions are believed to be the essential quantum
bits for topological quantum computation \cite{Nayak2008}. Therefore,
the pursuit for topological superfluids and Majorana fermions represents one of the most important challenges in fundamental science.
A number of settings have been proposed for the realization of topological
superfluids, including the fractional quantum Hall states at filling
$\nu=5/2$ \cite{Moore1991}, vortex states of $p_{x}+ip_{y}$ superconductors
\cite{Read2000,Mizushima2008}, and surfaces of three-dimensional
(3D) topological insulators in proximity to an $s$-wave superconductor
\cite{Sau2010}, and one-dimensional (1D) nanowires with strong spin-orbit
coupling coated also on an $s$-wave superconductor \cite{Orge2010}.
In the latter setting, indirect evidences of topological superfluid
and Majorana fermions have been reported \cite{Mourik2012}.
Here, we review briefly the possible realizations of topological superfluids,
in the context of a 1D spin-orbit coupled atomic Fermi gas \cite{liu3,Wei2012,impurity1d,liu4},
which can be prepared straightforwardly by loading a 3D spin-orbit
Fermi gas into deep 2D optical lattices. Later, we will discuss 2D
topological superfluids with Rashba SOC.

\begin{figure}[h]
\begin{centering}
\includegraphics[clip,width=0.5\textwidth]{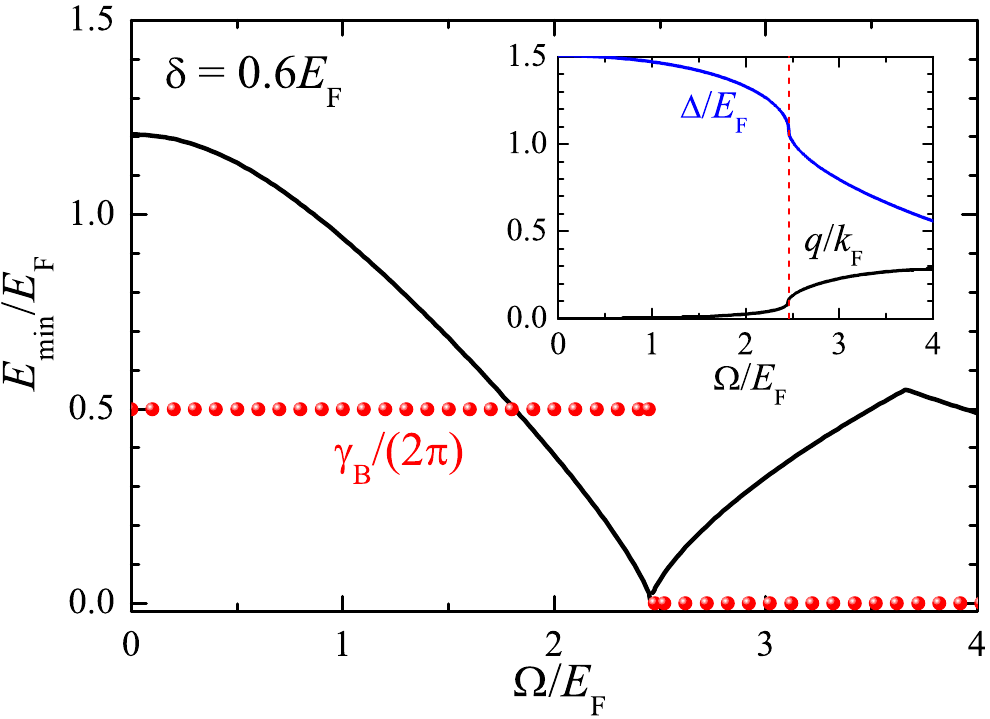} 
\par\end{centering}

\caption{ Theoretical examination of the topological phase transition at the
detuning $\delta=0.6E_{F}$ and $T=0$. The transition occurs at $\Omega\simeq2.46E_{F}$
, where the energy gap of the system (solid line) close and then open.
The Berry phase $\gamma_{B}$ is $\pi$ and 0 at the topologically
trivial and non-trivial regimes (circles). The insets shows the order
parameter and momentum of the FF superfluid, as a function of the
Rabi frequency. Figure taken from Ref. \cite{liu4} with modification.}

\label{fig13} 
\end{figure}

Consider first a homogeneous 1D Fermi gas with a nonzero detuning
$\delta\neq0$ \cite{liu4}. In this case, we actually anticipate
a topological inhomogeneous superfluid, where the order parameter
also varies in real space. Using the same theoretical technique as
in the previous subsection, we solve the BdG equation (\ref{bdgeq})
in 1D and then minimize the mean-field thermodynamic potential Eq. (\ref{eq:Omega})
to determine the pairing gap $\Delta_{0}$ and the FF momentum $q$.

In Fig. \ref{fig13}, we show the energy gap as a function of $\Omega$
at $\delta=0.6E_{F}$ and $T=0$. For this result, we use a Fermi
wavevector $k_{F}=0.8k_{r}$ and take a dimensionless interaction
parameter $\gamma\equiv-mg_{1D}/(n)=3$, where $g_{1D}$ is the 
strength of the 1D contact interaction and $n=2k_{F}/\pi$ is the 1D
linear density. Topological phase transition is associated with a
change of the topology of the underlying Fermi surface and therefore is accompanied 
with closing of the excitation gap at the transition point. In the main figure this feature is clearly evident.
To better characterize the change of topology, we may calculate the
Berry phase defined by \cite{Wei2012} 
\begin{equation}
\gamma_{B}=i\intop_{-\infty}^{+\infty}dk\left[W_{+}^{*}(k)\partial_{k}W_{+}(k)+W_{-}^{*}(k)\partial_{k}W_{-}(k)\right].\label{eq:BerryPhase}
\end{equation}
Here $W_{\eta}(k)\equiv[u_{k\eta\uparrow}e^{iqz/2},u_{k\eta\downarrow}e^{iqz/2},v_{k\eta\uparrow}
e^{-iqz/2},v_{k\eta\downarrow}e^{-iqz/2}]^{T}$
denotes the wave function of the upper $(\eta=+)$ and lower $(\eta=-)$ branch, respectively. In Fig. \ref{fig13}, the Berry phase is shown by circles. It jumps
from $\pi$ to $0$, right across the topological phase transition.
It is somewhat counter-intuitive that the $\gamma_{B}=0$ sector corresponds
to the topologically non-trivial superfluid state. It is important
to emphasize the inhomogeneous nature of the superfluid. Indeed, as
shown in the inset, the FF momentum $q$ increases rapidly across
the topological superfluid transition and reaches about $0.3k_{F}$
at $\Omega=4E_{F}.$

\begin{figure}[h]
\begin{centering}
\includegraphics[clip,width=0.5\textwidth]{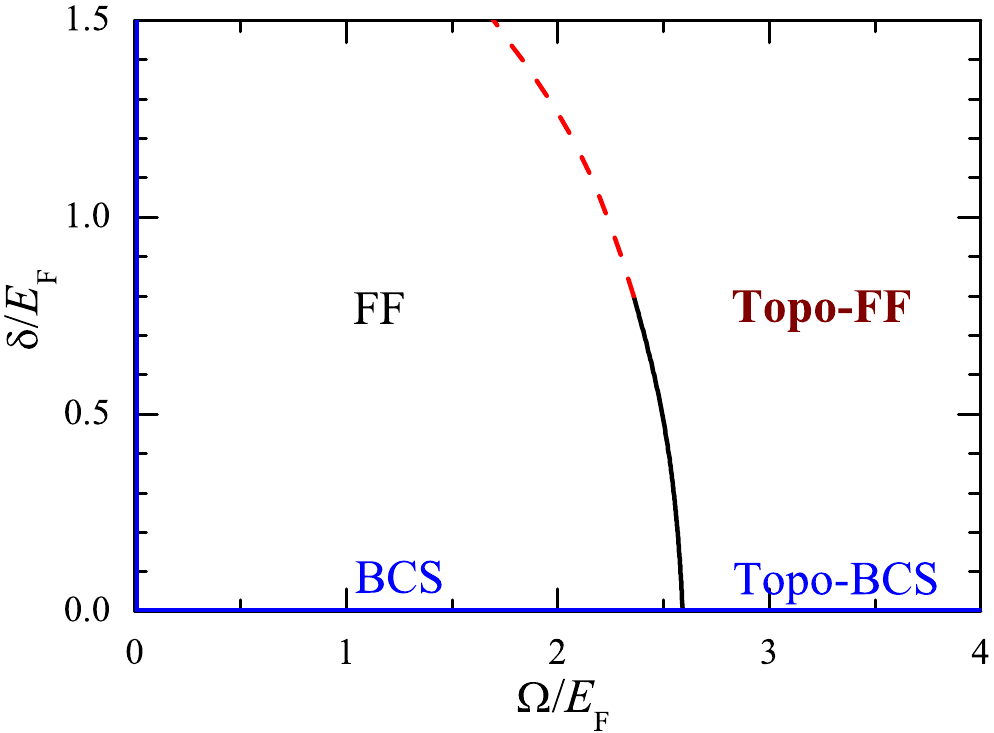} 
\par\end{centering}

\caption{Zero-temperature phase diagram. A topologically non-trivial FF superfluid
appears when the Raman coupling strength $\Omega$ is above a threshold
at finite detunings $\delta$. Depending on the detuning, the transition
could be either continuous (solid line) or of first order (dashed
line). The FF superfluid reduces to a BCS superfluid when $\Omega=0$
or $\delta=0$. Figure taken from Ref. \cite{liu4} with modification.}

\label{fig14} 
\end{figure}

In Fig. \ref{fig14}, we present the zero-temperature phase diagram
for the topological phase transition. The critical coupling strength
$\Omega_{c}$ decreases with the increase of the detuning $\delta$. At
zero detuning, $\Omega_{c}$ can be determined analytically, since
the expression for the BdG eigenenergy for single-particle excitations (after  dropping a constant energy shift
$E_{r}$)
is known \cite{jiang,Wei2012}, 
\begin{equation}
E_{k\eta}=\left[\left(\xi_{k}^{2}+\lambda^{2}k^{2}+\Omega^{2}/4+\Delta_{0}^{2}\pm\sqrt{4\xi_{k}^{2}\lambda^{2}k^{2}+\Omega^{2}\left(\xi_{k}^{2}+\Delta_{0}^{2}\right)}\right)\right]^{1/2},
\end{equation}
where $\xi_{k}=k^{2}/(2m)-\mu$ and $\lambda=k_{r}/m$. It is easy to see that
the excitation gap closes at $k=0$ for the lower branch (i.e., $\eta=-$),
leading to the well-known result \cite{Orge2010} 
\begin{equation}
\frac{\Omega_{c}}{2}=\sqrt{\mu^{2}+\Delta^{2}}.\label{eq: criterionTS}
\end{equation}
This criterion for topological superfluids is equivalent to the condition
that there are only two Fermi points on the Fermi surface \cite{zhang1},
under which the Fermi system behaves essentially like a 1D weak-coupling
\textit{p}-wave superfluid.

\begin{figure}[h]
\begin{centering}
\includegraphics[clip,width=0.4\textwidth]{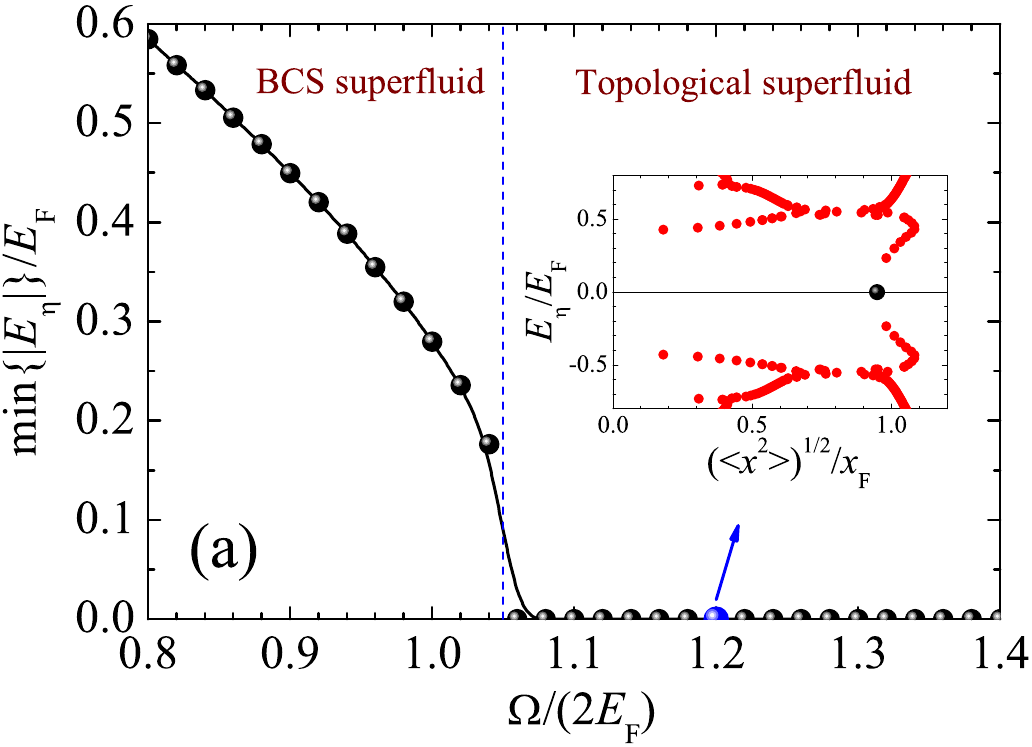}
\includegraphics[clip,width=0.4\textwidth]{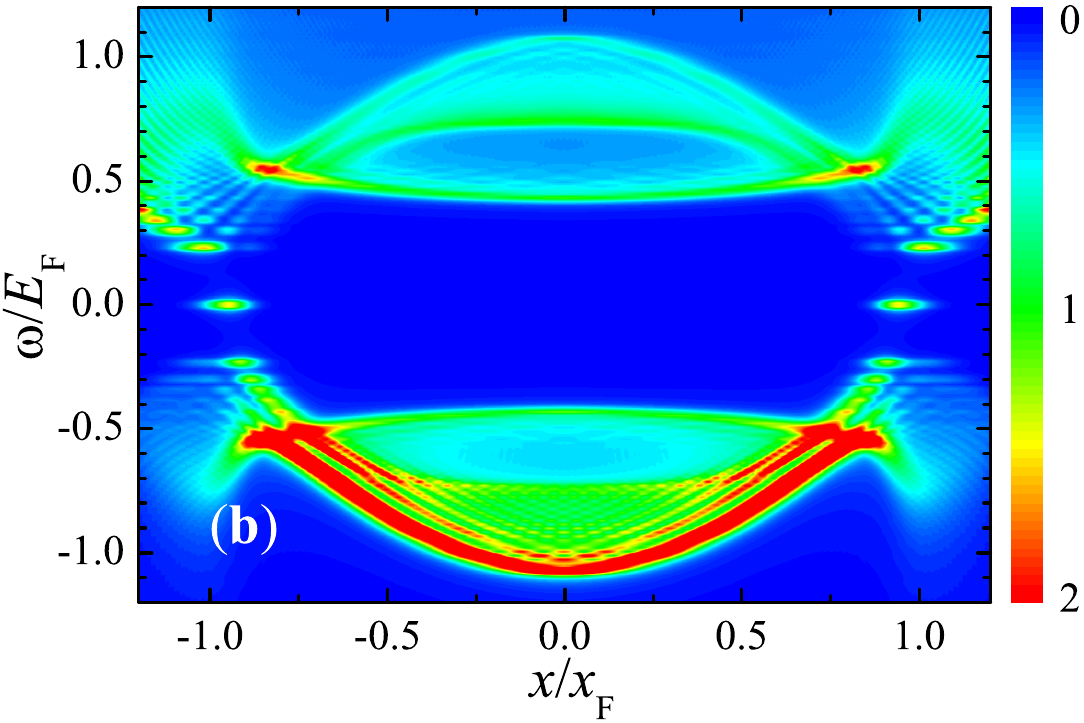} 
\par\end{centering}

\caption{ (a) Zero-temperature phase diagram of a trapped 1D spin-orbit coupled
Fermi gas, determined from the behavior of the lowest energy in quasiparticle
spectrum. The inset shows the energy spectrum at $\Omega=2.4E_{F}$
as a function of the position of quasiparticles. A zero-energy quasiparticle
(i.e., Majorana fermion) at the trap edge has been highlighted by
a big dark circle. Here, the position of a quasiparticle is approximately
characterized by: $\left\langle x^{2}\right\rangle =\int dxx^{2}\sum_{\sigma}[u_{\sigma}^{2}\left(x\right)+\nu_{\sigma}^{2}\left(x\right)]$.
$x_{F}$ is the Thomas-Fermi radius of the cloud. (b) Linear contour
plot of the local density of state at $\Omega=2.4E_{F}$. At each
trap edge, a series of edge states, including the zero-energy Majorana
fermion mode, are clearly visible. Figure taken from Ref. \cite{impurity1d}
with modification.}

\label{fig15} 
\end{figure}

Let us now turn to the experimentally realistic situation with a 1D
harmonic trap $V_{T}(x)=m\omega^{2}x^{2}/2$ and focus on the case
with $\delta=0$ \cite{liu3,Wei2012,impurity1d}. The BdG equation
(\ref{bdgeq}) can be solved self-consistently by expanding the Nambu
spinor wave function $\Phi_{\eta}\left(x\right)$ onto the eigenfucntion
basis of the harmonic oscillator. In this trapped environment, Majorana
fermions with zero energy are anticipated to emerge at the boundary,
if the Fermi gas stays in a topological superfluid state. The appearance
of Majorana fermions can be easily understood from the particle-hole
symmetry obeyed by the BdG equation, which states that every physical
state can be described either by a particle state with a positive
energy $E$ or a hole state with a negative energy $-E$. The Bogoliubov
quasiparticle operators associated with these two states therefore
satisfy $\Gamma_{E}=\Gamma_{-E}^{\dagger}$. At the boundary, Eq. (\ref{eq: criterionTS}) could be fulfilled
at some points and give locally the states with $E=0$. These states
are Majorana fermions, as the associated operators satisfy $\Gamma_{0}=\Gamma_{0}^{\dagger}$
- precisely the defining feature of a Majorana fermion \cite{Majorana1937,Wilczek2009}.

In Fig. \ref{fig15}(a), we present the zero-temperature phase diagram
of a trapped 1D Fermi gas at $k_{F}=2k_{r}$ and $\gamma=\pi$
\cite{impurity1d}. The transition from BCS superfluid to topological
superfluid is now characterized by the appearance of Majorana fermions,
whose energy is precisely zero and therefore the minimum of the quaisparticle
spectrum touches zero, $\min\{\left|E_{\eta}\right|\}=0$. In the
topological superfluid phase, as shown in Fig. \ref{fig15}(b) with
$\Omega=2.4E_{F}$, the Majorana fermions may be clearly identified
by using spatially-resolved rf spectroscopy. We note that for a trapped
Fermi gas with weak interatomic interaction and/or high density,
the upper branch of single-particle spectrum may be populated at the
trap center, leading to four Fermi points on the Fermi surface. This
violates Eq. (\ref{eq: criterionTS}). As a result, we may find a
phase-separation phase in which the topological superfluid occurs
only at the two wings of the Fermi cloud. This situation has been
discussed in Ref. \cite{liu3}.

\subsection{2D Rashba spin-orbit coupling}

Let us now discuss Rashba SOC, which takes the standard form $V_{\textrm{SO}}=\lambda(\hat{k}_{y}\hat{\sigma}_{x}-\hat{k}_{x}\hat{\sigma}_{y})$
\cite{footnote2}. The coupling between spin and orbital motions occurs
along two spatial directions and therefore we shall refer to it as 2D
Rashba SOC. This type of SOC is not realized experimentally yet, although
there are several theoretical proposals for its realization \cite{Sau2011,xu}.
The superfluid phase with 2D Rashba SOC at low temperatures shares
a lot of common features as its 1D counterpart as we reviewed in the
previous subsection. Here we focus on some specific features, for
example, the two-particle bound state at sufficiently strong SOC strength
- the rashbon \cite{vj2,vj3} - and the related crossover
to a BEC of rashbons. We will also discuss in greater detail the 2D
topological superfluid with Rasbha SOC in the presence of an out-of-plane
Zeeman field, since it provides an interesting platform to perform
topological quantum computation. We note that experimentally it is
also possible to create a 3D isotropic SOC, $V_{\textrm{SO}}=\lambda(\hat{k}_{x}\hat{\sigma}_{x}+\hat{k}_{y}\hat{\sigma}_{y}+\hat{k}_{z}\hat{\sigma}_{z})$,
where the spin and orbital degree of freedoms are coupled in all three
dimensions \cite{Anderson2012}. We note also that early theoretical
works on a Rashba spin-orbit coupled Fermi gas was reviewed very briefly
by Hui Zhai in Ref. \cite{zhai2}.

\begin{figure}[h]
\begin{centering}
\includegraphics[clip,width=0.4\textwidth]{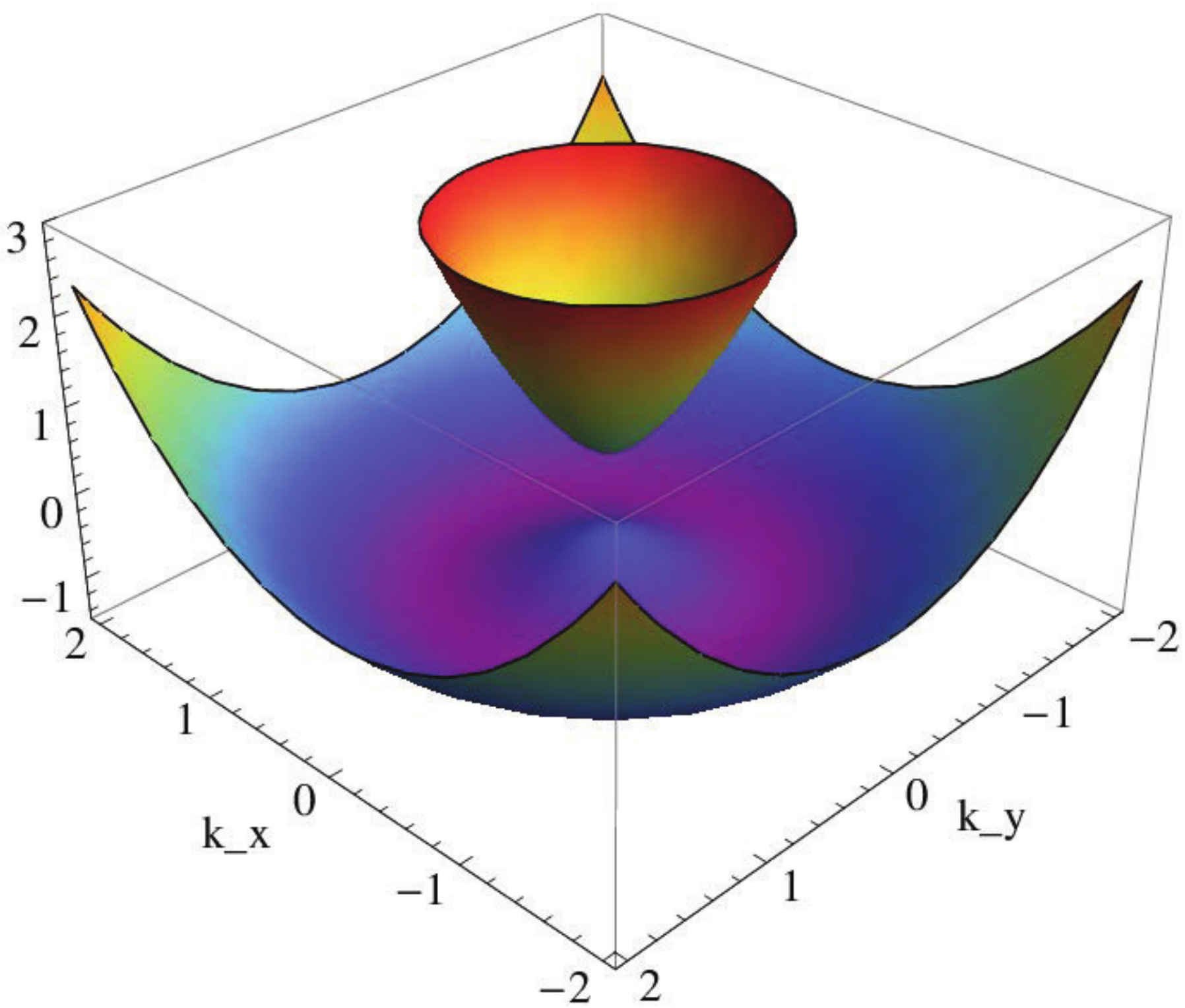}
\includegraphics[clip,width=0.4\textwidth]{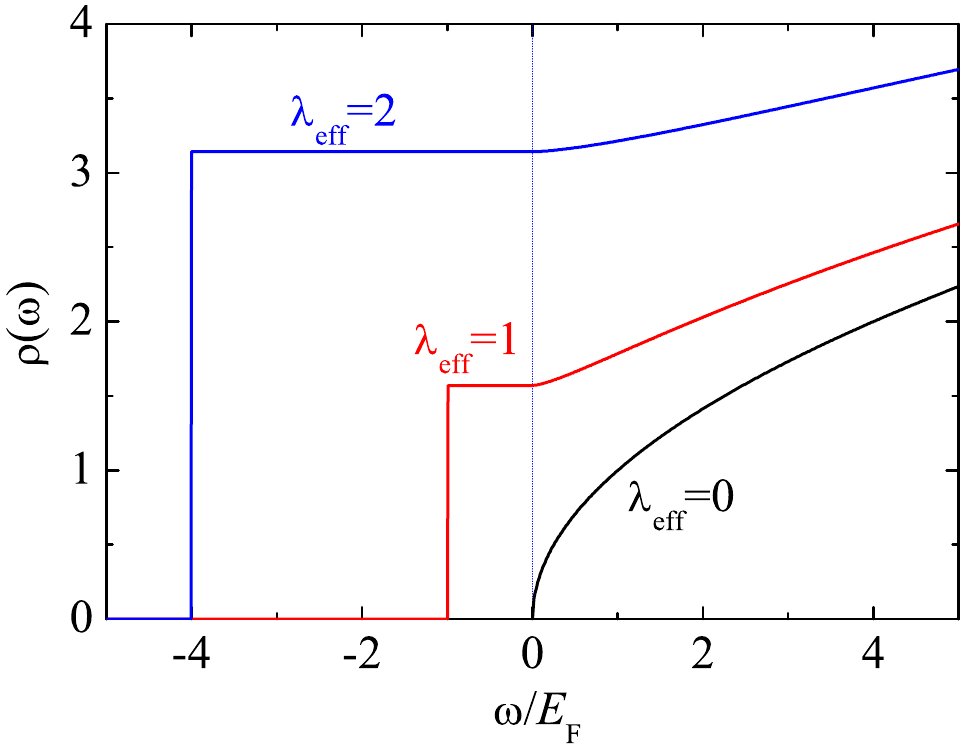} 
\par\end{centering}

\caption{ Left panel: schematic of the single-particle spectrum in the $k_{x}-k_{y}$
plane. A energy gap opens at $k=0$, due to a nonzero out-of-plane
Zeeman field $h$. Right panel: density of states of a 3D homogeneous
Rashba spin-orbit coupled system at several SOC strengths, in units
of $mk_{F}$. Right figure taken from Ref. \cite{hu4} with modification.}

\label{fig16} 
\end{figure}

\subsubsection{Single-particle spectrum}

In the presence of an out-of-plane Zeeman field $h\hat{\sigma}_{z}$,
the single-particle spectrum is given by, 
\begin{equation}
E{}_{\mathbf{k}\pm}=\frac{\mathbf{k}^{2}}{2m}\pm\sqrt{\lambda^{2}\left(k_{x}^{2}+k_{y}^{2}\right)+h^{2}}\,.
\end{equation}
The spectrum with
a nonzero $h$ is illustrated on the left panel of Fig. \ref{fig16}.
Compared with the single-particle spectrum with 1D equal-weight Rashba-Dresselhaus
SOC in Fig. \ref{fig2}, it is interesting that the two minima in
the lower energy branch now extend to form a ring structure. At low
energy, therefore, we may anticipate that in the momentum space the
particles will be confined along the ring. The effective dimensionality
of the system is therefore reduced. Indeed, it is not difficult to obtain the
density of states ($h=0$) \cite{hu4}:
\begin{equation}
\rho\left(\omega\right)=(mk_{F})\left\{ \begin{array}{ll}
0, & \left(\omega<-\lambda_{{\rm eff}}^{2}\right);\\
\lambda_{{\rm eff}}\pi/2, & (-\lambda_{{\rm eff}}^{2}\leq\omega<0);\\
\sqrt{\omega/E_{F}}+\lambda_{{\rm eff}}\left[\pi/2-\arctan\sqrt{\omega/\left(\lambda_{{\rm eff}}^{2}E_{F}\right)}\right], & (\omega\geq0).
\end{array}\right.,
\end{equation}
where we have defined a dimensionless SOC coupling strength $\lambda_{{\rm eff}}\equiv m\lambda/k_{F}$.
As can be seen from the right panel of Fig.~\ref{fig16}, $\rho (\omega)$ with Rashba SOC becomes a constant at low energy, which
is characteristic of a 2D system. This reduction in the effective
dimensionality will have interesting consequences
when the interatomic interaction comes into play, as we now disucss
in greater detail.

\subsubsection{Two-body physics}

We solve the two-body problem by calculating the two-particle vertex
function, following the general procedure outlined in the theoretical
framework (Sec.~\ref{general}). Focusing on the case without Zeeman fields, we have
\begin{equation}
\mathcal{G}_{0}\left(K\right)=\frac{\left(i\omega_{m}-\xi_{{\bf k}}\right)+\lambda\left(k_{y}\sigma_{x}-k_{x}\sigma_{y}\right)}{\left(i\omega_{m}-\xi_{{\bf k}}\right)^{2}-\lambda^{2}\left(k_{x}^{2}+k_{y}^{2}\right)}.
\end{equation}
By substituting it into Eq. (\ref{eq:VertexFunction}), it is straightforward
to obtain, 
\begin{equation}
\Gamma^{-1}=\frac{1}{U_{0}}+\frac{k_{B}T}{V}\sum_{{\bf k},i\omega_{m}}\left[\frac{1/2}{\left(i\omega_{m}-E_{{\bf k},+}\right)\left(i\nu_{n}-i\omega_{m}-E_{{\bf q}-{\bf k},+}\right)}+\frac{1/2}{\left(i\omega_{m}-E_{{\bf k},-}\right)\left(i\nu_{n}-i\omega_{m}-E_{{\bf q}-{\bf k},-}\right)}-A_{res}\right],
\end{equation}
with the single-particle energy 
\begin{eqnarray}
E_{{\bf k},\pm} & = & \xi_{{\bf k}}\pm \lambda \sqrt{k_x^2+k_y^2},
\end{eqnarray}
and
\begin{equation}
A_{res}\equiv\frac{\lambda^{2}k_{\bot}\sqrt{\left(q_{x}-k_{x}\right)^{2}+\left(q_{y}-k_{y}\right){}^{2}}+\lambda^{2}k_{x}\left(q_{x}-k_{x}\right)+\lambda^{2}k_{y}\left(q_{y}-k_{y}\right)}{\left(i\omega_{m}-E_{{\bf k},+}\right)\left(i\omega_{m}-E_{{\bf k},-}\right)\left(i\nu_{n}-i\omega_{m}-E_{{\bf q}-{\bf k},+}\right)\left(i\nu_{n}-i\omega_{m}-E_{{\bf q}-{\bf k},-}\right)}.
\end{equation}
By performing explicitly the summation over $i\omega_{m}$, replacing
${\bf k}$ by ${\bf q}/2+{\bf k}$ and re-arranging the terms, we
find that
\begin{eqnarray}
\Gamma^{-1} & = & \frac{m}{4\pi a_{s}}+\frac{1}{2V}\sum_{{\bf k}}\left[\frac{f\left(E_{{\bf q}/2+{\bf k},+}\right)+f\left(E_{{\bf q}/2-{\bf k},+}\right)-1}{i\nu_{n}-E_{{\bf q}/2+{\bf k},+}-E_{{\bf q}/2-{\bf k},+}}+\frac{f\left(E_{{\bf q}/2+{\bf k},-}\right)+f\left(E_{{\bf q}/2-{\bf k},-}\right)-1}{i\nu_{n}-E_{{\bf q}/2+{\bf k},-}-E_{{\bf q}/2-{\bf k},-}}-\frac{1}{\epsilon_{{\bf k}}}\right]\nonumber \\
 &  & -\frac{1}{4V}\sum_{{\bf k}}\left[1+\frac{q_{\bot}^{2}/4-k_{\bot}^{2}}{\sqrt{\left(q_{x}/2+k_{x}\right)^{2}+\left(q_{y}/2+k_{y}\right){}^{2}}\sqrt{\left(q_{x}/2-k_{x}\right)^{2}+\left(q_{y}/2-k_{y}\right){}^{2}}}\right]C_{res},
\end{eqnarray}
where 
\begin{eqnarray}
C_{res} & = & +\frac{\left[f\left(E_{{\bf q}/2+{\bf k},+}\right)+f\left(E_{{\bf q}/2-{\bf k},+}\right)-1\right]}{i\nu_{n}-E_{{\bf q}/2+{\bf k},+}-E_{{\bf q}/2-{\bf k},+}}+\frac{\left[f\left(E_{{\bf q}/2+{\bf k},-}\right)+f\left(E_{{\bf q}/2-{\bf k},-}\right)-1\right]}{i\nu_{n}-E_{{\bf q}/2+{\bf k},-}-E_{{\bf q}/2-{\bf k},-}}\nonumber \\
 &  & -\frac{\left[f\left(E_{{\bf q}/2+{\bf k},+}\right)+f\left(E_{{\bf q}/2-{\bf k},-}\right)-1\right]}{i\nu_{n}-E_{{\bf q}/2+{\bf k},+}-E_{{\bf q}/2-{\bf k},-}}-\frac{\left[f\left(E_{{\bf q}/2+{\bf k},-}\right)+f\left(E_{{\bf q}/2-{\bf k},+}\right)-1\right]}{i\nu_{n}-E_{{\bf q}/2+{\bf k},-}-E_{{\bf q}/2-{\bf k},+}}.
\end{eqnarray}
The above equation provides a starting point to investigate the fluctuation
effect due to interatomic interactions.

Here, for the two-body problem of interest, we discard the Fermi distribution
function and set $\mathbf{q}=0$, as the ground bound state has zero
center-of-mass momentum in the absence of Zeeman field. The two-body
vertex function is then given by,

\begin{equation}
\Gamma_{\textrm{2\textrm{b}}}^{-1}\left(\mathbf{q=0};i\nu_{n}\rightarrow\omega+i0^{+}\right)=\frac{m}{4\pi a_{s}}-\frac{1}{2V}\sum_{{\bf k}}\left[\frac{1}{\omega+i0^{+}-2E_{{\bf k},+}}+\frac{1}{\omega+i0^{+}-2E_{k,-}}+\frac{1}{\epsilon_{{\bf k}}}\right],
\end{equation}
The energy of the two-particle bound state $E$ can be obtained by
solving ${\rm Re}\{\Gamma_{{\rm 2b}}^{-1}\left[\mathbf{q=0};\omega=E\right]\}=0$
with $\mu=0$, as we already discussed in the theoretical framework.
More physically, we may calculate the phase shift 
\begin{equation}
\delta_{2\textrm{b}}\left(\mathbf{q=0};\omega\right)=-\textrm{Im}\ln\left[-\Gamma_{\textrm{2\textrm{b}}}^{-1}\left(\mathbf{q=0};i\nu_{n}\rightarrow\omega+i0^{+}\right)\right].
\end{equation}
Recall that the vertex function represents the Green function of Cooper
pairs. Thus, the phase shift defined above is simply $\int d\omega A(\mathbf{q},\omega)$,
where $A(\mathbf{q},\omega)$ is the spectral function of pairs. As
a result, a true bound state, corresponding to a delta peak in the
spectral function, will cause a $\pi$ jump in the phase shift at
the critical frequency $\omega_{c}=E$, from which we determine the
energy of the bound state.

\begin{figure}[h]
\begin{centering}
\includegraphics[clip,width=0.4\textwidth]{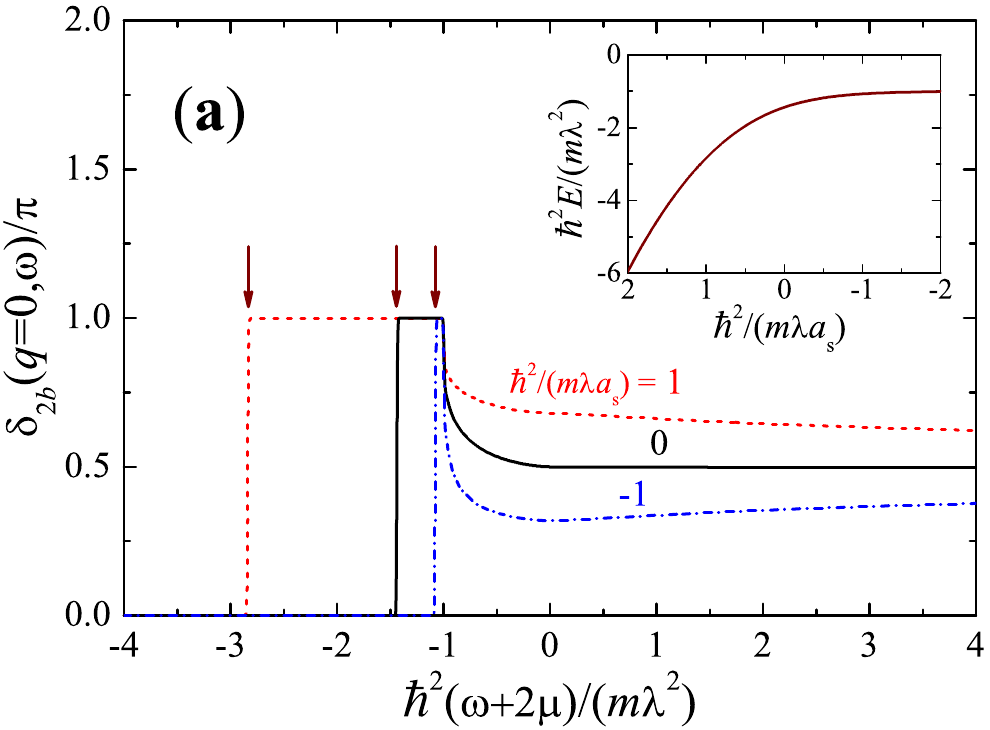}
\includegraphics[clip,width=0.4\textwidth]{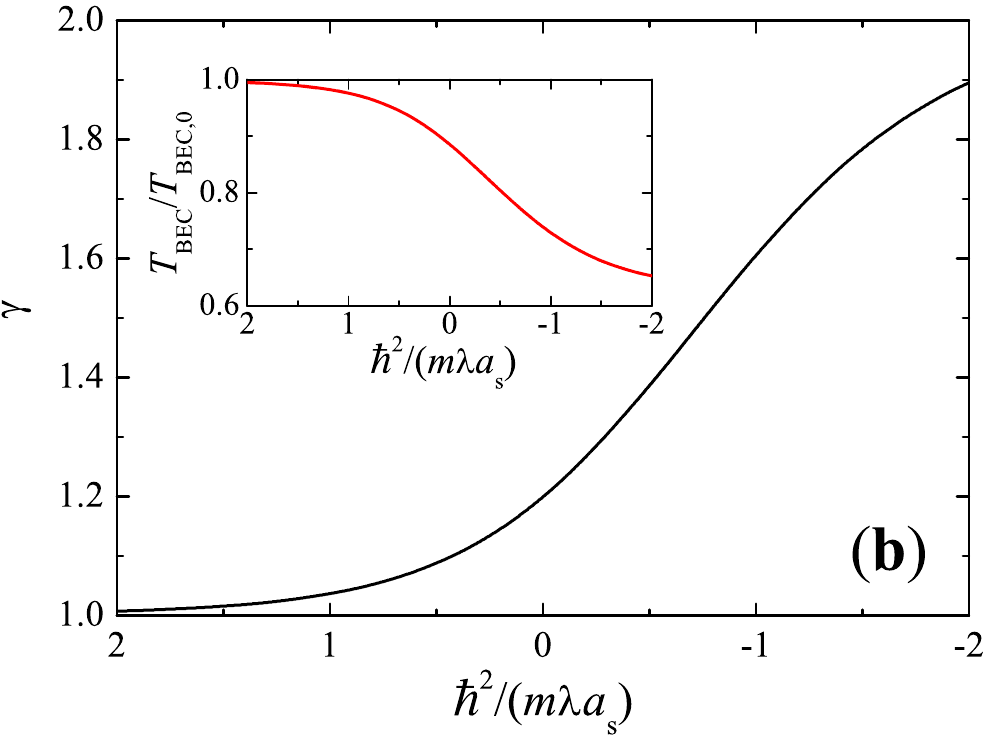} 
\par\end{centering}

\caption{(a) Two-body bound states as evidenced by the two-body phase shift
at three different scattering lengths, in the presence of Rashba SOC.
The arrows indicate the position of the bound state energy. The inset
shows the bound state energy as a function of the scattering length.
(b) Effective mass of the two-body bound state. The inset shows the
decreases of critical temperature due to the heavy mass of bound states.
Figure taken from Ref. \cite{hu1} with modification.}

\label{fig17} 
\end{figure}

In the main figure and inset of Fig. \ref{fig17}(a), we show the
two-body phase shift and the energy of the bound state of a Rashba
spin-orbit coupled Fermi gas, respectively. Interestingly, the bound
state exists even in the BCS limit, where the \textit{s}-wave scattering
length is small and negative \cite{vj1}. This is because
at the low energy the effective dimensionality of the Rashba system
reduces to two, as we mentioned earlier from the nature of the low-energy density
of states. In 2D, we know that any weak attraction can lead to a bound
state. We can calculate the effective mass of the bound state \cite{hu1,zhai1}, which is strongly renormalized by the SOC, by
determining the dispersion relation of the two-body bound state $E(\mathbf{q})$
at small momentum $\mathbf{q\sim0}$. The result is shown in Fig.
\ref{fig17}(b) for $\gamma\equiv M_{x}/(2m)=M_{y}/(2m)$. 

It is important to note that all the properties of the two-body bound
state, including its energy and effective mass, depend on a single
parameter $1/(m\lambda a_{s})$, which is the ratio of the only two
length scales $1/(m\lambda)$ and $a_{s}$ in the problem. Thus, in
the limit of sufficiently large SOC, the bound state becomes universal
and is identical to the one obtained at $1/(m\lambda a_{s})=0$. This
new kind of universal bound state has been referred to as rashbon
\cite{vj2,vj3}. The mass of rashbons (i.e., $\gamma\simeq1.2$ from
Fig. \ref{fig17}(b)) is notably heavier than the conventional molecules
$2m$ in the BEC limit. This causes a decrease in the condensation
temperature of rashbons in such a way that 
\begin{equation}
T_{\rm BEC}=\gamma^{-2/3}T_{\rm BEC}^{(0)}\simeq0.193T_{F},\label{eq: TcRashbonBEC}
\end{equation}
where $T_{\rm BEC}^{(0)}\simeq0.218T_{F}$ is the BEC temperature of conventional
molecules.

In the presence of out-of-plane Zeeman field $h$, the two-body problem
has been discussed in detail in Ref. \cite{jiang}.

\subsubsection{Crossover to rashbon BEC and anisotropic superfluidity }

\begin{figure}[h]
\begin{centering}
\includegraphics[clip,width=0.25\textwidth]{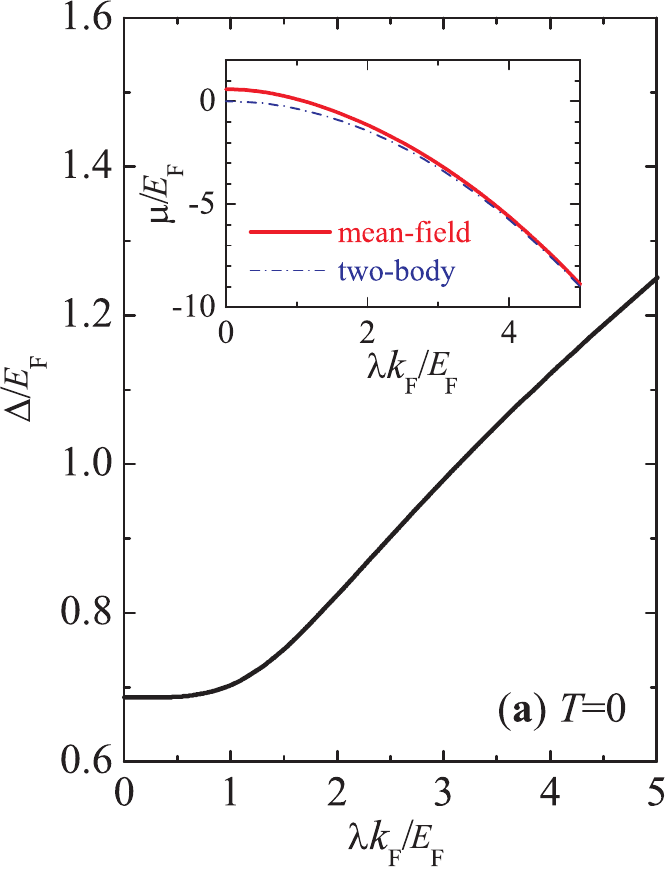}
\includegraphics[clip,width=0.5\textwidth]{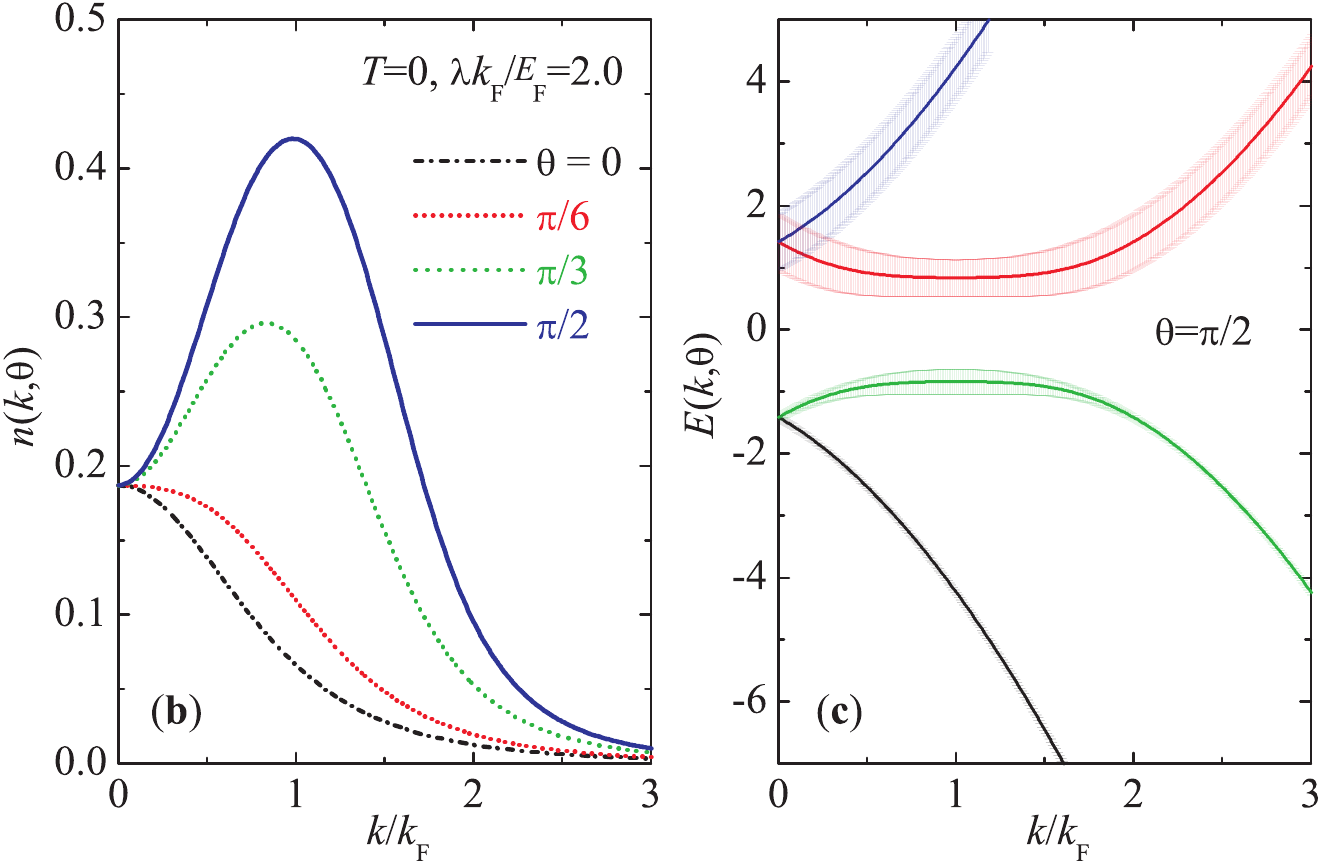} 
\par\end{centering}

\caption{(a) Mean-field order parameter as a function of the Rashba SOC for
a homogeneous unitary Fermi gas at zero temperature. The inset shows
the chemical potential and the half of bound state energy, both in
units of Fermi energy $E_{F}$. (b) Momentum distribution and (c)
single-particle spectral function for $\theta=\pi/2$ at $\lambda k_{F}/E_{F}=2$.
Here $\theta$ is the angle between ${\bf k}$ and the $z$-axis.
The width of the curves in (c) represents the weight factor $(1\pm\gamma_{{\bf k},\pm})/4$
for each of the four Bogoliubov excitations. Figure taken from Ref.
\cite{hu1}.}

\label{fig18} 
\end{figure}

Let us now discuss the crossover to a rashbon BEC. We focus on the
unitary limit with $a_{s}\rightarrow\infty$ and increase the 2D Rashba
SOC. At the mean-field saddle-point level, the single-particle Green
function Eq. (\ref{eq: GreenFunction}) takes the form ($h=0$) \cite{hu1},
\begin{equation}
G_{0}^{-1}=\left[\begin{array}{cc}
i\omega_{m}-\xi_{{\bf k}}-\lambda(\hat{k}_{y}\hat{\sigma}_{x}-\hat{k}_{x}\hat{\sigma}_{y}) & i\Delta_{0}\hat{\sigma}_{y}\\
-i\Delta_{0}\hat{\sigma}_{y} & i\omega_{m}+\xi_{{\bf k}}-\lambda(\hat{k}_{y}\hat{\sigma}_{x}+\hat{k}_{x}\hat{\sigma}_{y})
\end{array}\right].
\end{equation}
The inversion of the above matrix can be worked out explicitly, leading
to two single-particle Bogoliubov dispersions whose degeneracy is
lifted by the SOC, $E_{{\bf k},\pm}=[(\xi_{{\bf k}}\pm\lambda k_{\perp})^{2}+\Delta_{0}^{2}]^{1/2}$,
and the normal and anomalous Green functions from which we can immediately
obtain the momentum distribution $n\left({\bf k}\right)=1-\sum_{\alpha}[1/2-f(E_{{\bf k},\alpha})]\gamma_{{\bf k},\alpha}$
and the single-particle spectral function 
\begin{equation}
A_{\uparrow\uparrow}({\bf k},\omega)=A_{\downarrow\downarrow}({\bf k},\omega)=\frac{1}{4}\sum_{\alpha}\left[\left(1+\gamma_{{\bf k},\alpha}\right)\delta\left(\omega-E_{{\bf k},\alpha}\right)+\left(1-\gamma_{{\bf k},\alpha}\right)\delta\left(\omega+E_{{\bf k},\alpha}\right)\right],
\end{equation}
where $\gamma_{{\bf k},\pm}=(\xi_{{\bf k}}\pm\lambda k_{\perp}/E_{{\bf k},\pm})$.
The chemical potential and the order parameter are to be determined
by the number and the gap equations, $n=\sum_{{\bf k}}n({\bf k})$
and $\Delta_{0}=-U_{0}\Delta_{0}\sum_{\alpha}[1/2-f(E_{{\bf k},\alpha})]/(2E_{{\bf k},\alpha})$,
respectively. Fig.~\ref{fig18}(a) displays the chemical potential
$\mu$ and the order parameter as functions of the SOC strength. The increase
of the SOC strength leads to a deeper bound state. As a consequence,
in analogy with the BEC-BCS crossover, the order parameter and the critical
transition temperature are greatly enhanced at $\lambda k_{F}\sim\epsilon_{F}$.
In the large SOC limit, we have $\mu=(\mu_{B}+E)/2$, where $E$ is
the energy of the two-body bound state, and $\mu_{B}$ is positive
due to the repulsion between rashbons and decreases with increasing
coupling as shown in the inset of Fig.~\ref{fig18}(a). By assuming
an \textit{s}-wave repulsion with scattering length $a_{B}$ between
rashbons, where $\mu_{B}\simeq(n/2)4\pi a_{B}/M$, we estimate within
mean-field that in the unitarity limit, $a_{B}\simeq3/(m\lambda)$,
comparable to the size of rashbons. Figs.~\ref{fig18}(b) and (c) illustrate the
momentum distribution and the single-particle spectral function, respectively.
These quantities exhibit anisotropic distribution in momentum space
due to the SOC and can be readily measured in experiment.

\begin{figure}[h]
\begin{centering}
\includegraphics[clip,width=0.5\textwidth]{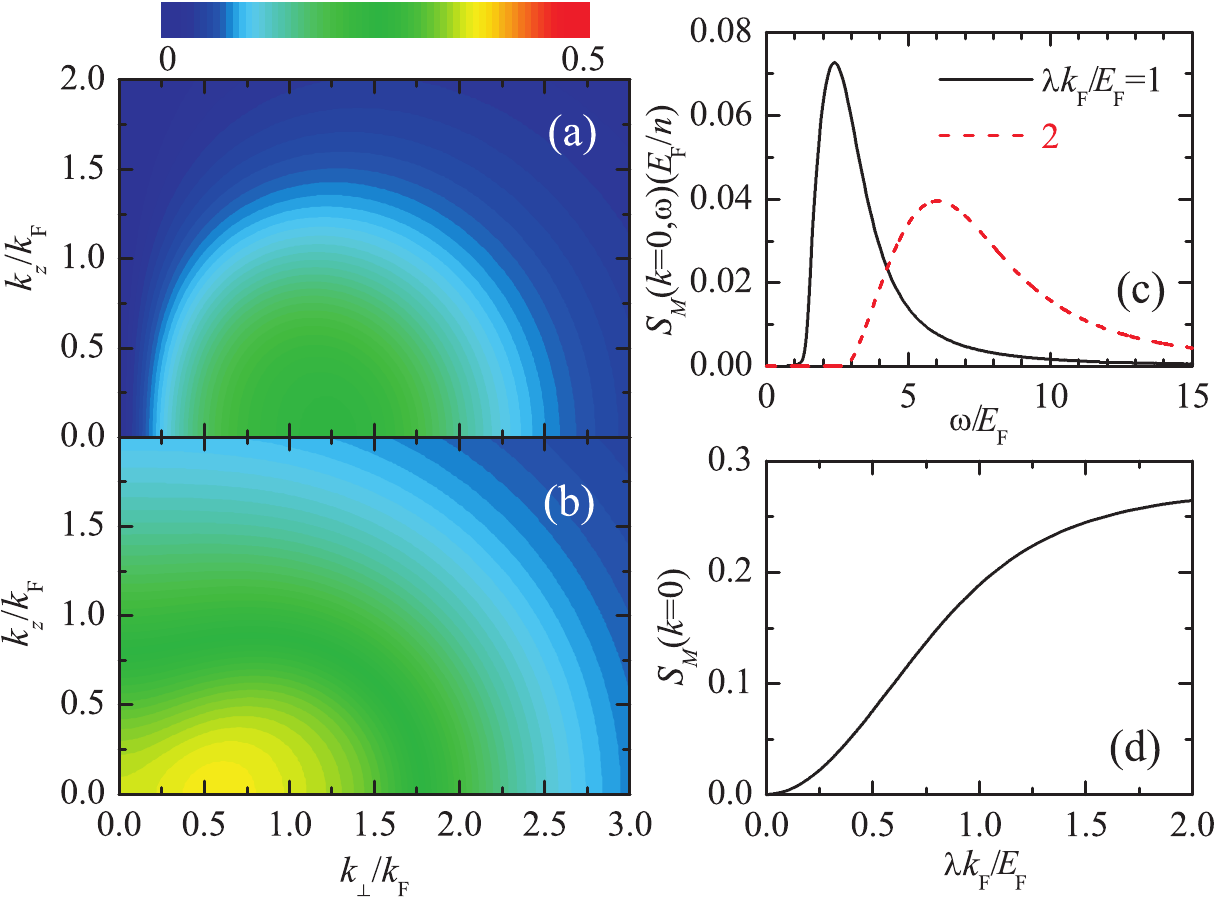} 
\par\end{centering}

\caption{ Linear contour plot for the triple pairing correlation $|\left\langle \psi_{{\bf k}\uparrow}\psi_{-{\bf k}\uparrow}\right\rangle |$
between like spins (a) and the singlet pairing correlation $|\left\langle \psi_{{\bf k}\uparrow}\psi_{-{\bf k}\downarrow}\right\rangle |$
between un-like spins (b) for a homogeneous unitary Fermi gas at zero
temperature with $\lambda k_{F}/E_{F}=2$. The zero-momentum dynamic
and static spin structure factor are shown in (c) and (d), respectively.
Figure taken from Ref.~\cite{hu1}.}

\label{fig19} 
\end{figure}

Another interesting feature of the crossover to rashbon BEC is that
the pairing field contains both a singlet and a triplet component
\cite{Gorkov2001}. For the system under study, it is straightforward
to show that the triplet and singlet pairing fields are given by $\left\langle \psi_{{\bf k}\uparrow}\psi_{-{\bf k}\uparrow}\right\rangle =-i\Delta_{0}e^{-i\varphi_{{\bf k}}}\sum_{\alpha}\alpha[1/2-f\left(E_{{\bf k},\alpha}\right)]/(2E_{{\bf k},\alpha})$
and $\left\langle \psi_{{\bf k}\uparrow}\psi_{-{\bf k}\downarrow}\right\rangle =\Delta_{0}\sum_{\alpha}[1/2-f\left(E_{{\bf k},\alpha}\right)]/(2E_{{\bf k},\alpha})$,
respectively, where $e^{-i\varphi_{{\bf k}}}\equiv(k_{x}-ik_{y})/k_{\perp}$.
The magnitude of the pairing fields are shown in Fig.~\ref{fig19}(a)
and (b). The weight of the triplet component increases
and approaches that of the singlet component as the SOC strength increases.
In Fig.~\ref{fig19}(c) and (d), we plot the zero-momentum dynamic
and static spin structure factor, respectively. In the absence of
the SOC, both these quantities vanish identically. Hence a nonzero
spin structure factor is a direct consequence of triplet pairing \cite{Gorkov2001}.
Note that spin structure factor can be measured using the Bragg spectroscopy
method as demonstrated in recent experiments \cite{Hoinka2012}.

The condensate fraction and superfluid density of the rashbon system
have also been studied \cite{Zhou2012,he3}, and have been found to
exhibit unusual behaviors: The condensate fraction is generally enhanced
by the SOC due to the increase of the pair binding; while the superfluid
density is suppressed because of the nontrivial effective mass of
rashbons.

\begin{figure}[h]
\begin{centering}
\includegraphics[clip,width=0.5\textwidth]{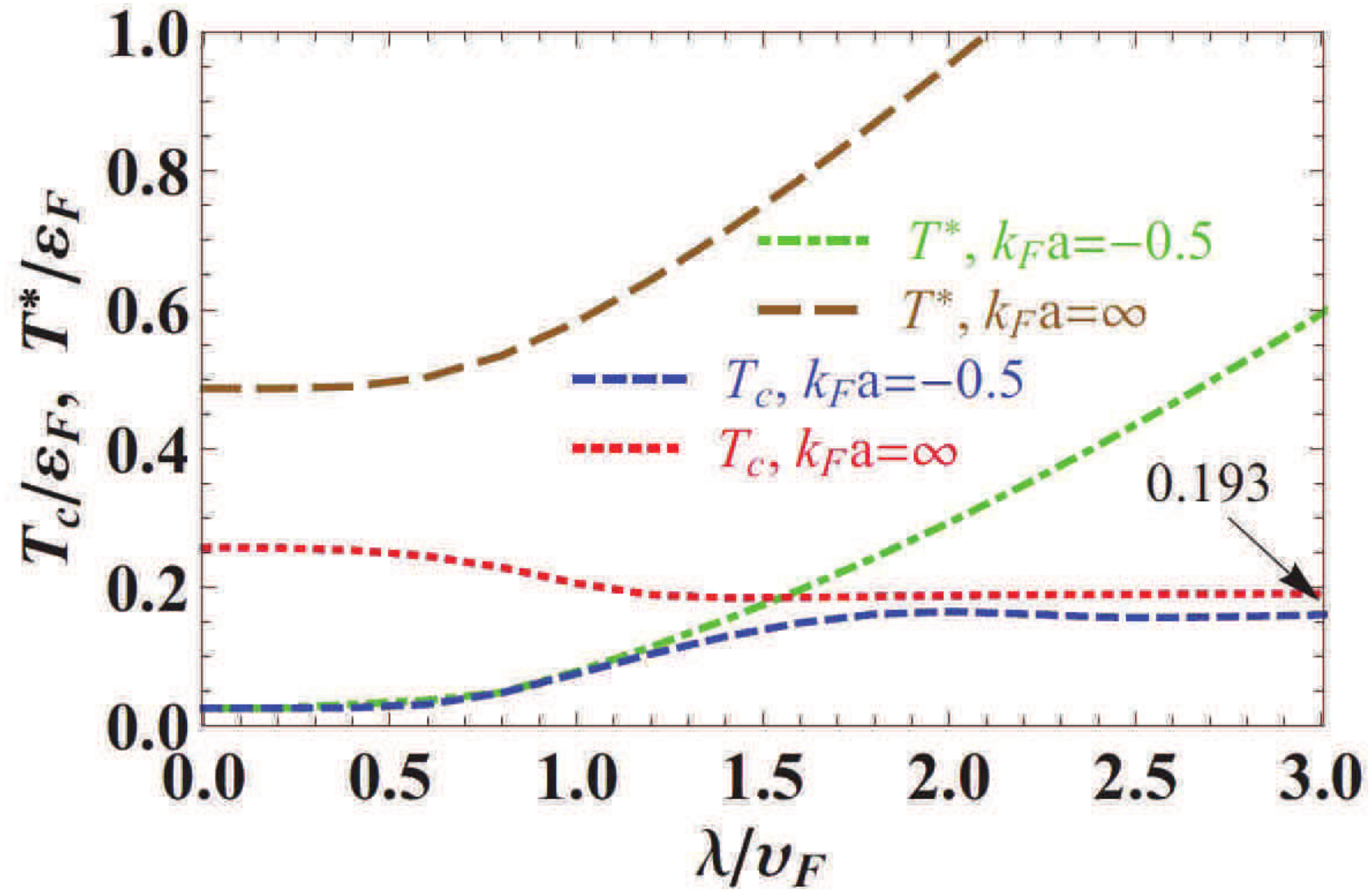} 
\par\end{centering}

\caption{Critical temperature $T_{c}$ and the dissociation temperature $T^{\ast}$
scaled by the Fermi energy $\lyxmathsym{\textgreek{e}}_{F}$ as a
function of the SOC strength $\lambda/v_{F}$ for fixed gas parameters
$ $$1/(k_{F}a_{s})=\lyxmathsym{\textminus}2$ and $1/(k_{F}a_{s})=0$.
Here, we measure the SOC strength in units of Fermi velocity $v_{F}=k_{F}/m$.
Figure taken from Ref. \cite{he1} with modification.}

\label{fig20} 
\end{figure}

To understand the finite-temperature properties of rashbons, the mean-field
approach becomes less reliable. So far, a careful analysis based on
the pair-fluctuation theory as outlined in the theoretical framework
is yet to be performed. In Fig. \ref{fig20}, we show the superfluid
transition temperature as a function of the Rashba SOC strength, predicted
by the approximate many-body \textit{T}-matrix theory - pseudogap
theory \cite{he1}. At sufficiently large SOC strength, $T_{c}$ tends
to the critical temperature of a rashbon BEC given by Eq. (\ref{eq: TcRashbonBEC})
- $T_{c}\simeq0.193T_{F}$ - regardless of the dimensionless interaction
parameter $1/(k_{F}a_{s})$, as we may anticipate. For a more detailed
discussion of the crossover from BCS to rashbon BEC, we refer to Ref.
\cite{vj3}.

\subsubsection{2D Topological superfluidity}

Here we consider 2D topological superfluidity with 2D Rashba SOC,
in the presence of an out-of-plane Zeeman field $h$. It is of particular
interest, considering the possibility of performing topological quantum
computation. This is because each vortex core in a 2D topological
superfluid can host a Majorana fermion. Thus, by properly interchanging
two vortices and thus braiding Majorana fermions, fault-tolerant quantum
information stored non-locally in Majorana fermions may be processed
\cite{Ivanov2001,Nayak2008}. In the context of ultracold atoms, the
use of 2D Rashba SOC to create a 2D topological superfluid was first
proposed by Zhang and co-workers \cite{Zhang2008}, and later considered
by a number of researchers \cite{Zhu2011,liu2,wan,sademelo2,he2,iskin2,zhang2,impurity}.
In free space, the criterion to enter topological superfluid phase
is given by $h>\sqrt{\mu^{2}+\Delta^{2}}$, above which the system
behaves like a 2D weak-coupling \textit{p}-wave superfluid, as we
already discussed in the previous subsection (see, for example, Eq.
(\ref{eq: criterionTS})). Here, we are interested in the nature of
2D topological superfluids for the experimentally relevant situation
with the presence of harmonic traps \cite{liu2}.

Theoretically, we solve numerically the BdG equation (\ref{bdgeq}).
In the presence of a single vortex at trap center, we take $\Delta_{0}({\bf r})=\Delta_{0}(r)e^{-i\varphi}$
and decouple the BdG equation into different angular momentum channels
indexed by an integer $m$. The quasiparticle wave functions take
the form, $[u_{\uparrow\eta}(r)e^{-i\varphi},u_{\downarrow\eta}(r),v_{\uparrow\eta}(r)e^{i\varphi},v_{\downarrow\eta}(r)]e^{i(m+1)\varphi}/\sqrt{2\pi}$.
We have solved self-consistently the BdG equations using the basis
expansion method. For the results presented below, we have taken $N=400$
and $T=0$. We have used $E_{a}=0.2E_{F}$ and $\lambda k_{F}/E_{F}=1$,
where the binding energy $E_{a}$ is a useful parameter to characterize
the interatomic interaction in 2D. These are typical parameters that
can be readily realized in a 2D $^{40}$K Fermi gas.

\begin{figure}[h]
\begin{centering}
\includegraphics[clip,width=0.7\textwidth]{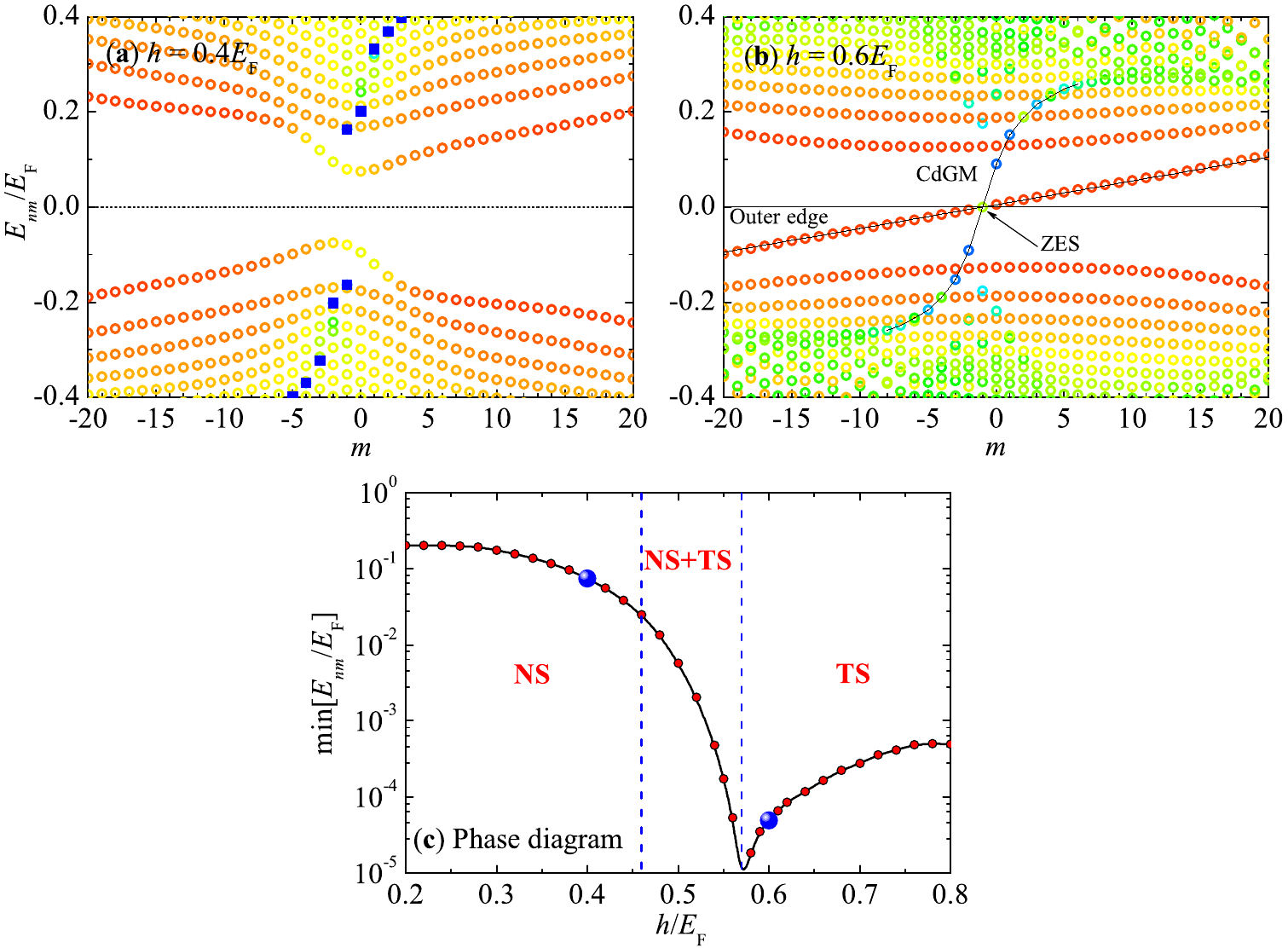} 
\par\end{centering}

\caption{(a) and (b) Energy spectrum at $h/E_{F}=0.4$ and $0.6$ (where $E_{F}=k_{F}^{2}/(2M)=\sqrt{N}\omega_{\perp}$
is the Fermi energy) in the presence of a single vortex. The color
of symbols indicates the mean radius $\sqrt{\left\langle r^{2}\right\rangle }/r_{F}$
(where $r_{F}=(4N)^{1/4}\sqrt{1/(M\omega_{\perp})}$ is the Fermi
radius) of the eigenstate, which is defined by $\left\langle r^{2}\right\rangle =\int r^{2}[\left|u_{\uparrow}\right|^{2}+\left|u_{\downarrow}\right|^{2}+\left|v_{\uparrow}\right|^{2}+\left|v_{\downarrow}\right|^{2}]d{\bf r}$.
The color of symbols changes from blue when the excited state is localized
at the trap center to red when its mean radius approaches the Thomas-Fermi
radius. In (a) and (b), the CdGM states are indicated by blue squares
and a solid line, respectively. (c) Phase diagram, along with the
lowest eigenenergy of Bogoliubov spectrum. Figure taken from Ref.
\cite{liu2} with modification.}

\label{fig21} 
\end{figure}

Figs. \ref{fig21} reports the phase diagram {[}Fig. \ref{fig21}(c){]}
along with the quasiparticle energy spectrum of different phases {[}Figs.
\ref{fig21}(a) and \ref{fig21}(b){]} in the presence of a single
vortex. By increasing the Zeeman field, the system evolves from a
non-topological state (NS) to a topological state (TS), through an
intermediate mixed phase in which NS and TS coexist. The topological
phase transition into TS is well characterized by the low-lying quasiparticle
spectrum, which has the particle-hole symmetry $E_{m+1}=-E_{-(m+1)}$.
As shown in Fig. \ref{fig21}(a), the spectrum of the NS is gapped. While in the TS, two branches of mid-gap states
with small energy spacing appear: One is labeled by ``Outer edge''
and another ``CdGM'' which refers to localized states at the vortex
core, i.e., the so--called Caroli-de Gennes-Matricon (CdGM) states
\cite{CdGM}. The eigenstates with nearly zero energy at $m=-1$ could
be identified as the zero-energy Majorana fermions in the thermodynamic
limit.

\begin{figure}[h]
\begin{centering}
\includegraphics[clip,width=0.8\textwidth]{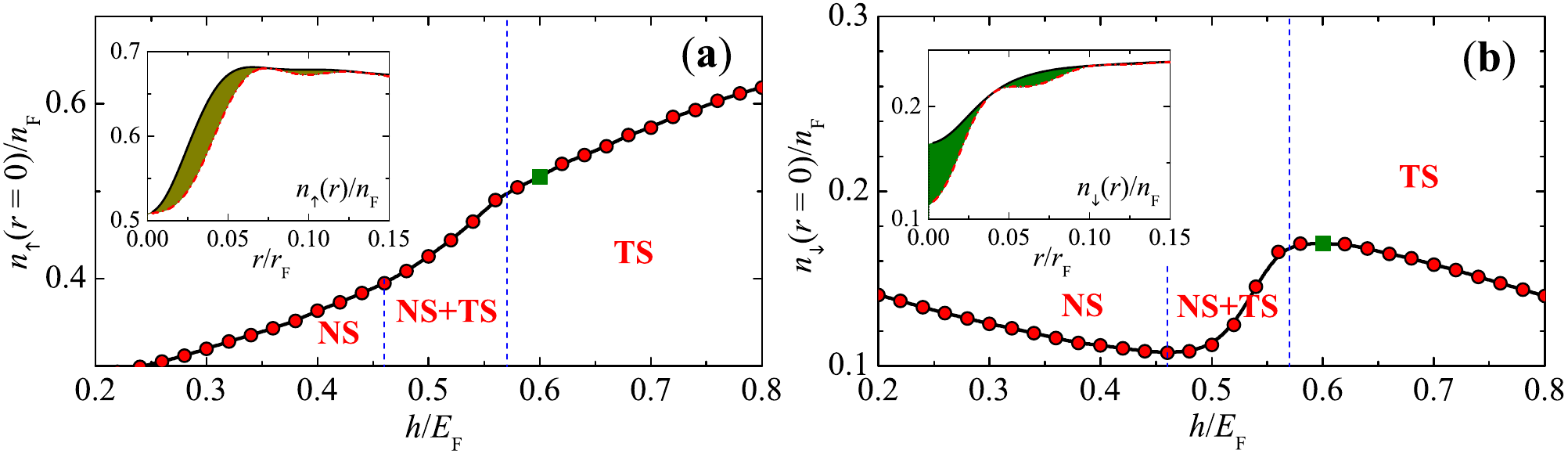} 
\par\end{centering}

\caption{Zeeman field dependence of spin-up (a) and spin-down (b) densities
at the vortex core. The density is normalized by the Thomas-Fermi
density $n_{F}=(\sqrt{N}/\pi)\sqrt{M\omega_{\perp}}$. The insets
show the core density distributions at $h=0.6E_{F}$. The red dot-dashed
lines show the result by excluding artificially the Majorana vortex
core state, whose contribution is shown by the shaded area. Figure
taken from Ref. \cite{liu2} with modification.}

\label{fig22} 
\end{figure}

In the TS, the occupation of the Majorana vortex-core state affects
significantly the atomic density and the local density of states (LDOS) of the Fermi gas near
the trap center, which in turn gives a strong experimental signature
for observing Majorana fermions. Fig. \ref{fig22} presents the spin-up
and -down densities at the trap center, $n_{\uparrow}(0)$ and $n_{\downarrow}(0)$,
as a function of the Zeeman field. In general, $n_{\uparrow}(0)$
and $n_{\downarrow}(0)$ increases and decreases respectively with
increasing field. However, we find a sharp increase of $n_{\downarrow}(0)$
when the system evolves from the mixed phase to the full TS. Accordingly,
a change of slope or kink appears in $n_{\uparrow}(0)$. The increase
of $n_{\downarrow}(0)$ is associated with the \emph{gradual} formation
of the Majorana vortex-core mode, whose occupation contributes notably
to atomic density due to the \emph{large} amplitude of its localized
wave function. We plot in the inset of Fig.~\ref{fig22}(b) $n_{\downarrow}(0)$
at $h=0.6E_{F}$, with or without the contribution of the Majorana
mode, which is highlighted by the shaded area. This contribution is
apparently absent in the NS. Thus, a sharp increase of $n_{\downarrow}(0)$,
detectable in {\em in situ} absorption imaging, signals the topological
phase transition and the appearance of the Majorana vortex-core mode.
This feature persists at typical experimental temperature, i.e., $T=0.1T_{F}$.

In the presence of impurity scattering, topological superfluid can
also host a universal impurity-induced bound state \cite{impurity,impurity1d}.
That is, regardless of the type of impurities, magnetic or non-magnetic,
the impurity will always cause the same bound state within the pairing
gap, provided the scattering strength is strong enough. The observation
of such a universal impurity-induced bound state will give a clear
evidence for the existence of topological superfluids.

\section{Experiments}

We now review the experimental work, focusing on the ones carried
out at Shanxi University. The apparatus and cooling scheme in the
experiment have been described in previous papers \cite{shanone,shantwo,shanthree,shanfour,shanfive}
and briefly introduced here (see Fig.~\ref{fig1}). An atomic mixture
sample of $^{87}$Rb and $^{40}$K atoms in hyperfine state $|F=2,m_{F}=2\rangle$
and $|F=9/2,m_{F}=9/2\rangle$, respectively, are first pre cooled
to 1.5 $\mu K$ by radio-frequency evaporative cooling in a quadrupole-Ioffe
configuration (QUIC) trap. The QUIC trap consists of a pair of anti-Helmholtz
coils and a third coil in perpendicular orientation. To gain larger
optical access, the atoms are first transported from the QUIC trap
to the center of the quadrupole coils (glass cell) by lowering the
current passing through quadrupole coils and increasing the current
in the Ioffe coil, and then are transferred into an crossed optical
trap in the horizontal plane, created by two off-resonance laser beams,
at a wavelength of 1064 nm. A degenerate Fermi gas of about $N\simeq2\times10^{6}$
$^{40}$K atoms in the $|9/2,9/2\rangle$ internal state at $T/T_{F}\simeq0.3$
is created inside the crossed optical trap. Here $T$ is the temperature
and $T_{F}$ is the Fermi temperature defined by $T_{F}=E_{F}/k_{B}=(6N)^{1/3}\overline{\omega}/k_{B}$
with a geometric mean trapping frequency $\overline{\omega}\simeq2\pi\times130$
Hz. A $780$ nm laser pulse of $0.03$ ms is used to remove all the
$^{87}$Rb atoms in the mixture without heating $^{40}$K atoms.

To create SOC, a pair of Raman laser beams are extracted from a continuous-wave
Ti-sapphire single frequency laser. The two Raman beams are frequency-shifted
by two single-pass acousto-optic modulators (AOM) respectively. In
this way the relative frequency difference $\Delta\omega$ between
the two laser beams is precisely controlled. At the output of the
optical fibers, the two Raman beams each has a maximum intensity $I=130$
$mW$, counter-propagating along the ${x}$-axis with a $1/e^{2}$
radius of 200 $\mu m$ and are linearly polarized along the ${z}$-
and ${y}$-axis, respectively, which correspond to $\pi$ ($\sigma$)
and $\sigma$ ($\pi$) of the quantization axis $\hat{z}$ ($\hat{y}$).
The momentum transferred to atoms during the Raman process is $2k_{0}=2k_{r}\sin(\theta/2)$,
where $k_{r}=2\pi/\lambda$ is the single-photon recoil momentum,
$\lambda$ is the wavelength of the Raman beam, and $\theta$ is the
intersecting angle of two Raman beams. Here, $k_{r}$ and $E_{r}=k_{r}^{2}/2m$
are the units of momentum and energy. The optical transition wavelengths
of the D1 and D2-line are 770.1 nm and 766.7 nm, respectively. The
wavelengths of the Raman lasers are about 772 $\sim$ 773 nm.

The two internal states involved in SOC are chosen as follows. In
the case of noninteracting system, the two states are magnetic sublevels
$\mid\uparrow\rangle=|9/2,9/2\rangle$ and $\mid\downarrow\rangle=|9/2,7/2\rangle$.
These two spin states are stable and are weakly interacting with a
background $s$-wave scattering length $a_{\text{s}}=169a_{0}$. We
use a pair of Helmholtz coils along the ${y}$-axis (as shown in Fig.~\ref{fig1})
to provide a homogeneous bias magnetic field, which gives a Zeeman
shift between the two magnetic sublevels. A Zeeman shift of $\omega_{{\rm Z}}=2\pi\times10.27$
MHz between these two magnetic sublevels is produced by a homogeneous
bias magnetic field of $31$ G. When the Raman coupling is at resonance
(at $\Delta\omega=2\pi\times10.27$ MHz and two-photon Raman detuning
$\delta=\Delta\omega-\omega_{{\rm Z}}\approx0$), the detuning between
$|9/2,7/2\rangle$ and other magnetic sublevels like $|9/2,5/2\rangle$
is about $2\pi\times170$ kHz, which is one order of magnitude larger
than the Fermi energy. Hence all the other states can be safely neglected.
In the case of the strongly interacting spin-orbit coupled Fermi gas,
two magnetic sublevels $\mid\downarrow\rangle=|9/2,-9/2\rangle$ and
$\mid\uparrow\rangle=|9/2,-7/2\rangle$ are chosen. To create strong
interaction, the bias field is ramped from 204 G to a value near the
$B_{0}=202.1$ G Feshbach resonance at a rate of about 0.08 G/ms.
We remark that due to a decoupling of the nuclear and electronic spins,
the Raman coupling strength decreases with increasing of the bias
field \cite{Wei2013}. When working at a large bias magnetic field,
we have to use a smaller detuning of the Raman beams with respect
to the atomic D1 transition in order to increase the Raman coupling
strength.

In order to control the magnetic field precisely and reduce the magnetic
field noise, the power supply (Delta SM70-45D) has been operated in
remote voltage programming mode, whose voltage is set by an analog
output of the experiment control system. The current through the coils
is controlled by the external regulator relying on a precision current
transducer (Danfysik ultastable 867-60I). The current is detected
with the precision current transducer, then the regulator compares
the measured current value to a set voltage value from the computer.
The output error signal from the regulator actively stabilize the
current with the PID (proportional-integral-derivative) controller
acting on the MOSFET (metal-oxide-semiconductor field-effect transistor).
In order to reduce the current noise and decouple the control circuit
from the main current, a conventional battery is used to power the
circuit.

We use the standard time-of-flight technique to perform our measurement.
To this end, the Raman beams, optical dipole trap and the homogeneous
bias magnetic field are turned off abruptly at the same time, and
a magnetic field gradient along the ${y}$-axis provided by the Ioffe
coil is turned on. The two spin states are separated along the ${y}$-direction,
and imaging of atoms along the ${z}$-direction after $12$ ms expansion
gives the momentum distribution for each spin component.

\subsection{ The noninteracting spin-orbit coupled Fermi gas}

In this section, we review the experiment on non-interacting system.

\subsubsection{Rabi oscillation}

\begin{figure}
\includegraphics[width=0.5\textwidth]{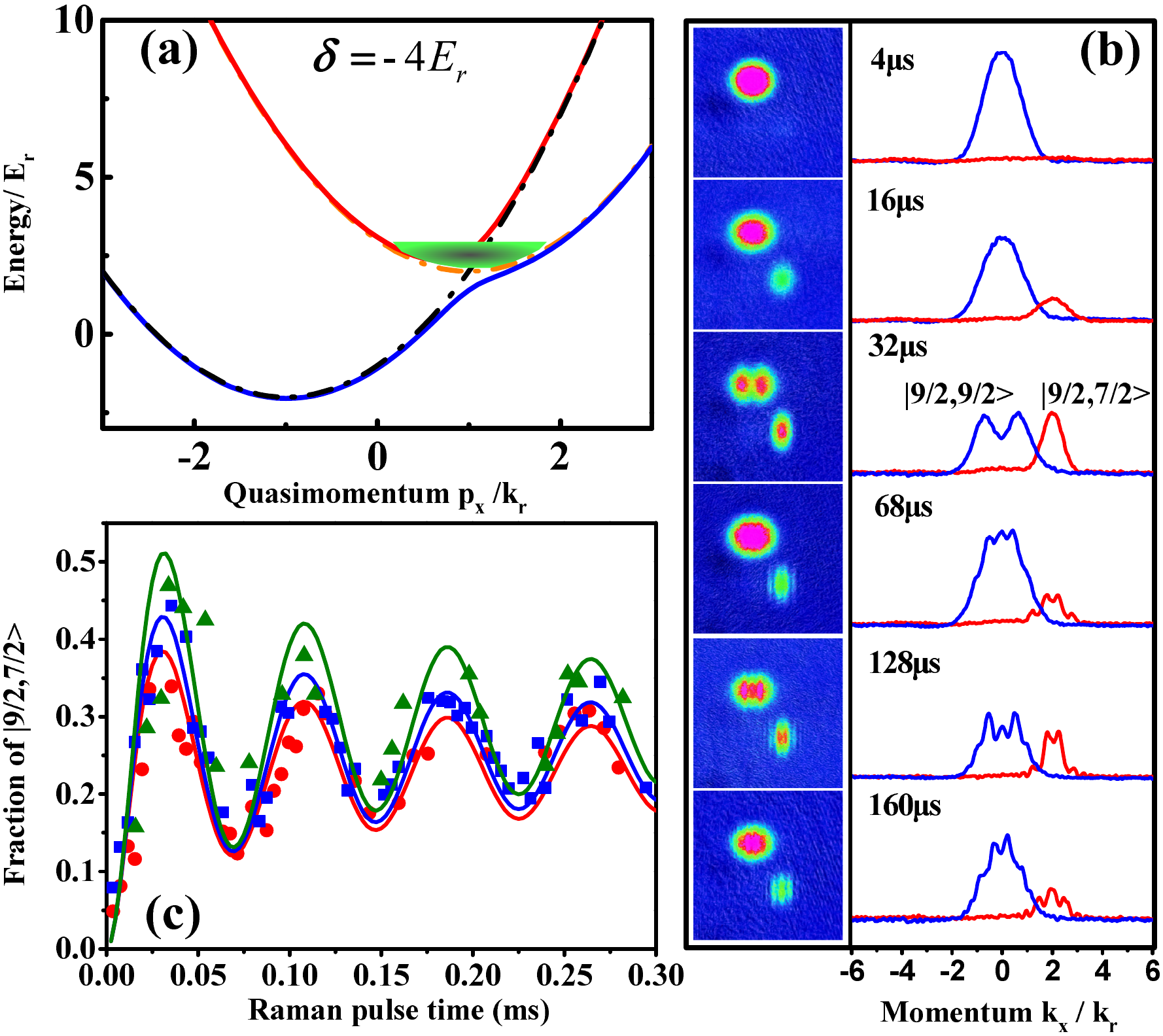} \caption{Raman-induced Rabi oscillation: (a) The energy dispersion with $\delta=-4E_{\text{r}}$.
The system is initially prepared with all atoms in $|9/2,9/2\rangle$
state. (b) Time-of-flight image (left) and integrated time-of-flight
image (integrated along $\hat{y}$) at different duration time for
$|\uparrow\rangle$ (blue) and $|\downarrow\rangle$ (red). The parameters
are $k_{\text{F}}=1.35k_{\text{r}}$ and $T/T_{\text{F}}=0.35$. (c)
The population in $|9/2,7/2\rangle$ as a function of duration time
of Raman pulse. $k_{\text{F}}=1.9k_{\text{r}}$ and $T/T_{\text{F}}=0.30$
for red circles, $k_{\text{F}}=1.35k_{\text{r}}$ and $T/T_{\text{F}}=0.35$
for blue squares, $k_{\text{F}}=1.1k_{\text{r}}$ and $T/T_{\text{F}}=0.29$
for green triangles. The solid lines are theory curves with $\Omega=1.52E_{\text{r}}$.
Figure taken from Ref.~\cite{Jing_fermionic}. \label{Rabi} }
\end{figure}

We first study the Rabi oscillation between the two spin states induced
by the Raman coupling. All atoms are initially prepared in the $\mid\uparrow\rangle$
state. The homogeneous bias magnetic field is ramped to a certain
value so that $\delta=-4E_{\text{r}}$, that is, the ${\bf k}=0$
component of state $\mid\uparrow\rangle$ is at resonance with ${\bf k}=2k_{\text{r}}\hat{x}$
state of $\mid\downarrow\rangle$ component, as shown in Fig. \ref{Rabi}(a).
Then we apply a Raman pulse to the system, and measure the spin population
for different duration time of the Raman pulse. Similar experiment
in bosonic system yields an undamped and completely periodic oscillation,
which can be well described by a sinusoidal function with frequency
$\Omega$ \cite{lin1}. This is because for bosons, macroscopic number
of atoms occupy the resonant ${\bf k}=0$ mode, and therefore there
is a single Rabi frequency determined by the Raman coupling only.
While for fermions, atoms occupy different momentum states. Due to
the effect of SOC, the coupling between the two spin states and the
resulting energy splitting are momentum dependent, and atoms in different
momentum states oscillate with different frequencies. Hence, dephasing
naturally occurs and the oscillation will be inevitably damped after
several oscillation periods. In our case, the spin-dependent momentum
distribution shown in Fig.~\ref{Rabi}(b) clearly shows the out-of-phase
oscillation for different momentum states.

For a non-interacting system, the population of $\mid\downarrow\rangle$
component is given by 
\begin{equation}
n_{\downarrow}({\bf k}+2k_{\text{r}}\hat{x},{\bf r},t)=n_{\uparrow}({\bf k},{\bf r},0)\frac{\sin^{2}\sqrt{(k_{x}k_{\text{r}}/m)^{2}+\Omega^{2}/4}\, t}{1+\left(\frac{2k_{x}k_{\text{r}}}{\Omega m}\right)^{2}},\label{Rabitheory}
\end{equation}
where $t$ is the duration time of Raman pulse, $n_{\uparrow}({\bf k},{\bf r},0)$
is the equilibrium distribution of the initial state in local density
approximation. From Eq.~(\ref{Rabitheory}) one can see that the
momentum distribution along the ${x}$-axis of the $\mid\downarrow\rangle$
component is always symmetric respect to $2k_{\text{r}}$ at any time,
which is clearly confirmed by the experimental data as shown in Fig.~\ref{Rabi}(b).
The total population in the $\mid\downarrow\rangle$ component is
given by $N_{\downarrow}(t)=\int{d{\bf k}d{\bf r}}\, n_{\downarrow}({\bf k},{\bf r},t)$,
and in Fig.~\ref{Rabi}(c), one can see that there is an excellent
agreement between the experiment data and theory, from which we determine
$\Omega=1.52(5)E_{{\rm r}}$.

\begin{figure}
\includegraphics[width=0.5\textwidth]{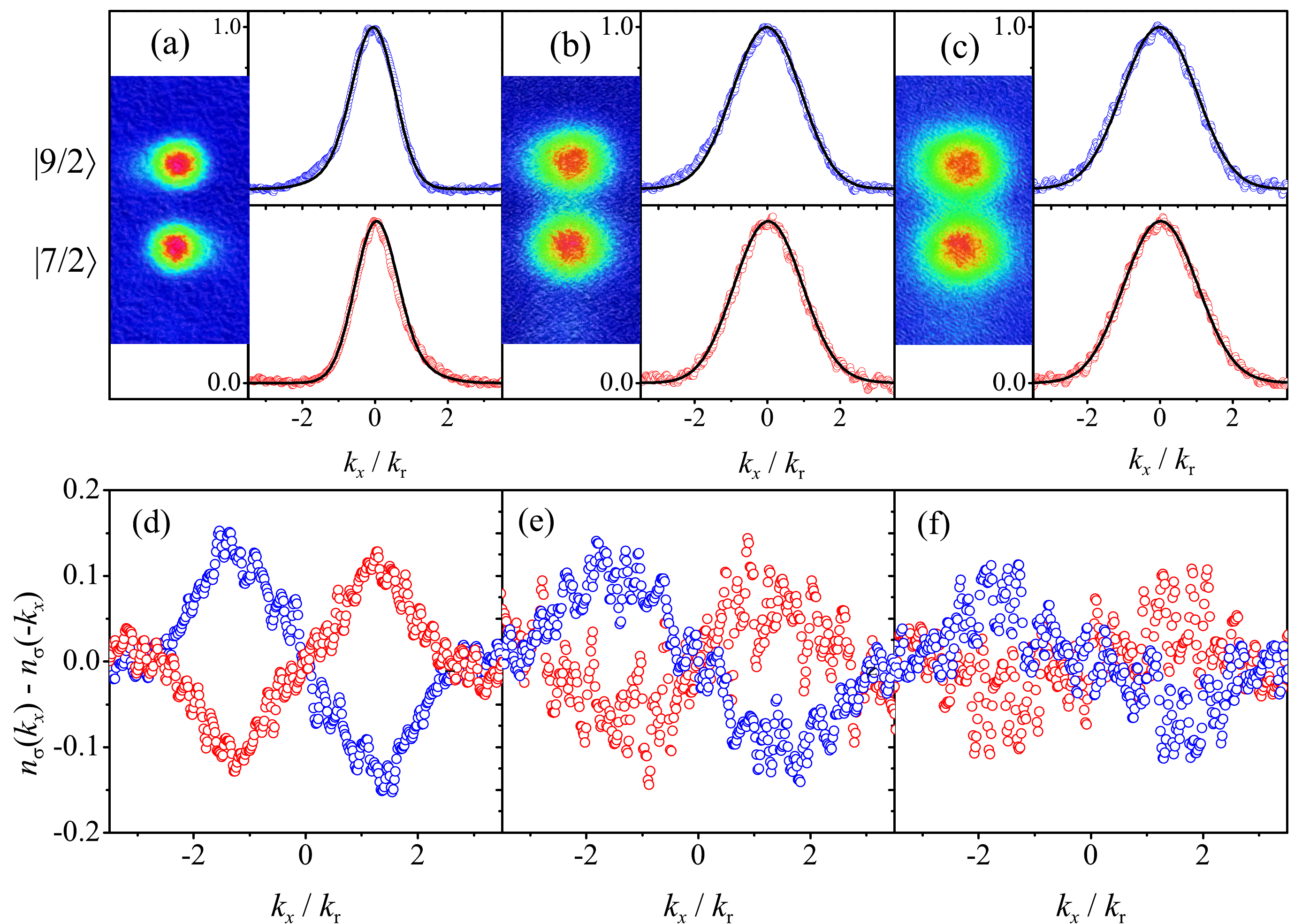}
\caption{Momentum distribution asymmetry as a hallmark of SOC: (a-c) time-of-flight
measurement of momentum distribution for both $|\uparrow\rangle$
(blue) and $|\downarrow\rangle$ (red). Solid lines are theory curves.
(a) $k_{\text{F}}=0.9k_{\text{r}}$ and $T/T_{\text{F}}=0.8$ (b)
$k_{\text{F}}=1.6k_{\text{r}}$ and $T/T_{\text{F}}=0.63$; (c) $k_{\text{F}}=1.8k_{\text{r}}$
and $T/T_{\text{F}}=0.57$. (d-f): plot of integrated momentum distribution
$n_{\sigma}({\bf k})-n_{\sigma}({\bf -k})$ for the case of (a-c).
Figure taken from Ref.~\cite{Jing_fermionic}. \label{distribution} }
\end{figure}

\subsubsection{Momentum distribution}

We focus on the case with $\delta=0$, and study the momentum distribution
in the equilibrium state. We first transfer half of $^{40}$K atoms
from $\mid\downarrow\rangle$ to $\mid\uparrow\rangle$ using radio
frequency sweep within $100$ ms. Then the Raman coupling strength
is ramped up adiabatically in $100$ ms from zero to its final value
and the system is held for another $50$ ms before time-of-flight
measurement. Since SOC breaks spatial reflection symmetry ($x\rightarrow-x$
and $k_{x}\rightarrow-k_{x}$), the momentum distribution for each
spin component will be asymmetric, i.e. $n_{\sigma}({\bf k})\neq n_{\sigma}({\bf -k})$,
with $\sigma=\uparrow,\downarrow$. On the other hand, when $\delta=0$
the system still satisfies $n_{\uparrow}({\bf k})=n_{\downarrow}(-{\bf k})$.
The asymmetry can be clearly seen in the spin-resolved time-of-flight
images and integrated distributions displayed in Fig.~\ref{distribution}(a)
and (b), where the fermion density is relatively low. While it becomes
less significant when the fermion density becomes higher, as shown
in Fig.~\ref{distribution}(c), because the strength of the SOC is
relatively weaker compared to the Fermi energy. Although the presence
of the Raman lasers cause additional heating to the cloud, we find
that the temperature is within the range of $0.5-0.8T_{\text{F}}$,
which is still below degenerate temperature. In Fig.~\ref{distribution}(d-f),
we also show $n_{\sigma}(k_{x})-n_{\sigma}(-k_{x})$ to reveal the
momentum distribution asymmetry more clearly.

\subsubsection{Lifshitz transition }

\begin{figure}
\includegraphics[width=0.5\textwidth]{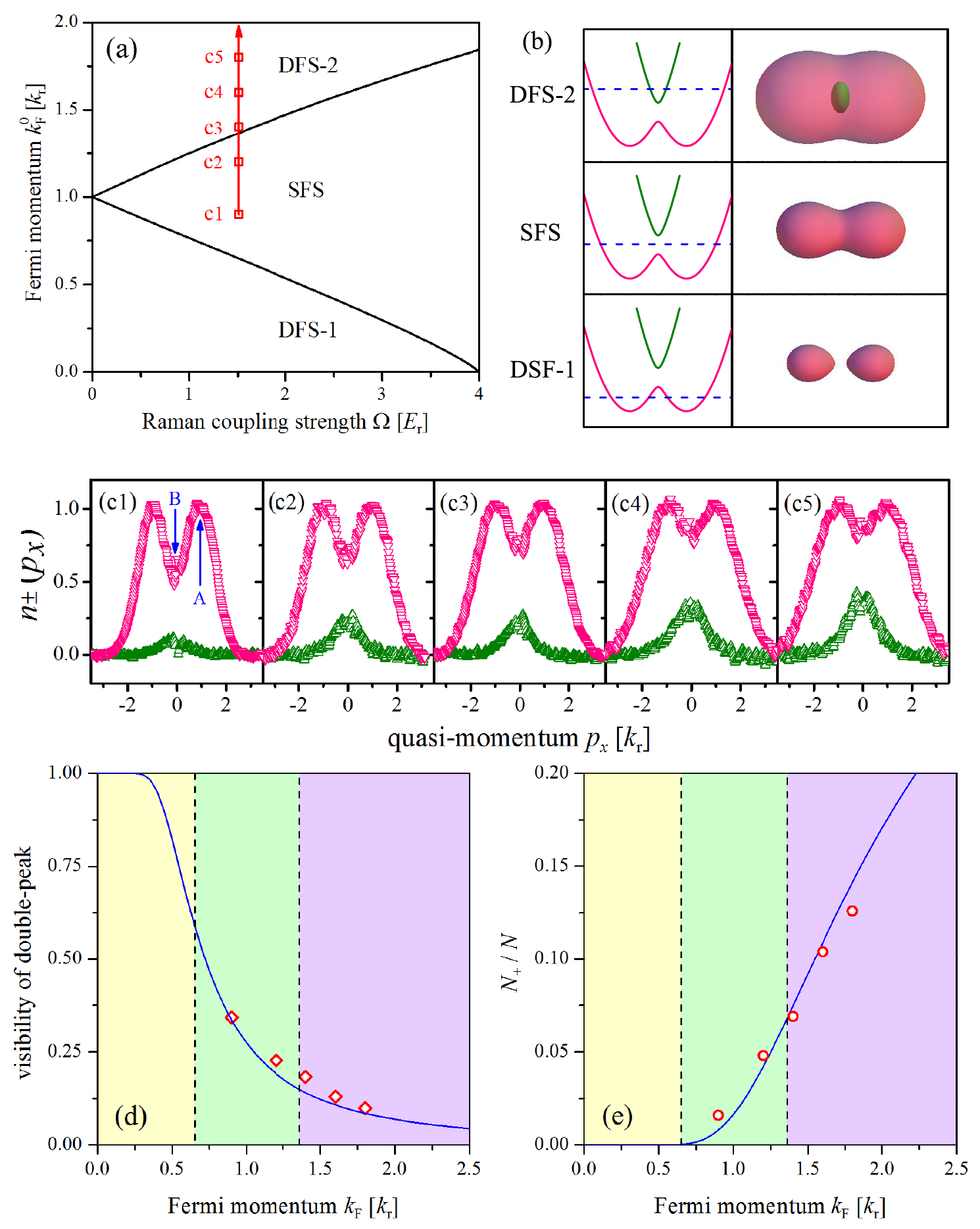} \caption{Topological change of Fermi surface and Lifshitz transition: (a) Theoretical
phase diagram at $T=0$. $k_{\text{F}}^{0}=(3\pi^{2}n)^{1/3}$. ``SFS''
means single Fermi surface. ``DFS'' means double Fermi surface.
(b) Illustration of different topology of Fermi surfaces. The single
particle energy dispersion is drawn for small $\Omega$. Dashed blue
line is the chemical potential. (c) Quasi-momentum distribution in
the helicity bases. Red and green points are distributions for $s=-1$
and $s=1$ helicity branches, respectively. $k_{\text{F}}=0.9k_{\text{r}}$,
$T/T_{\text{F}}=0.80$ for (c1); $k_{\text{F}}=1.2k_{\text{r}}$,
$T/T_{\text{F}}=0.69$ for (c2); $k_{\text{F}}=1.4k_{\text{r}}$,
$T/T_{\text{F}}=0.61$ for (c3); $k_{\text{F}}=1.6k_{\text{r}}$,
$T/T_{\text{F}}=0.63$ for (c4); $k_{\text{F}}=1.8k_{\text{r}}$,
$T/T_{\text{F}}=0.57$ for (c5). All these points are marked on phase
diagram in (a). (d) Visibility $v=(n_{\text{A}}-n_{\text{B}})/(n_{\text{A}}+n_{\text{B}})$
decreases as $k_{\text{F}}/k_{\text{r}}$ increases (A and B points
are marked in (c1)). (e) Atom number population in $s=1$ helicity
branch $N_{+}/N$ increases as $k_{\text{F}}/k_{\text{r}}$ increases
increases. In both (d) and (e), the blue solid line is a theoretical
curve with $T/T_{\text{F}}=0.65$, and the background color indicates
three different phases in the phase diagram. Figure taken from Ref.~\cite{Jing_fermionic}.
\label{transition} }
\end{figure}

With SOC, the single particle spectra of Eq.~(\ref{HSO}) are dramatically
changed from two parabolic dispersions into two helicity branches
as shown in Fig. \ref{transition}(b). Here, two different branches
are eigenstates of ``helicity'' $\hat{s}$ and the ``helicity''
operator describes whether spin $\boldsymbol{\sigma}_{{\bf p}}$ is
parallel or anti-parallel to the ``effective Zeeman field'' ${\bf h}=(-\Omega,0,k_{\text{r}}p_{x}/m+\delta)$
at each momentum, i.e. $\hat{s}=\boldsymbol{\sigma}_{{\bf p}}\cdot{\bf h}/|\boldsymbol{\sigma}_{{\bf p}}\cdot{\bf h}|$.
$s=1$ for the upper branch and $s=-1$ for the lower branch. The
topology of Fermi surface exhibits two transitions as the atomic density
varies. At sufficient low density, it contains two disjointed Fermi
surfaces with $s=-1$, and they gradually merge into a single Fermi
surface as the density increases to $n_{\text{c}1}$. Finally a new
small Fermi surface appears at the center of large Fermi surface when
density further increases and fermions begin to occupy $s=1$ helicity
branch at $n_{\text{c}2}$. A theoretical ground state phase diagram
for the uniform system is shown in Fig.~\ref{transition}(a), and
an illustration of the Fermi surfaces at different density are shown
in Fig.~\ref{transition}(b). Across the phase boundaries, the system
experiences Lifshitz transitions as density increases \cite{Lifshitz},
which is a unique property in a Fermi gas due to Pauli principle.

We fix the Raman coupling and vary the atomic density at the center
of the trap, as indicated by the red arrow in Fig.~\ref{transition}(a).
In Fig.~\ref{transition}(c1-c5), we plot the quasi-momentum distribution
in the helicity bases for different atomic density. At the lowest
density, the $s=1$ helicity branch is nearly unoccupied, which is
consistent with that the Fermi surface is below $s=1$ helicity branch.
The quasi-momentum distribution of the $s=-1$ helicity branch exhibits
clearly a double-peak structure, which reveals that the system is
close to the boundary of having two disjointed Fermi surfaces at $s=-1$
helicity branch. As density increases, the double-peak feature gradually
disappears, indicating the Fermi surface of $s=-1$ helicity branch
finally becomes a single elongated one, as the top one in Fig.~\ref{transition}(b).
Here we define a quality of visibility $v=(n_{\text{A}}-n_{\text{B}})/(n_{\text{A}}+n_{\text{B}})$,
where $n_{\text{A}}$ is the density of $s=-1$ branch at the peak
and $n_{\text{B}}$ is the density at the dip between two peaks. Theoretically
one expects $v$ approaches unity at low density regime and approaches
zero at high density regime. In Fig.~\ref{transition}(d) we show
that our data decreases as density increases and agrees very well
with a theoretical curve with a fixed temperature of $T/T_{\text{F}}=0.65$.
Moreover, across the phase boundary between SFS and DFS-1, one expects
a significant increase of population on $s=1$ helicity branch. In
Fig.~\ref{transition}(e), the fraction of atom number population
at $s=1$ helicity branch is plotted as a function of Fermi momentum
$k_{{\rm F}}$, which grows near the critical point predicted in zero-temperature
phase diagram. The blue solid line is a theoretical calculation for
$N_{+}/N$ with $T/T_{\text{F}}=0.65$, and the small deviation between
the data and this line is due to the temperature variation between
different measurements. Because the temperature is too high, the transition
is smeared out. For both $v$ and $N_{+}/N$ we observe only a smooth
decreasing or growth across the regime where it is supposed to have
a sharp transition, however, the agreement with theory suggests that
with better cooling a sharper transition should be observable.

\subsubsection{Momentum-resolved rf spectrum}

\begin{figure}
\includegraphics[width=0.5\textwidth]{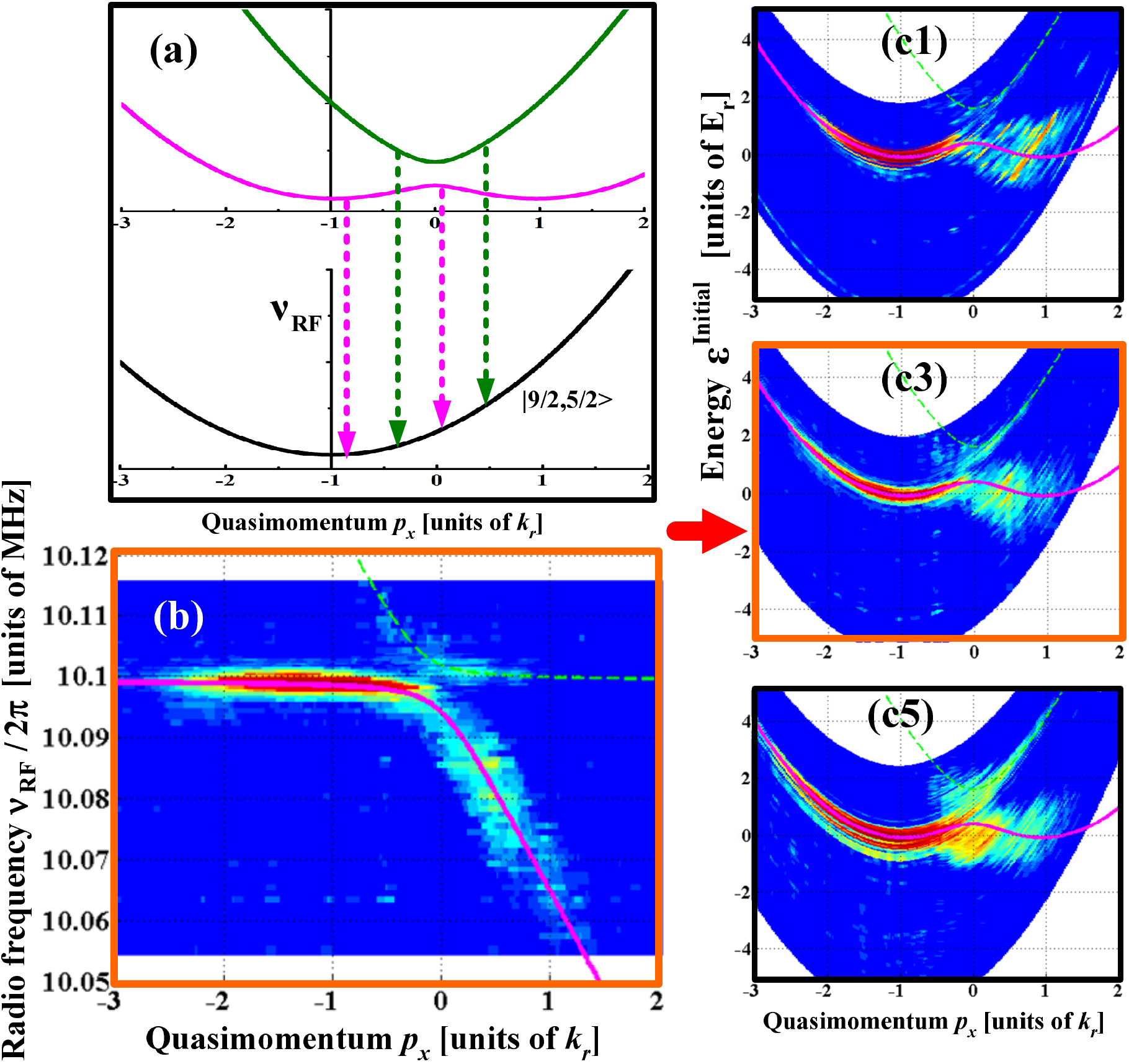}
\caption{Momentum-resolved rf spectroscopy of a spin-orbit coupled Fermi gas:
(a) Schematic of momentum-resolved rf spectroscopy of SO coupled Fermi
gases. Green and pink solid lines are two helicity branches in which
the eigenstates are all superposition of $|9/2,9/2\rangle$ and $|9/2,7/2\rangle$.
Thus both can undergo rf transition from $|9/2,7/2\rangle$ to $|9/2,5/2\rangle$,
as indicated by dashed lines. (b) Intensity map of the atoms in $|9/2,5/2\rangle$
state as a function of ($\nu_{RF},k_{x}$) plane. (c) Single particle
dispersion and atom population measured for (c1), (c3) and (c5) in
Fig.~\ref{transition}. Figure taken from Ref.~\cite{Jing_fermionic}.
\label{spectra} }
\end{figure}

The effect of SOC is further studied with momentum resolved rf spectroscopy
\cite{Stewart2008}, which maps out the single-particle dispersion relation. A Gaussian shaped pulse of rf field is applied
for 200 $\mu s$ to transfer atoms from $|9/2,7/2\rangle$ ($\left|\downarrow\right\rangle $)
state to the final state $|9/2,5/2\rangle$, as shown in Fig.~\ref{spectra}(a),
and then the spin population at $|9/2,5/2\rangle$ is measured with
time-of-flight at different rf frequencies. In Fig.~\ref{spectra}(b)
we plot an example of the final state population as a function of
momentum $p_{x}$ and the frequency of rf field $\nu_{\text{RF}}$,
from which one can clearly see the back-bending feature and the gap
opening at the Dirac point. Both are clear evidences of SOC.

For an occupied state, the initial state dispersion $\epsilon_{i}({\bf k})$
can be mapped out by 
\begin{eqnarray}
\epsilon_{i}({\bf k})=\nu_{RF}-E_{Z}+\epsilon_{f}({\bf k}).
\end{eqnarray}
where $\epsilon_{f}({\bf k})=\textbf{k}^{2}/2m$ is the dispersion
of the final $|9/2,5/2\rangle$ state, and $E_{Z}$ is the energy
difference between $|9/2,7/2\rangle$ and $|9/2,5/2\rangle$ state.
Here, the momentum of the rf photon is neglected, thus the rf pulse
does not impart momentum to the atom in the final state. In Fig.~\ref{spectra}(c)
we show three measurements corresponding to (c1), (c3) and (c5) in
Fig.~\ref{transition}. For (c1), clearly only $s=-1$ branch is
populated. For (c3), the population is slightly above the $s=1$ helicity
branch. And for (c5), there are already significant population at
$s=1$ helicity branch. In (c5) one can also identify the chiral nature
of two helicity branches: For $s=-1$ branch, most left-moving states
are dominated by $\left|\downarrow\right\rangle $ state; while for
$s=1$ branch, right-moving states are mostly dominated by $\left|\downarrow\right\rangle $
states.

The theoretical simulation of momentum-resolved rf spectroscopy has
been performed and discussed in Sec.~\ref{sp} (see, in particular, Fig.~\ref{fig3}). We note that, the
definition of momentum and rf frequency is different. These are related
by, $k_{x}=-p_{x}-k_{r}$ and $\omega=-\nu_{RF}$.

The single-particle spectrum is also measured using the technique
of spin injection spectroscopy in a spin-orbit coupled Fermi gas of
$^{6}$Li by the MIT group \cite{MIT_fermionic}. In that work, the
following four lowest hyperfine states are chosen $|3/2,-1/2\rangle$,
$|3/2,-3/2\rangle$, $|1/2,-1/2\rangle$, $|1/2,1/2\rangle$, which
are labelled as $\left|\uparrow\right\rangle _{i}$, $\left|\uparrow\right\rangle _{f}$,
$\left|\downarrow\right\rangle _{f}$, $\left|\downarrow\right\rangle _{i}$.
The Raman process couples $\left|\uparrow\right\rangle _{f}$ to $\left|\downarrow\right\rangle _{f}$
to induce SOC between these two states. For momentum-resolved rf spectroscopy,
the state $\left|\downarrow\right\rangle _{i}$ is coupled via rf
field to the state $\left|\downarrow\right\rangle _{f}$, as this
connects the first and second lowest hyperfine states. Similarly,
an atom in state $\left|\uparrow\right\rangle _{i}$ is coupled to
$\left|\uparrow\right\rangle _{f}$. Since the dispersion for initial
states $\left|\uparrow\right\rangle _{i}$ and $\left|\downarrow\right\rangle _{i}$
($\epsilon_{i}({\bf k})=\textbf{k}^{2}/2m$) are known, the spectra
of the final states, which is subject to the SOC, are obtained.

The dispersion investigated above is the simplest case for a spin-orbit
coupled system. An even richer band structure involving multiple spinful
bands separated by fully insulating gaps can arise in the presence
of a periodic lattice potential. This has been realized for Bose-Einstein
condensates by adding rf coupling between the Raman-coupled states
$\left|\uparrow\right\rangle _{f}$ and $\left|\downarrow\right\rangle _{f}$~\cite{jime2012peierls}.
Using a similar method, a spinful lattice for ultracold fermions is
created, and one can use spin-injection spectroscopy to probe the
resulting spinful band structure \cite{MIT_fermionic}, see, for example,
Fig. \ref{fig4}.

\subsection{The strongly interacting spin-orbit coupled Fermi gas}

We now consider the Femi gas where interaction cannot be neglected.
In particular, we focus on the effect of SOC on fermionic pairing.

\subsubsection{Integrated radio-frequency spectrum }

To create a strongly interacting Fermi gas with spin-orbit coupling,
first, the bias magnetic field is tuned from high magnetic field above
Feshbach resonance to a final value $B$ (which is varied) below Feshbach
resonance. Thus, Feshbach molecules are created in this process. Then,
we ramp up adiabatically the Raman coupling strength in $15$ ms from
zero to its final value $\Omega=1.5E_{r}$ with Raman detuning $\delta=0$.
The temperature of the Fermi cloud after switching on the Raman beams
is at about $0.6T_{F}$ \cite{Jing_fermionic}. The Fermi energy is
$E_{F}\simeq2.5E_{r}$ and the corresponding Fermi wavevector is $k_{F}\simeq1.6k_{r}$.
To characterize the strongly-interacting spin-orbit coupled Fermi
system, we apply a Gaussian shaped pulse of rf field with a duration
time about 400 $\mu$s and frequency $\omega$ to transfer the spin-up
fermions to an un-occupied third hyperfine state $|3\rangle=\left|F=9/2,m_{F}=-5/2\right\rangle $.

\begin{figure}
\includegraphics[width=0.6\textwidth]{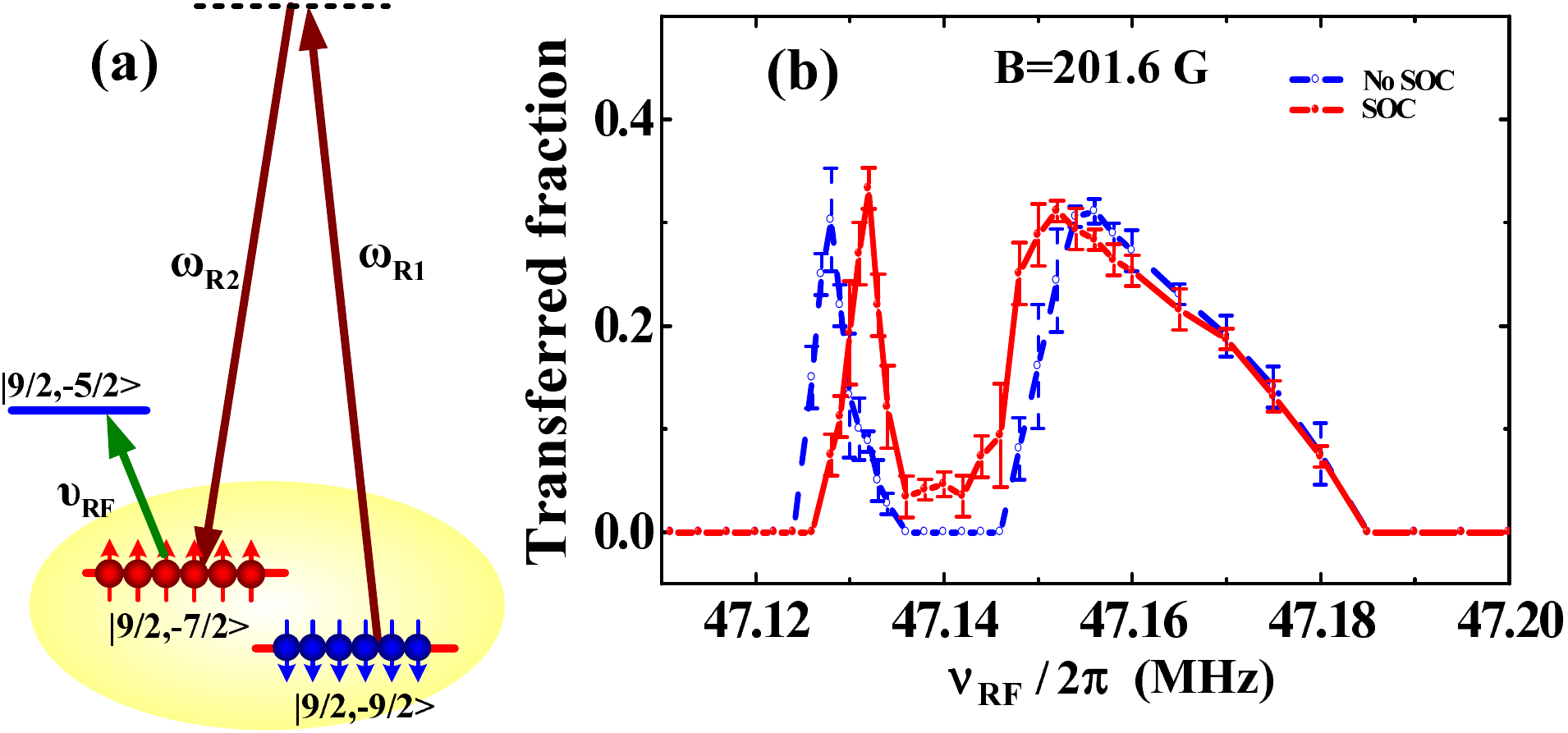}
\caption{(a) Energy level of a strongly-interacting Fermi gas of $^{40}$K
atoms with SOC. (b) The integrated rf-spectroscopy below the Feshbach
resonance (at $B=201.6$ G and $a_{s}\simeq2215.6a_{0}$, where $a_{0}$
is the Bohr radius), in the presence (solid circles) and absence (empty
circles) of the spin-orbit coupling. The Raman detuning is $\delta=0$.
The dimensionless interaction parameter $1/(k_{F}a_{s})\simeq0.66$.
The fraction is defined as $N_{-5/2}/(N_{-5/2}+N_{-7/2})$, where
$N_{-5/2}$ and $N_{-7/2}$ are obtained from the TOF absorption image.
Figure taken from Ref.~\cite{Jing_RF}.}

\label{fig27} 
\end{figure}

In Fig.~\ref{fig27}(b), we show that the integrated rf-spectrum
of an interacting Fermi gas below the Feshbach resonance, with or
without spin-orbit coupling. Here, we carefully choose the one photon
detuning of the Raman lasers to avoid shifting Feshbach resonance
by the Raman laser on the bound-to-bound transition between the ground
Feshbach molecular state and the electronically excited molecular
state. We also make sure that the single-photon process does not affect
the rf spectrum. The narrow and broad peaks in the spectrum should
be interpreted respectively as the rf-response from free atoms and
fermionic pairs. With spin-orbit coupling, we find a systematic blue
shift in the atomic response and a red shift in the pair response.
The latter is an unambiguous indication that the properties of fermionic
pairs are strongly affected by spin-orbit coupling~\cite{Jing_RF}.
The red shift of the response from the pairs may be understood from
the binding energy of pairs in the two-body limit. As mentioned below
Eq.~(\ref{HSO}), the Raman coupling may be regarded as an effective
Zeeman field. The stronger the effective Zeeman field, the smaller
the binding energy of the two-particle bound states \cite{jiang,dong1}.

\begin{figure}
\begin{centering}
\includegraphics[clip,width=0.7\textwidth]{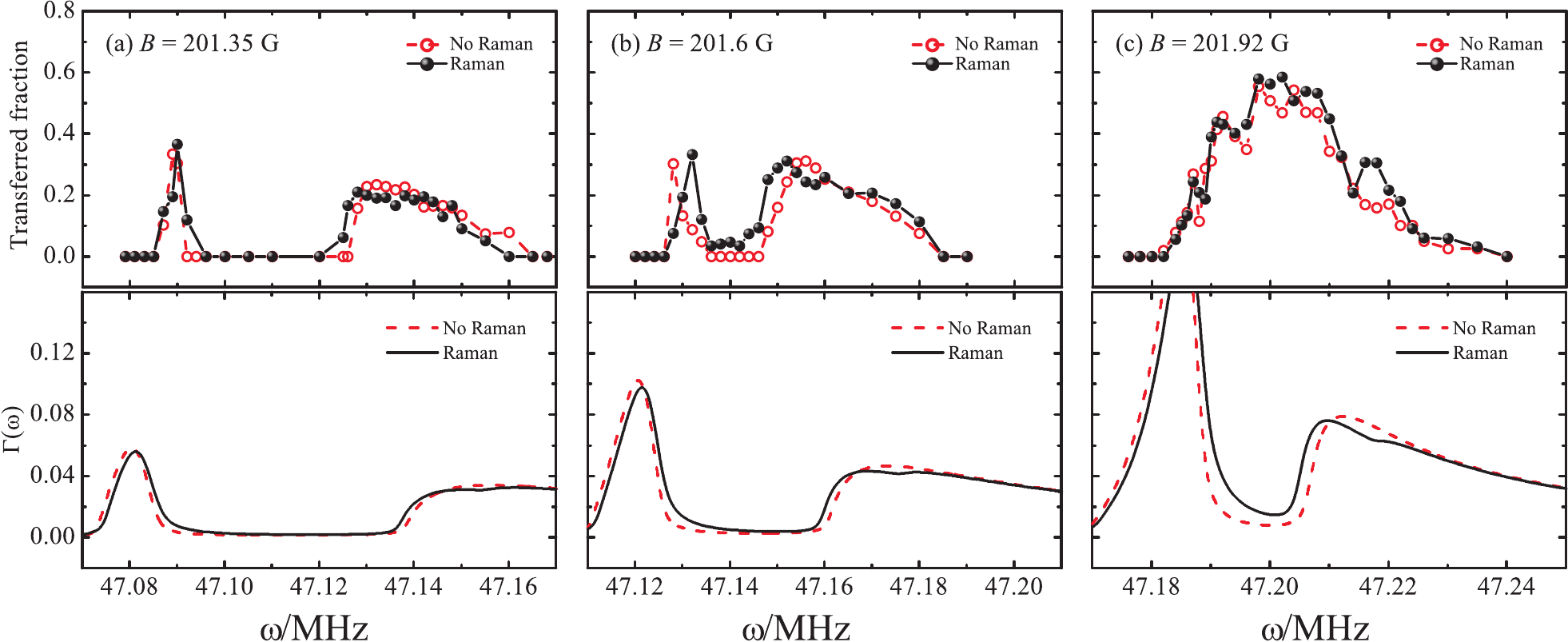} 
\par\end{centering}

\caption{ The integrated rf-spectrum for a spin-orbit coupled Fermi gas. The
red solid circles (red lines) and dark empty circles show respectively
the experimental data in the presence and absence of spin-orbit coupling
with Raman detuning $\delta=0$. The upper panel represents experimental
data and the lower panel represents the theoretical calculation. The
dimensionless interaction parameter $1/(k_{F}a_{s})$ in (a), (b),
and (c) are $0.89$, $0.66$, and $0.32$, respectively. Figure taken
from Ref.~\cite{Jing_RF}.}

\label{fig28} 
\end{figure}

In Fig.~\ref{fig28}, we compare the experimentally measured rf-spectrum
with the many-body \textit{T}-matrix prediction, which is obtained
within the pseudogap approximation \cite{Jing_RF} (see the discussion in Sec.~\ref{tm} and \ref{rf}). In the calculation,
at a qualitative level, we do not consider the trap effect and take
the relevant experimental parameters at the trap center. Otherwise,
there are no adjustable free parameters used in the theoretical calculations.
As shown in Fig.~\ref{fig28}, we find a qualitative agreement between
theory and experiment, both of which show the red shift of the response
from fermionic pairs. Note that, near Feshbach resonances our many-body
pseudogap theory is only qualitatively reliable. It cannot explain
well the separation of atomic and pair peaks in the observed integrated
rf-spectrum. More seriously, it fails to take into account properly
the strong interactions between atoms and pairs.

\subsubsection{Coherent formation of Feshbach molecules by spin-orbit coupling}

In a recent experiment, we studied the formation of Feshbach molecules
from an initially spin-polarized Fermi gas \cite{Jing_FeshbachMolecule}.
For simplicity, let us consider two atoms both prepared in the $|\downarrow\rangle$
state. We label this state as $|\downarrow\rangle_{1}|\downarrow\rangle_{2}$,
which is obviously a spin-symmetric state. Under the $s$-wave interaction,
the Feshbach molecule is spin-antisymmetric singlet state. Hence to
form Feshbach molecule from this initial state, a spin-antisymmetric
coupling is required. To this end, we apply two Raman laser beams
that effectively couples the hyperfine states $|\uparrow\rangle$
and $|\downarrow\rangle$. The effective Hamiltonian arising from
the Raman beams can be written as $H_{R}=H_{R}^{(1)}+H_{R}^{(2)}$
with 
\begin{eqnarray}
H_{R}^{(j)}=-\frac{\delta}{2}\sigma_{z}^{(j)}+\frac{\Omega}{2}e^{2ik_{0}x_{j}}\sigma_{+}^{(j)}+\frac{\Omega}{2}e^{-2ik_{0}x_{j}}\sigma_{-}^{(j)},\label{hr}
\end{eqnarray}
for $j=1,2$. Here we have $\sigma_{z}^{(j)}=(|\uparrow\rangle_{j}\langle\uparrow|-|\uparrow\rangle_{j}\langle\uparrow|)/2$,
$\sigma_{+}^{(j)}=|\uparrow\rangle_{j}\langle\downarrow|$, and $\sigma_{-}^{(j)}=\sigma_{+}^{(j)\dagger}$.
In Eq.~(\ref{hr}), $\Omega$ is the Raman coupling intensity, $x_{j}$
is the position of the $j$-th atom in the $x$-direction, and $k_{0}=k_{r}\sin(\theta/2)$,
with $k_{r}$ the single-photon recoil momentum and $\theta$ the
angle between the two Raman beams. It is apparent that $H_{R}$ can
be written as $H_{R}=H_{R}^{(+)}+H_{R}^{(-)}$ with 
\begin{eqnarray}
H_{R}^{(\pm)}=\frac{\Omega}{4}\left(e^{i2k_{0}x_{1}}\pm e^{i2k_{0}x_{2}}\right)\left(\sigma_{+}^{(1)}\pm\sigma_{+}^{(2)}\right)+h.c..
\end{eqnarray}
Obviously, $H_{R}^{(-)}$ and $H_{R}^{(+)}$ are \textit{anti-symmetric}
and \textit{symmetric} under the exchange of the hyperfine state of
the two atoms, respectively. Therefore, only $H_{R}^{(-)}$ can create
spin-antisymmetric state out of the initially polarized state $|\downarrow\rangle_{1}|\downarrow\rangle_{2}$,
and as a consequence make the formation of Feshbach molecule possible.
When the two Raman beams propagate along the same direction, i.e.,
$\theta=0$, we have $k_{0}=0$ and thus $H_{R}^{(-)}=0$. Then the
Feshbach molecule cannot be produced from the polarized atoms. In
contrast, when the angle $\theta$ between the two Raman beams is
non-zero, we have $H_{R}^{(-)}\neq0$ and Feshbach molecule can thus
be created.

\begin{figure}
\begin{centering}
\includegraphics[width=0.5\textwidth]{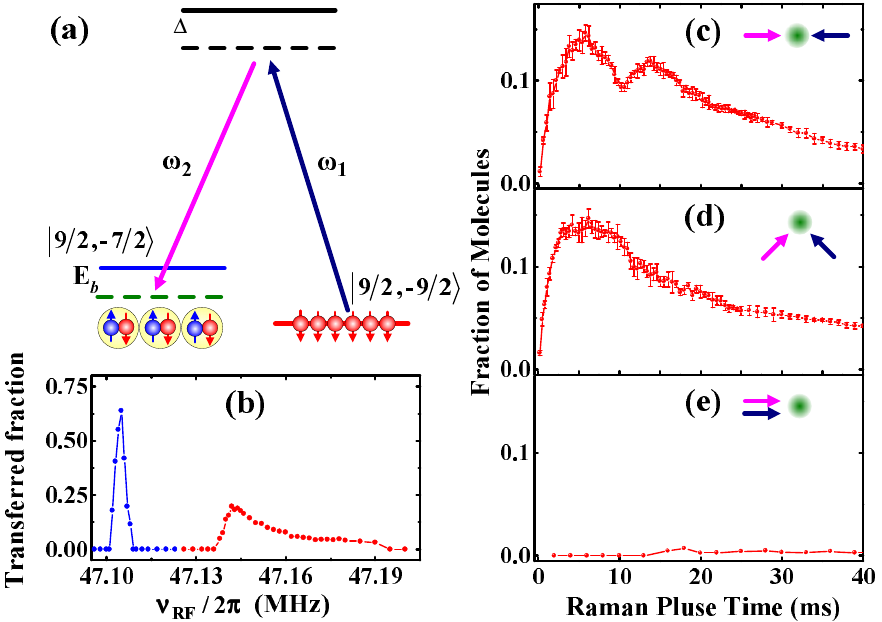} 
\par\end{centering}

\caption{Energy level diagram and spin-orbit coupling induced Feshbach molecules.
\textbf{(a)} Schematic diagram of the energy levels. A pair of Raman
lasers couples spin polarized state $|9/2,-9/2\rangle$ to Feshbach
molecules in Fermi gases $^{40}$K. \textbf{(b)} Radio-frequency spectrum
$|9/2,-7/2\rangle$ to $|9/2,-5/2\rangle$ transition applied to a
mixture of Feshbach molecules and scattering atoms in $|9/2,-7/2\rangle$.
\textbf{(c-e)} The population of Feshbach molecules detected by the
rf pulse as a function of duration time of the Raman pulse. The angle
of two Raman beams is $\theta=180^{\circ}$ (c), $\theta=90^{\circ}$
(d) and $\theta=0^{\circ}$ (e). The Raman coupling strength is $\Omega=1.3E_{r}$
and the two-photon Raman detuning is $\delta=-E_{b}=-3.59E_{\text{r}}$.
Figure taken from Ref.~\cite{Jing_FeshbachMolecule}. }

\label{fig29} 
\end{figure}

This picture is exactly confirmed by our data. Our experiment is performed
with the spin polarized $^{40}$K gas in $|F,m_{{\rm F}}\rangle=|9/2,-9/2\rangle$
state, at $201.4$ G, below the Feshbach resonance located at $202.1$
G, which corresponds to a binding energy of $E_{b}=2\pi\times30$
kHz (corresponding to 3.59$E_{\text{r}}$) for the Feshbach molecules
and $1/(k_{\text{F}}a_{\text{s}})\approx0.92$ for our typical density.
After applying the Raman lasers for certain duration time, we turn
off the Raman lasers and measure the population of Feshbach molecule
and atoms in $|9/2,-7/2\rangle$ state with an rf pulse. This rf field
drives a transition from $|9/2,-7/2\rangle$ to $|9/2,-5/2\rangle$.
For a mixture of $|9/2,-7/2\rangle$ and Feshbach molecules, as a
function of rf frequency $\nu_{\text{RF}}$, we find two peaks in
the population of $|9/2,-5/2\rangle$, as shown in Fig.~\ref{fig29}(b).
The first peak (blue curve) is attributed to free atom-atom transition
and the second peak (red curve) is attributed to molecule-atom transition.
Thus, in the following, we set $\nu_{\text{RF}}/2\pi$ to $47.14$
MHz to measure Feshbach molecules.

When the two-photon Raman detuning $\delta$ is set to $\delta=-E_{b}=-3.59E_{\text{r}}$,
as shown in Fig.~\ref{fig29}(a), we measure the population of Feshbach
molecule as a function of duration time for three different angles,
$\theta=180^{\circ}$, $\theta=90^{\circ}$, and $\theta=0^{\circ}$,
as shown in Fig. \ref{fig29}(c), (d) and (e). We find for $\theta=180^{\circ}$,
Feshbach molecules are created by Raman process and the coherent Rabi
oscillation between atom-molecule can be seen clearly. For $\theta=90^{\circ}$,
production of Feshbach molecules is reduced a little bit and the atom-molecule
Rabi oscillation becomes invisible. For $\theta=0^{\circ}$, no Feshbach
molecule is created even up to $40$ ms, which means the transition
between Feshbach molecules and a fully polarized state is prohibited
if Raman process imparts no momentum transfer, i.e., no SOC.

In a related work, the NIST group recently carried out an experiment
in which they swept a magnetic field on the BEC side of the Feshbach
resonance \cite{NIST-fermionic}. It is shown that the number of remaining
atoms exhibits a dip as a function of the magnetic field strength.
This dip represents the loss of atom due to the formation of the Feshbach
molecules. The position of the dip moves towards the lower field (to
the BEC limit) as the Raman detuning $\delta$ is increased. The phenomenon
can also be explained by the fact that the effective Zeeman field
(in this case, the detuning $\delta$) disfavors the formation of
bound molecules. Hence at larger $\delta$, a larger $a_{s}^{-1}$
(i.e., stronger attraction between unlike spins) is required to form
molecules \cite{dong1}. This is in full agreement with the theoretical discussion concerning the two-body physics for the equal-weight Rashba-Dresselhaus SOC presented in Sec.~\ref{2b}.

\section{Conclusion}

In this chapter, we described the properties of a spin-orbit coupled
Fermi gas. Recent progress, both theoretical and experimental, were
reviewed. As we have shown, spin-orbit coupled Fermi gases possess
a variety of intriguing properties. The diverse configuration of the
synthetic Gauge field and the extraordinary controllability of atomic
systems provide new opportunities to explore quantum many-body systems
and quantum topological matter. We note that this article by no means
is a comprehensive review. For example, we only focused on a continuum
system and neglected many interesting theoretical works on lattice
systems.

So far only one particular scheme (equal-weight Rashba-Dresselhaus)
of SOC has been realized in the experiment, which is based on the
Raman transition between two hyperfine ground states of the atom.
One drawback of the laser-based SOC generating scheme is that the
application of the laser fields inevitably induce additional heating.
For certain atoms, this heating may be severe enough to prevent the
system from becoming quantum degenerate. Furthermore, many interesting
physics requires a strong interaction strength which is induced by
applying a fairly strong magnetic field via the Feshbach resonance.
Due to a decoupling of the nuclear and electronic spins in large magnetic
fields, Raman coupling efficiency quickly reduces with increasing
of the magnetic field \cite{Wei2013}. This poses another severe experimental
challenge. Due to these reasons, no superfluid spin-orbit coupled
Fermi gas has been realized yet. As a result, many interesting theoretical
proposals (e.g., topological superfluids, Majorna fermion, etc.) are still waiting to be experimentally realized. Nevertheless, we want to remark that despite the relatively high temperature of the experimental system, the effects of SOC have been clearly revealed in single-particle properties as well as the two- and many-body properties on the BEC side of the resonance, as such properties are not easily washed out by finite temperature effects. Very recently,
a scheme to synthesize a general SOC is proposed, which is based on
purely magnetic field pulses and involves no laser fields \cite{mag,mag1}.
Whether this scheme will overcome the problems mentioned above remains
to be seen.

\begin{acknowledgments}
We are deeply appreciative for discussions with Congjun Wu, Wei Yi,
Hui Zhai, Chuanwei Zhang, and many others; as well as the students
and postdocs in our groups: Lin Dong, Lei Jiang, Shi-Guo Peng, Pengjun
Wang, and Zhengkun Fu. JZ is supported by NFRP-China (Grant No. 2011CB921601),
NSFC Project for Excellent Research Team (Grant No. 61121064), NSFC
(Grant No. 11234008), Doctoral Program Foundation of Ministry of Education
China (Grant No. 20111401130001). XJL and HH are supported by the
ARC DP0984637 and DP0984522. HP is supported by the NSF, the DARPA
OLE program and the Welch Foundation (Grant No. C-1669). \end{acknowledgments}


\begin{thebibliography}{100}
\bibitem{note}The quantum version of Maxwell's theory is described by quantum electrodynamics, which is a dynamic gauge theory in the sense that the gauge symmetries generate dynamics by giving rise to interaction couplings between particles mediated by photons. The artificial gauge fields for cold atoms, by contrast, are static, i.e., not dynamically coupled to the atomic fields.
\bibitem{footnote1}Strictly speaking, we set the chemical potential
to be the minimum energy in the two-particle continuum $2E_{{\rm min}}$,
where $E_{{\rm min}}$ is the single-particle ground state energy,
such that the Fermi surface shrinks and vanishes. Thus, the Fermi
distribution functions in the vertex function disappear. Note that,
in the vertex function the energy $\omega$ is measured with respect
to the chemical potential. This procedure is exactly identical to
what we have taken in the text by setting $\mu=0$ and measuring the
energy $\omega$ from zero.
\bibitem{footnote2}Another equivalent form of Rashba SOC, used extensively
in the literature, is $V_{\textrm{SO}}=\lambda(\hat{k}_{x}\hat{\sigma}_{x}+\hat{k}_{y}\hat{\sigma}_{y})$.
These are related by a spin-rotation.
\end{thebibliography}
\end{document}